\documentclass{aa}  

\usepackage{graphicx}
\usepackage{siunitx}
\usepackage{ctable}
\usepackage{textcomp}
\usepackage{txfonts}
\usepackage[colorlinks=true, linkcolor=blue, citecolor=blue, urlcolor=blue]{hyperref}

\newcommand{\Pa}{Pa$\rm \alpha$}
\newcommand{\Brg}{Br$\rm \gamma$}
\newcommand{\Brd}{Br$\rm \delta$}
\newcommand{\sisix}{[SiVI]}
\newcommand{\hd}{H$_2~$}
\newcommand{\hdone}{$\rm H_2$1-0S(1)}
\newcommand{\hdtwo}{$\rm H_2$1-0S(2)}
\newcommand{\hdthree}{$\rm H_2$1-0S(3)}
\newcommand{\hdfour}{$\rm H_2$1-0S(4)}
\newcommand{\hdfive}{$\rm H_2$1-0S(5)}

\newcommand{\flux}{W/$\rm m^2$/\textmu m.$\rm km~s^{-1}$}
\newcommand{\noise}{W/$\rm m^2$/\textmu m}
\newcommand{\barolo}{$\rm^{3D}$BAROLO}
\newcommand{\kms}{$\rm km~s^{-1}$}
\newcommand{\ergs}{$\rm erg~s^{-1}$}
\newcommand{\msun}{$\rm M_{\odot}$}
\newcommand{\sfr}{$\rm M_{\odot}~yr^{-1}$}

\defcitealias{ramosalmeida2022}{RA22}
\defcitealias{speranza2024}{SP24}
\newcommand{\RA}{\citetalias{ramosalmeida2022}}
\newcommand{\SP}{\citetalias{speranza2024}}

\begin{document} 

\title{Unveiling the warm molecular outflow component of type-2 quasars with SINFONI}

   \author{M.V. Zanchettin\inst{\ref{inst1}, \ref{inst2a}, \ref{inst2},}\thanks{Corresponding author, \email{mazanch@sissa.it, maria.zanchettin@inaf.it}}
   \and C. Ramos Almeida\inst{\ref{inst3}, \ref{inst4}}
   \and A. Audibert\inst{\ref{inst3}, \ref{inst4}}
   \and J. A. Acosta-Pulido\inst{\ref{inst3}, \ref{inst4}}   
    \and P. H. Cezar\inst{\ref{inst3}, \ref{inst4}}
    %\and C. Harrison \inst{\ref{inst8}}
    \and E. Hicks \inst{\ref{inst9}, \ref{inst10}, \ref{inst11}}
    \and A. Lapi \inst{\ref{inst1}, \ref{inst5}, \ref{inst6}, \ref{inst7}}
    \and J. Mullaney \inst{\ref{inst12}}
   }
   
\institute{SISSA, Via Bonomea 265, I-34136 Trieste, Italy\label{inst1} \and INAF – Osservatorio Astrofisico di Arcetri, Largo E. Fermi 5, 50125 Firenze, Italy\label{inst2a} \and INAF Osservatorio Astronomico di Trieste, via G.B. Tiepolo 11, 34143 Trieste, Italy\label{inst2} \and Instituto de Astrofísica de Canarias, Calle Vía Láctea, s/n, 38205 La Laguna, Tenerife, Spain\label{inst3} \and Departamento de Astrofísica, Universidad de La Laguna, 38206 La Laguna, Tenerife, Spain\label{inst4}  
\and Department of Physics \& Astronomy, University of Alaska Anchorage, Anchorage, AK 99508-4664, USA \label{inst9} \and Department of Physics, University of Alaska, Fairbanks, Alaska 99775-5920, USA \label{inst10}
\and  Department of Physics and Astronomy, The University of Texas at San Antonio, 1 UTSA Circle, San Antonio, Texas, 78249-0600, USA \label{inst11}
\and IFPU - Institute for fundamental physics of the Universe, Via Beirut 2, 34014 Trieste, Italy \label{inst5} \and INAF, Istituto di Radioastronomia, Italian ARC, Via Piero Gobetti 101, I-40129 Bologna, Italy\label{inst6} \and INFN, Sezione di Trieste, via Valerio 2, Trieste I-34127, Italy\label{inst7} \and Astrophysics Research Cluster, School of Mathematical and Physical Sciences, University of Sheffield, Sheffield S3 7RH, UK \label{inst12}
} 

  \date{Received November 29, 2024; accepted February 17, 2025}

  \abstract{We present seeing-limited (\SI{0.8}{\arcsecond}) near-infrared integral field spectroscopy data of the type-2 quasars (QSO2s) SDSS J135646.10+102609.0 (J1356) and SDSS J143029.89+133912.1 (J1430, the Teacup), both belonging to the Quasar Feedback (QSOFEED) sample. The nuclear K-band spectra (1.95-2.45 \textmu m) of these radio-quiet QSO2s reveal several \hd emission lines, indicative of the presence of a warm molecular gas reservoir (T$\geq$1000 K). We measure nuclear masses of $\rm M_{H_2}$=5.9, 4.1, and $1.5 \times 10^3 $ \msun~in the inner \SI{0.8}{\arcsecond} diameter region of the Teacup ($\sim$1.3 kpc), J1356 north (J1356N), and south nuclei ($\sim$1.8 kpc), respectively. The total warm \hd mass budget is $\sim 4.5 \times 10^4$ \msun~in the Teacup and $\sim 1.3 \times 10^4$ \msun~in J1356N, implying warm-to-cold molecular gas ratios of $10^{-6}$. The warm molecular gas kinematics, traced with the \hdone~and S(2) emission lines, is consistent with that of the cold molecular phase, traced by ALMA CO emission at higher angular resolution (\SI{0.2}{\arcsecond} and \SI{0.6}{\arcsecond}). In J1430, we detect the blue- and red-shifted sides of a compact warm molecular outflow extending up to 1.9 kpc and with velocities of 450 \kms. In J1356 only the red-shifted side is detected, with a radius of up to 2.0 kpc and velocity of 370 \kms. The outflow masses are 2.6 and 1.5$\times 10^3$ \msun~for the Teacup and J1356N, and the warm-to-cold gas ratios in the outflows are 0.8 and 1 $\times 10^{-4}$, implying that the cold molecular phase dominates the mass budget. We measure warm molecular mass outflow rates of 6.2 and 2.9 $\times 10^{-4}$ \sfr~for the Teacup and J1356N, which are approximately 0.001\% of the total mass outflow rate (ionized + cold and warm molecular). We find an enhancement of velocity dispersion in the \hdone~residual dispersion map of the Teacup, both along and perpendicular to the compact radio jet direction. This enhanced turbulence can be reproduced by simulations of jet-ISM interactions.}

 %\abstract
  {} %leave it empty if necessary  

 \keywords{Galaxies: active -- Galaxies: ISM -- Galaxies: evolution -- Galaxies: nuclei -- Individual: Teacup}

   \maketitle
%
%-------------------------------------------------------------------

\section{Introduction}

Strong observational evidence demonstrated that a tight relation exists between the properties of accreting super massive black holes (SMBHs) and those of their hosting galaxies. Scaling relations have been identified among the SMBH mass and host galaxy properties, galaxy luminosity, bulge mass, velocity dispersion, and total stellar mass \citep[e.g.][]{marconihunt2003, haringrix2004, gultekin2009, kormendyho2013}. 
These scaling relations are due to the coevolution of accreting SMBHs (i.e. active galactic nuclei; AGN) and the host galaxy. However, the mechanisms driving this coevolution are not yet fully understood.
The AGN can power strong winds and jets impacting on the galaxy interstellar medium (ISM), altering both further star formation (SF) and nuclear gas accretion \citep[see][for a recent review]{harrisonramos2024}. SMBH growth and nuclear activity are then stopped, until new cold gas replenishes the nucleus, thus starting a new AGN phase, which gives rise to the feeding and feedback cycle \citep[e.g.][]{garcia-burillo2021, garcia-burillo2024}. 
Indeed, cosmological simulations require AGN feedback to regulate galaxy growth and reproduce the observed properties of massive galaxies \citep[e.g.][]{schaye2015,dubois2016,nelson2018}.  
Hence, the interplay between AGN-driven outflows and the host galaxy ISM lies at the core of the processes governing the coevolution of SMBHs and galaxies.

The primary feedback mechanisms are known as the quasar or radiative mode and the radio or kinetic mode. The quasar mode is believed to operate in AGN with high accretion rates, with the AGN energy producing gas outflows \citep{fabian2012}, while the radio mode is typically linked to powerful radio galaxies with lower accretion rates, where jets mechanically compress and accelerate the gas \citep{mcnamaranulsen2007}. However, the separation between the two feedback mechanisms is overly simplistic, as both modes can occur simultaneously.
Not only powerful jets in radio-loud objects can accelerate powerful outflows \citep[e.g.][]{nesvadba2008,vayner2017,colomapuga2023,colomapuga2024}, but outflows can also be accelerated by low power compact radio jets
in galaxies typically classified as radio-quiet \citep{combes2013, garcia-burillo2014, harrison2015, morganti2015,jarvis2019,audibert2019,audibert2023,venturi2021,girdhar2022}.
Moreover, AGN-driven outflows are almost ubiquitous and occur on a wide range of physical scales \citep{fiore2017, lutz2020}. 

The ISM consists of a mixture of different gas phases, from the cold molecular and atomic to the warm ionized and hot X-ray emitting gas, and so are the outflows \citep{cicone2018, harrisonramos2024}.
However, the relationships between the different gas phases involved in the outflows, their relative weight, and impact on the galaxy ISM are largely unconstrained \citep{bischetti2019b, fluetsch2019}. To date, there are only a few sources for which multiphase and multiscale outflows have been constrained simultaneously \citep[e.g.][]{garcia-burillo2014,cresci2015,ramosalmeida2017,ramosalmeida2019,venturi2018,alonso-herrero2018,garcia-burillo2019,rosario2019,shimizu2019,feruglio2020,garcia-bernete2021,speranza2022,zanchettin2023}.
In some cases, molecular and ionized winds have similar velocities and are nearly co-spatial \citep{feruglio2018, alonso-herrero, zanchettin2021}, while more commonly, AGN show ionized winds that are faster than molecular ones \citep[see][and references therein]{veilleux2020}.

Among the different gas phases involved in galactic outflows, the molecular phase is of particular significance, as it has been shown to dominate the mass of the outflow \citep{feruglio2010,rupke2013,cicone2014,carniani2015,fiore2017,fluetsch2021,speranza2024}.
Furthermore, since the hydrogen molecule $\rm H_2$ is the fuel required to form stars and feed the SMBH, the impact of the outflows on this gaseous phase could really affect how systems evolve. Thanks to the synergy of the Atacama Large Millimeter/submillimeter Array (ALMA) with near- and mid-infrared integral observations such as VLT/SINFONI and the James Webb Space Telescope (JWST), it is now possible to probe the cold (CO) and warm (H$_2$) molecular gas phases
\citep[e.g.][]{combes2014,smajic2015,pereirasantaella2016,ramosalmeida2017,ramosalmeida2019,shimizu2019,ramosalmeida2022,alonsoherrero2023,liuphangs2023,zanchettin2024}.
In the near-infrared, we can find both rotational and vibrational \hd lines that trace molecular gas at temperatures of $\sim$1000 K, whereas in the mid-infrared, the rotational \hd lines trace gas at hundreds of kelvin. Before JWST, only a few works studied warm molecular outflows using near-infrared observations \citep[e.g.][]{rupke2013,ramosalmeida2017,ramosalmeida2019,speranza2022,riffel2023}, and only in a few sources were warm molecular outflows detected \citep{rupke2013,ramosalmeida2019,riffel2023}. In fact, some AGN with reported ionized and CO outflows do not show a warm molecular gas counterpart \citep{ramosalmeida2017,riffel2023}. \citet{riffel2023} found that 94\% of their sample of 33 low-luminosity AGN present ionized outflows, while warm outflows were observed for 76\% of them.
Here we report deep SINFONI observations of two type-2 quasars with the aim of investigating their warm molecular gas content and kinematics. 

The paper is organized as follows. Section \ref{sec:targets} presents the main properties of the targets. Section \ref{sec:obs} describes the VLT/SINFONI observational set-up and data reduction. Section \ref{sec:results} presents the main results, in particular the nuclear spectra, the warm molecular gas properties, and the \hd kinematics.
In Section \ref{sec:discussion} we discuss our results, and in Section \ref{sec:conclusions} we summarize the main results and present our conclusions.
Throughout this work, we assume the following cosmology: $H_0$ = 70.0  $\rm km \ s^{-1} \, Mpc^{-1}$ , $\Omega_{m}$ = 0.3, and $\Omega_{\Lambda}$ = 0.7. We adopt the same cosmology and redshifts as those used by \citet{speranza2024}, hereafter \citetalias{speranza2024}, of z = 0.1232 and 0.0851 for J1356 and the Teacup. The spatial scales are 2.213 and 1.597 kpc/\SI{}{\arcsecond}, respectively.

\section{The targets}\label{sec:targets}

The two targets studied here, SDSS J135646.10+102609.0 (J1356) and SDSS J143029.89+133912.1 (J1430, the Teacup), are drawn from a complete sample of 48 type-2 quasars (QSO2s; that is, obscured quasars), the Quasar Feedback (\href{https://research.iac.es/galeria/cristina.ramos.almeida/qsofeed/}{QSOFEED}) sample \citep[see][]{ramosalmeida2022,Pierce23,bessiere2024}.  
The QSO2s in the parent sample are luminous ($\rm L_{[O III]} > 10^{8.5}L_{\odot}$) 
and nearby (z $<$ 0.14), and hosted in massive galaxies ($\rm M_{*}\sim 10^{11} M_{\odot}$). 

Our two targets have been studied across a wide range of wavelengths, showing evidence of outflowing gas in the cold molecular and warm ionized gas phases.
\citet{ramosalmeida2022}, hereafter \citetalias{ramosalmeida2022}, analyzed ALMA CO(2-1) data at $\sim$ \SI{0.2}{\arcsecond} resolution (370 pc at z$\sim$0.1), finding diverse CO morphologies.
The molecular mass outflow rates reported in RA22 for J1356 and J1430, of 7.8 and 15.8 \sfr, are lower than those expected from observational empirical relations \citep{cicone2014,fiore2017, fluetsch2019}, considering their bolometric luminosities of $10^{45.54}$ and $10^{45.83}$ \ergs~(\SP), suggesting that AGN luminosity alone does not guarantee the presence of powerful outflows. The ionized outflows of J1356 and J1430 have been studied using data from different facilities operating in both optical \citep{greene2012,keel2012,harrison2014, fischer2018} and near-infrared (NIR) \citep{ramosalmeida2017,ramosalmeida2019}. 
\SP~analyzed Integral Field Spectroscopy (IFS) observations of the [O III]$\lambdaup$5007\AA\ emission line from the MEGARA 
%(Multi-Espectrógrafo en GTC de Alta Resolución para Astronomía) 
instrument on the 10.4 m GTC. 
%(Gran Telescopio CANARIAS).
In these two QSO2s \SP~reported the detection of the approaching and receding sides of the outflows.
In the case of J1430 and J1356, which have extended radio emission despite being radio-quiet\footnote{Less than 7\% of these two and other nearby QSO2s radio emission is associated with star formation, with non thermal AGN emission being responsible for the bulk of radio emission \citep{jarvis2019}.}, the ionized outflows are well aligned with the radio, suggesting that low-power jets ($\rm P_{jet} \sim 10^{43-44}$ \ergs) could be compressing and accelerating the ionized gas, although it is also possible, especially in the case of J1356, that the extended radio emission is due to shocks induced by the ionized outflow as it shocks the surrounding ISM \citep{Fischer2023}. In the case of Teacup, \citet{audibert2023} studied the jet-ISM interaction using data from the Very Large Array (VLA) and ALMA CO, and they concluded, after comparing with tailored simulations of jet-ISM interactions, that the jet, almost coplanar with the CO disc, shocks the ISM and drives the molecular gas outflow. 
%The total molecular gas content and global CO excitation of the sample were partially studied by \citet{molyneux2024} and \citet{jarvis2020}, combining the APEX and ALMA ACA observations.\citet{molyneux2024} reported a median line ratio $r_{21}$ (that is, $r_{21} = L'_{CO(2-1)}/L'_{CO(1-0)}$) equal to 1.06 and low excitation in the CO (6-5) and CO (7-6) emission lines. These findings further support the idea that the quasar feedback does not have a significant instantaneous impact on the global molecular gas content. However, impact on gas excitation occurs on local scales (i.e., kiloparsec scales), as shown by \citet{audibert2023} for the case of the Teacup galaxy, which cannot be probed by their APEX observations (\SI{10}{\arcsecond}-\SI{30}{\arcsecond} spatial resolution).

 \subsection{The Teacup}

J1430, known as the Teacup \citep{keel2012}, has been the subject of numerous studies across multiple wavelength ranges, including X-rays \citep{lansbury2018}, optical \citep{keel2012, harrison2015, villarmartina2018, venturi2023}, near-infrared \citep{ramosalmeida2017}, submillimeter \citep[\RA,][]{audibert2023}, and radio \citep{harrison2015,jarvis2019}.
The host galaxy is a bulge-dominated system that shows evidence of past interactions, such as concentric shell structures and nuclear dust lanes \citep{keel2015}.
\citet{ramosalmeida2017} detected ionized and coronal nuclear outflows lacking a warm molecular gas counterpart (i.e., broad component in the \hd line profile) based on the analysis of near-infrared IFS data obtained with VLT/SINFONI.
Among the five QSO2s with detected cold molecular CO(2-1) outflows reported by RA22, the Teacup exhibits the most peculiar CO(2-1) morphology and disturbed kinematics. The CO outflow is compact ($\rm r_{out} \sim 0.5$ kpc) and contains a cold molecular mass of 1.4 $\rm \times 10^7$ \msun, corresponding to a cold molecular mass outflow rate of 15.8 \sfr.
By combining ALMA CO(2–1) data with additional archival ALMA CO(3–2) observations, \citet{audibert2023} identified evidence of jet-induced molecular gas excitation and turbulence perpendicular to the compact radio jet (spanning $\sim$0.8 kpc, PA = \ang{60}) observed in VLA data at \SI{0.25}{\arcsecond} resolution \citep{harrison2015}.
\citet{venturi2023} observed an enhanced ionized gas velocity dispersion perpendicular to the jet head using optical IFS data from VLT/MUSE. Furthermore, \SP, analyzing optical data from GTC/MEGARA, reported ionized outflow sizes of 3.1 and 3.7 kpc for the receding and approaching outflow sides, and outflow mass rates of 1.6-3.3 \sfr~depending on the tracer used to estimate the electron density.

\subsection{J1356}

J1356 is hosted in a merger system that shows distorted optical morphology. The two nuclei (North and South, hereafter N and S) are separated by \SI{1.1}{\arcsecond} (2.4 kpc). The N nucleus is the QSO2, whose host galaxy is the dominant member of the merger and the brightest in both optical \citep{greene2012, harrison2014} and molecular gas (\citealt{sun2014}; \RA). 
Although J1356 is a massive eary-type galaxy (ETG) with a stellar population dominated by old stars \citep{greene2009}, the star formation rate (SFR) of 69 \sfr~\RA~is probably a consequence of the ongoing merger. In addition to N and S nuclei, a huge stellar feature named the western arm (W arm) was identified from the ALMA and HST morphologies. This merging component represents $\sim$32\% of the total molecular mass in the system (\citealt{sun2014}; \RA).
J1356 was reported to host multiphase AGN driven outflows. \RA~detected a compact CO outflow in the inner $\sim$ 0.4 kpc with a mass outflow rate of approximately 8 \sfr. Furthermore, \SP~reported ionized outflow sizes of 6.8 and 12.6 kpc for the receding and approaching outflow sides, and outflow mass rates of 1.2-6.1 \sfr~depending on the tracer used to estimate the electron density.

\begin{figure}[ht]
\resizebox{\hsize}{!}{\includegraphics{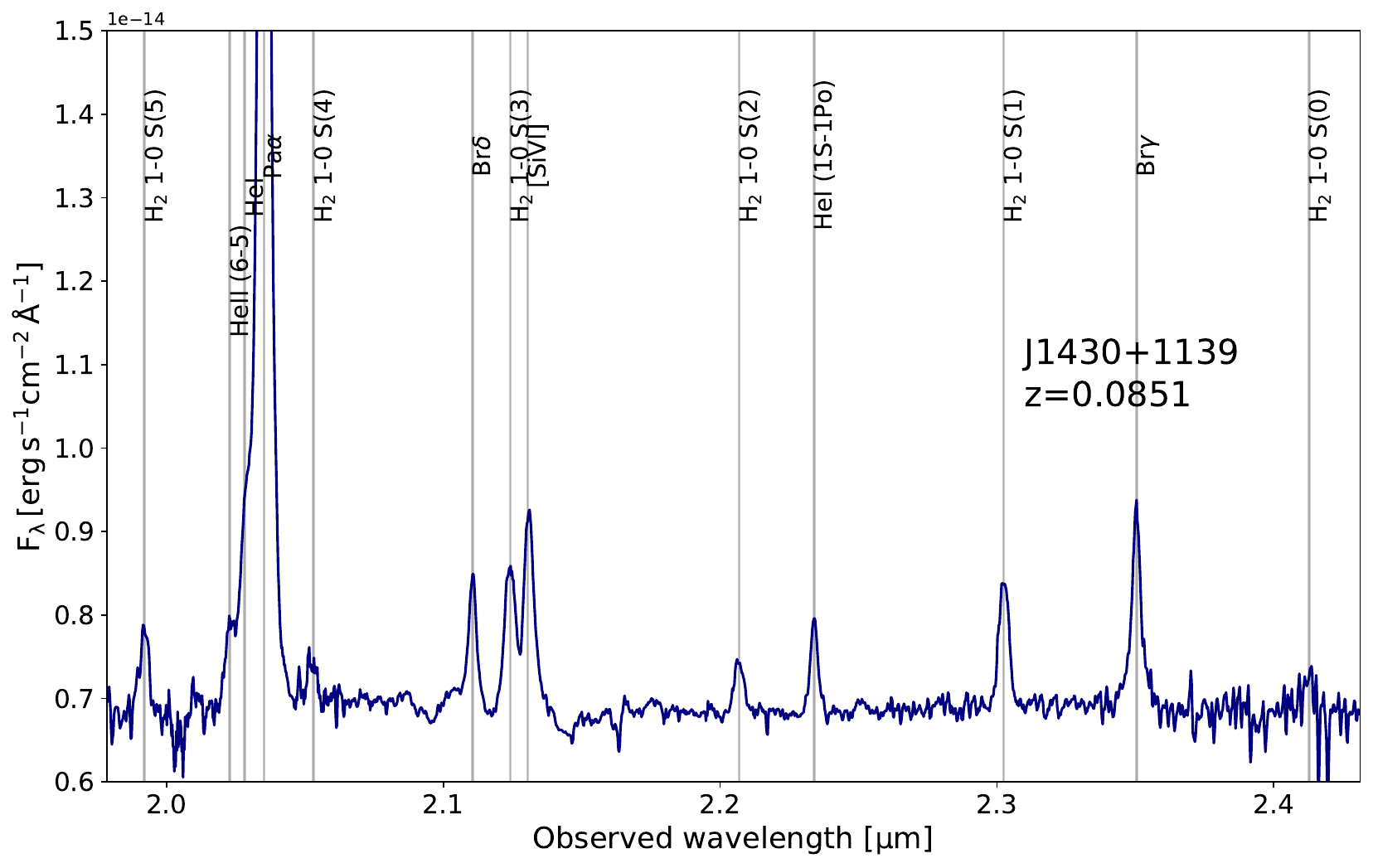}}
\resizebox{\hsize}{!}{\includegraphics{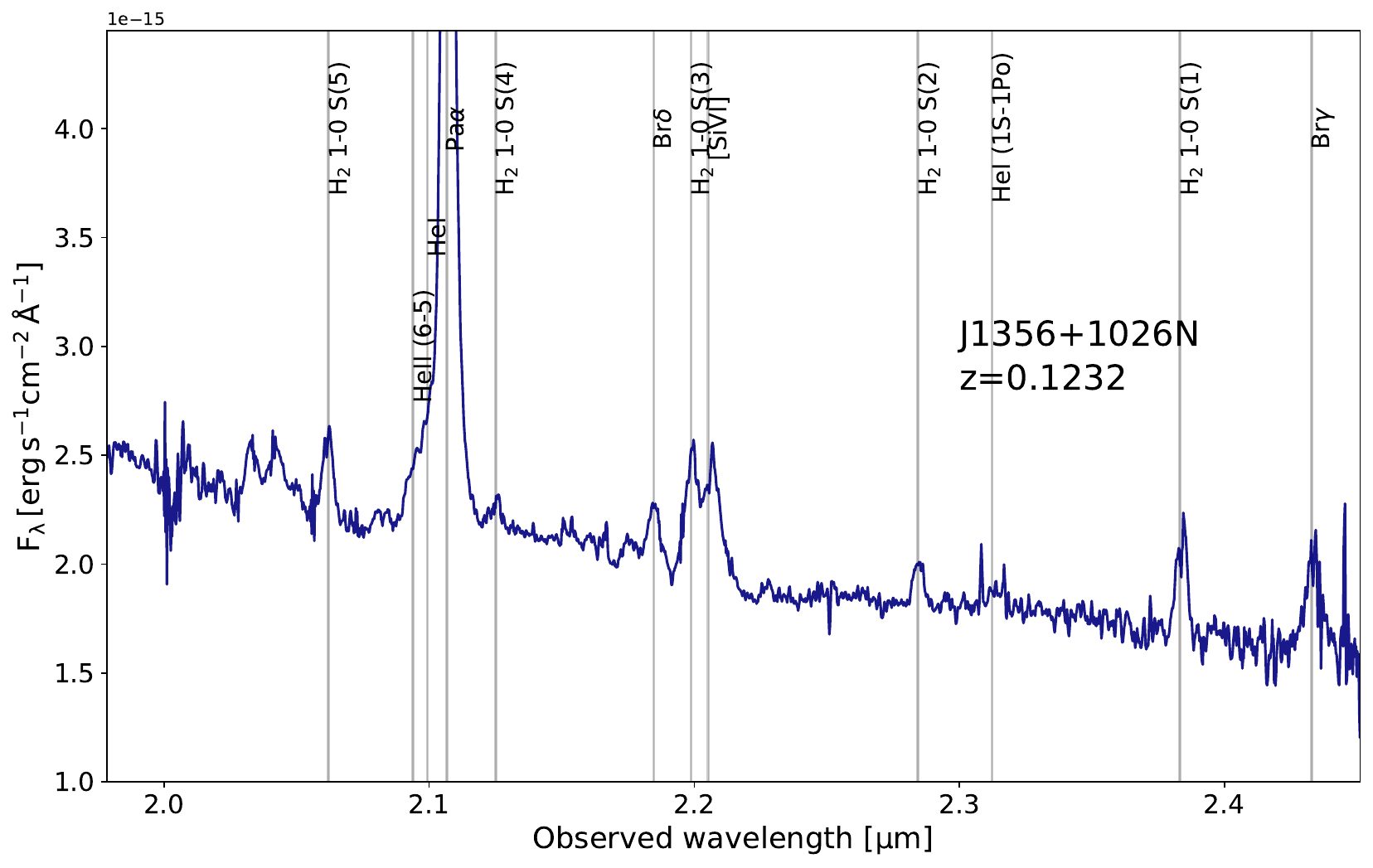}}
\resizebox{\hsize}{!}{\includegraphics{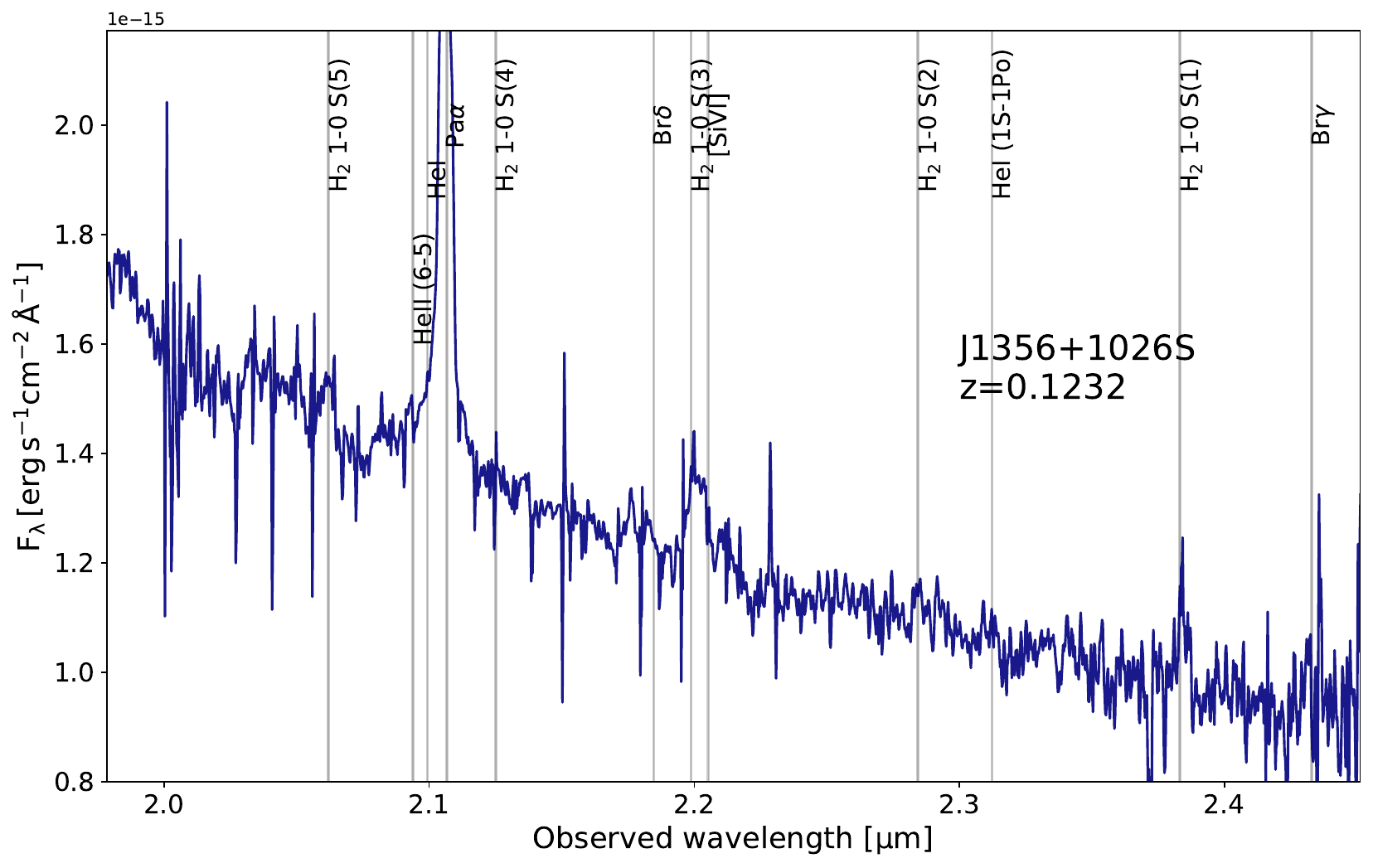}}
\caption{Flux-calibrated nuclear spectra of the Teacup, J1356N and J1356S extracted in a circular aperture of \SI{0.8}{\arcsecond} diameter and smoothed using a 3 pixels boxcar. The most prominent emission lines are labeled.}
  \label{fig:nuclear-spectra}
\end{figure} 

\section{Observations and data reduction}\label{sec:obs}

The Teacup and J1356 were observed in the K-band (1.95-2.45 \textmu m) with the Very Large Telescope/Spectrograph for INtegral Field Observations in the Near Infrared (VLT/SINFONI; \citealt{Eisenhauer2003,Bonnet2004}).
Observations were carried out between April 2016 and August 2017 (Program ID: 097.B-0923(A); PI: J. Mullaney) in service mode.
J1356 observations consist of 24 integrations of 147\,s on-target, each carried out in 10 nights, for a total exposure time of 3528\,s.
The Teacup's observations consist of 13 integrations of 147\,s on-source, each carried out in 5 nights, for a total exposure time of 1911\,s.
To increase the signal-to-noise ratio of the Teacup's dataset, we combined these data with those analyzed in \citet{ramosalmeida2017} (Program ID: 094.B-0189(A); PI: M. Villar-Martín), obtaining a total exposure time of 3711 s, comparable to the exposure time of the J1356 observations.
The observations are split into short exposures because of the strong and rapid variation of the infrared sky emission, following a jittering O-S-S-O pattern for object and sky frames.
The observing conditions were clear and the seeing variation over the on-source observing periods was small and fullfilled the requested observing conditions (seeing $<$ \SI{0.8}{\arcsecond} and an airmass $<$ 2.0)
according to the ESO astronomical site monitor\footnote{\href{https://www.eso.org/asm/ui/publicLog}{https://www.eso.org/asm/ui/publicLog}} and to our own measurements. To do so we took the cubes of the standard stars observed during the same nights as the science targets, and we collapsed them along the spectral axis. We then fitted a 2D Gaussian to the collapsed cubes to estimate the seeing FWHM and obtained average values of FWHM = (\SI{0.62}{\arcsecond}$\pm$\SI{0.09}{\arcsecond})$\times$(\SI{0.51}{\arcsecond}$\pm$\SI{0.09}{\arcsecond}) for the Teacup and FWHM = (\SI{0.71}{\arcsecond}$\pm$\SI{0.15}{\arcsecond})$\times$(\SI{0.59}{\arcsecond}$\pm$\SI{0.14}{\arcsecond}) for J1356.

We used the configuration \SI{0.125}{\arcsecond} $\times$ \SI{0.250}{\arcsecond} pixel$^{-1}$, which produces a field-of-view (FOV) of \SI{8}{\arcsecond} $\times$ \SI{8}{\arcsecond} per single exposure. Thanks to the jittering process, the effective FOV in the case of Teacup and J1356, respectively, are \SI{9}{\arcsecond} $\times$  \SI{8.5}{\arcsecond} (14.4 kpc $\times$ 13.6 kpc) and \SI{8.25}{\arcsecond} $\times$ \SI{10.25}{\arcsecond} (18.3 kpc $\times$ 22.7 kpc). 
The channel width is 2.45 \AA\ ($\sim$30 \kms) and the average spectral resolution in the \textit{K}-band is $\sim$3300, which translates into a spectral resolution of $\sim$ 75 \kms.

We adopted the already reduced and calibrated data from the 094.B-0189(A) program \citep[see][for the details of the data observations and reduction]{ramosalmeida2017} of the Teacup.
For the other dataset 097.B-0923(A) we used the ESO Recipe Flexible Execution Workbench \textit{EsoReFlex} (version 2.11.5), and the ESO Recipe Execution TooL \textit{EsoRex} (version 3.13.8) to reduce the SINFONI data, and our own \texttt{IDL} routines for telluric correction and flux calibration.
The ESO pipeline applies the usual calibration corrections of dark subtraction, flat-fielding, detector linearity, geometrical distortion, wavelength calibration, and subtraction of the sky emission to the individual frames.  
The flux-calibrated individual frames from multiple exposures were combined using the pipeline recipe \texttt{sinfo\_utl\_cube\_combine} applying sigma clipping before coaddition (\texttt{--ks\_clip=TRUE}), and by scaling the sky (using \texttt{--scale\_sky=TRUE}) within individual exposures.
If \texttt{--scale\_sky=TRUE}, the spatial median of each cube plane is subtracted from all contributing cube planes before cube coaddition. This step removes sky background residuals that may not have been fully corrected during earlier data reduction stages (e.g. frame stacking), potentially caused by temporal sky variations.

\section{Results}\label{sec:results}

\subsection{Nuclear spectra}\label{sec:nuclearspectra}

\begingroup
\setlength{\tabcolsep}{6pt} % Default value: 6pt
\renewcommand{\arraystretch}{1.3} % Default value: 1
\begin{table}
     \caption[]{Emission lines detected in the nuclear spectrum of J1356N (\SI{0.8}{\arcsecond} diameter).}
         \label{tab:J1356N_fit}
\centering                          
\begin{tabular}{l c c c}        
\hline  
%Line & \multicolumn{3}{c}{Nuclear Spectrum}\\
  Line   & FWHM & $\rm V_{s}$ & Line Flux $\times$ $\rm 10^{16}$ \\ 
 & (\kms) & (\kms) & ($\rm erg~cm^{-2}~s^{-1}$) \\ 
\hline
\Pa~blue & 399 $\pm$ 16 & -130 $\pm$ 15 & 9.69 $\pm$ 1.84 \\ 
\Pa~red & 395 $\pm$ 14 & 229 $\pm$ 15 &  9.61 $\pm$ 1.70 \\ 
\Pa~(b) & 1280 $\pm$ 145 & 175 $\pm$ 260 & 13.47 $\pm$ 5.36 \\ 
\Brd & 582 $\pm$ 23 & 2 $\pm$ 16 & 1.37 $\pm$ 0.22 \\
\Brg & 465 $\pm$ 89 & 27 $\pm$ 35 & 1.16 $\pm$ 0.43\\
\Brg~(b) & 1411 $\pm$ 283 & 96 $\pm$ 114 & 1.86 $\pm$ 1.07\\
He I & 750 $\pm$ 939 & 50 $\pm$ 337 & 3.34 $\pm$ 5.05 \\
He II & 607 $\pm$ 260 & -50 $\pm$ 419 & 1.43 $\pm$ 1.97 \\
$\rm[Si VI]$ & 889 $\pm$ 26 & 230 $\pm$ 17 & 3.49 $\pm$ 0.54 \\
\hdone~blue & 454 $\pm$ 78 & -70 $\pm$ 67 & 1.37 $\pm$ 0.35 \\
\hdone~red & 265 $\pm$ 36 & 260 $\pm$ 28 & 0.91 $\pm$ 0.31 \\
\hdtwo~& 497 $\pm$ 35 & 23 $\pm$ 23 & 0.75 $\pm$ 0.14 \\
\hdthree & 737 $\pm$ 22 & 47 $\pm$ 15 & 3.03 $\pm$ 0.47 \\
\hdfour & 494 $\pm$ 30  & -3 $\pm$ 20  &  0.09 $\pm$ 0.02 \\
\hdfive & 668 $\pm$ 25 & -36 $\pm$ 17 & 1.90 $\pm$ 0.31 \\
\hline
\end{tabular}
\flushleft 
\tablefoot{The columns report the full width half maximum (FWHM), the velocity shift ($\rm V_s$), and the line flux obtained from fits with a maximum of three Gaussian components. FWHMs are corrected for instrumental broadening, and the velocity shifts are calculated using $z=0.1232$ to define the systemic velocity.}
\end{table}

\endgroup

\begin{figure*}[ht]
\resizebox{\hsize}{!}{
\includegraphics{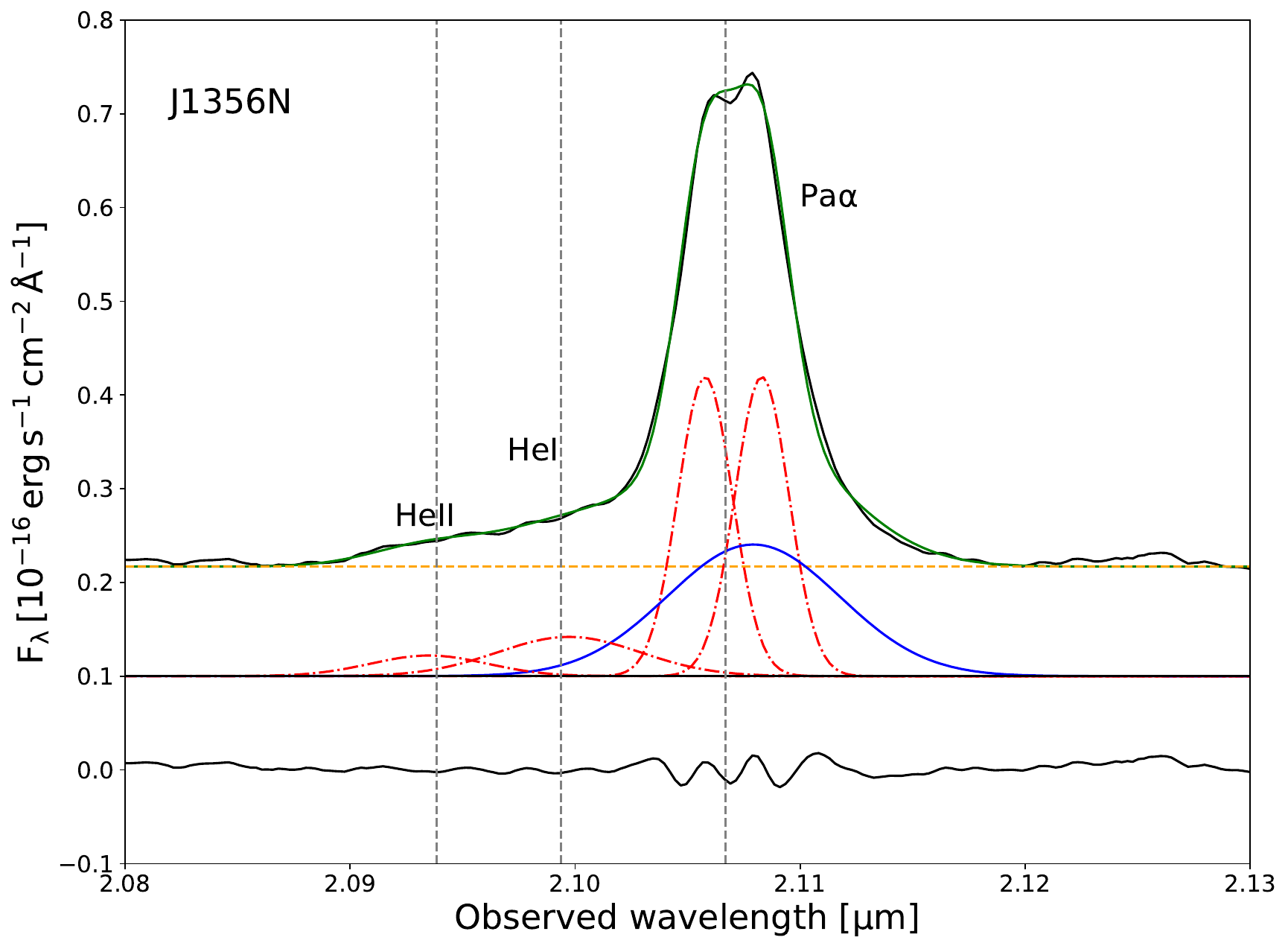}
\includegraphics{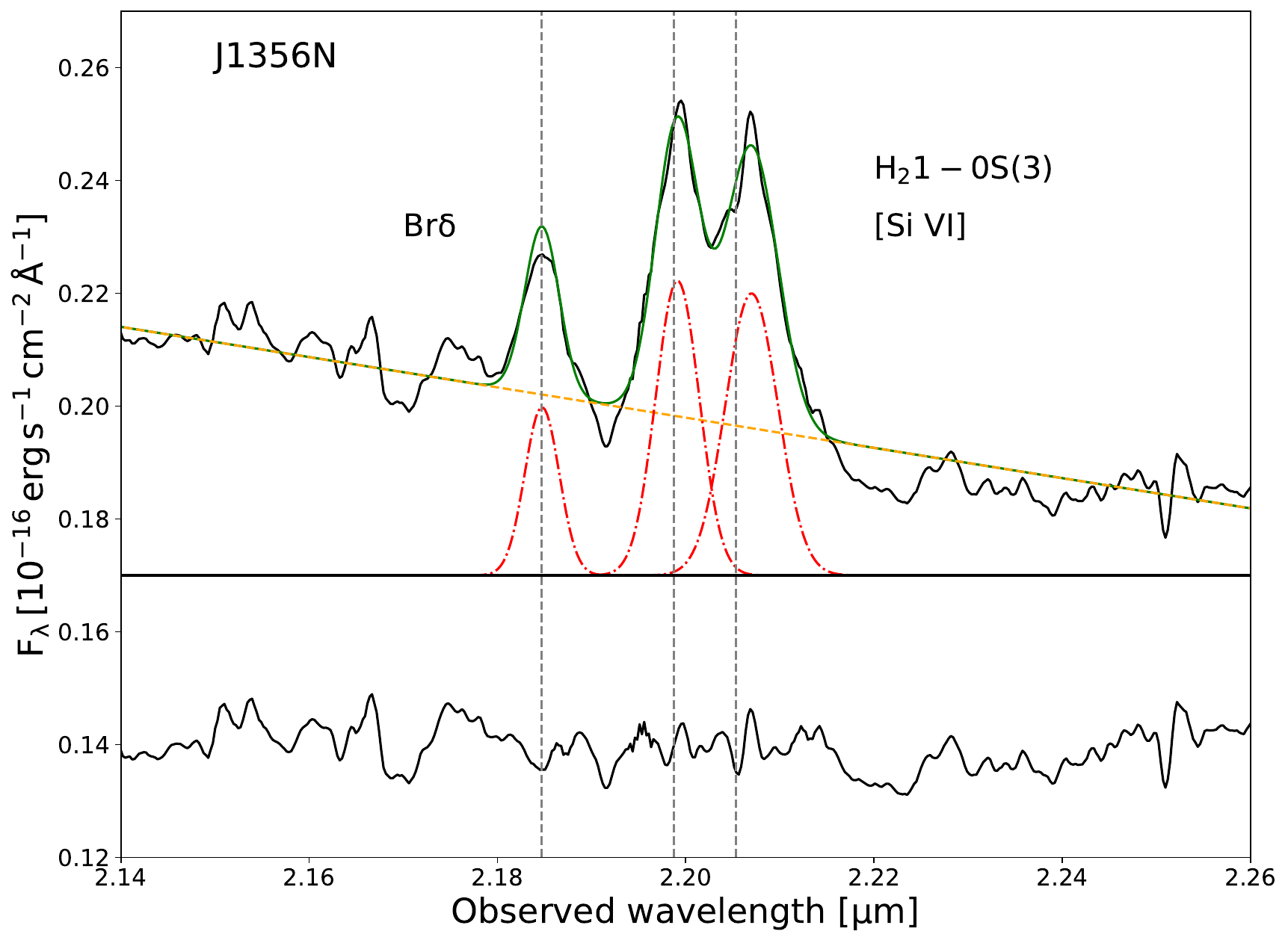}
\includegraphics{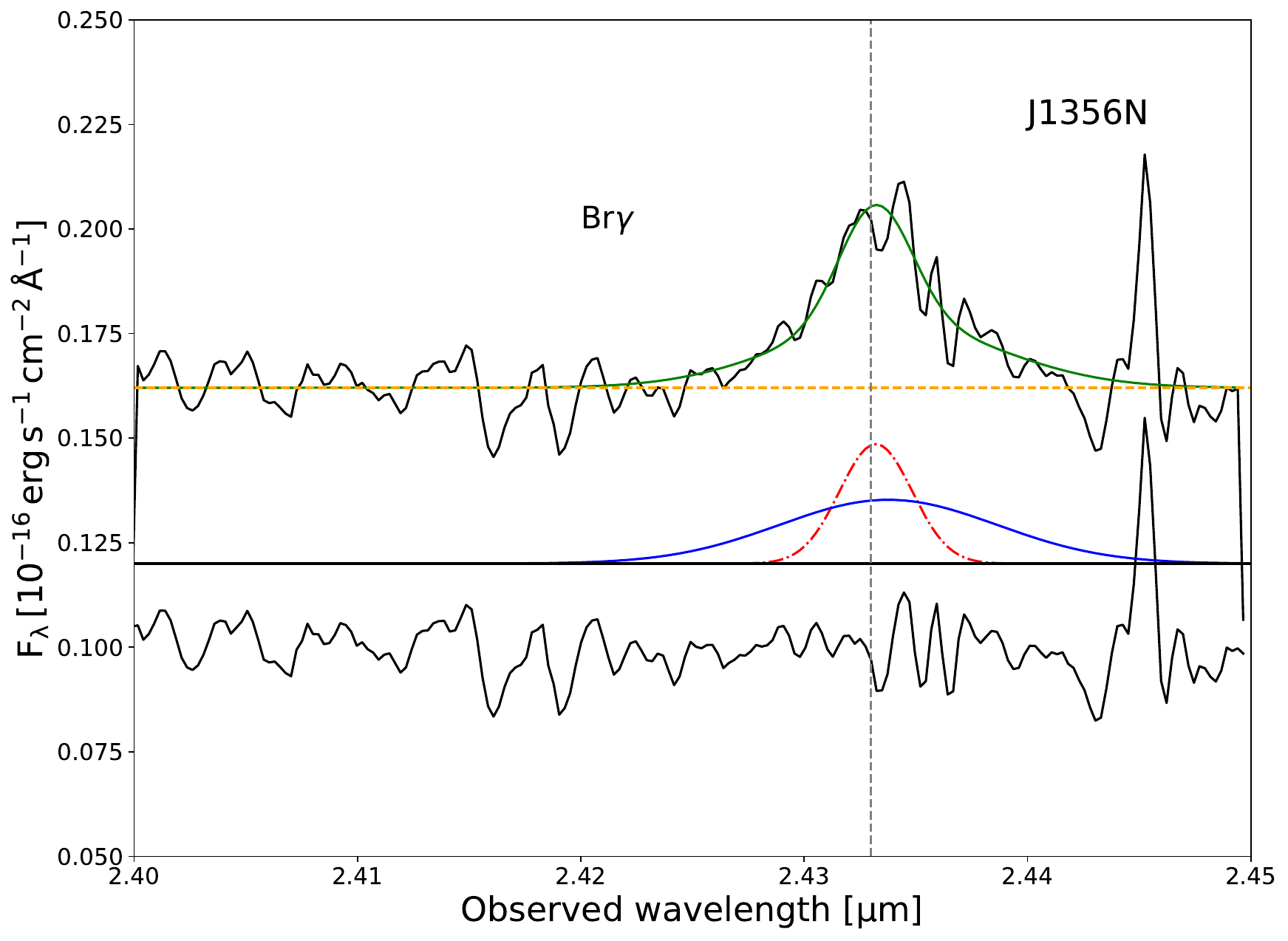}}
\resizebox{\hsize}{!}{
\includegraphics{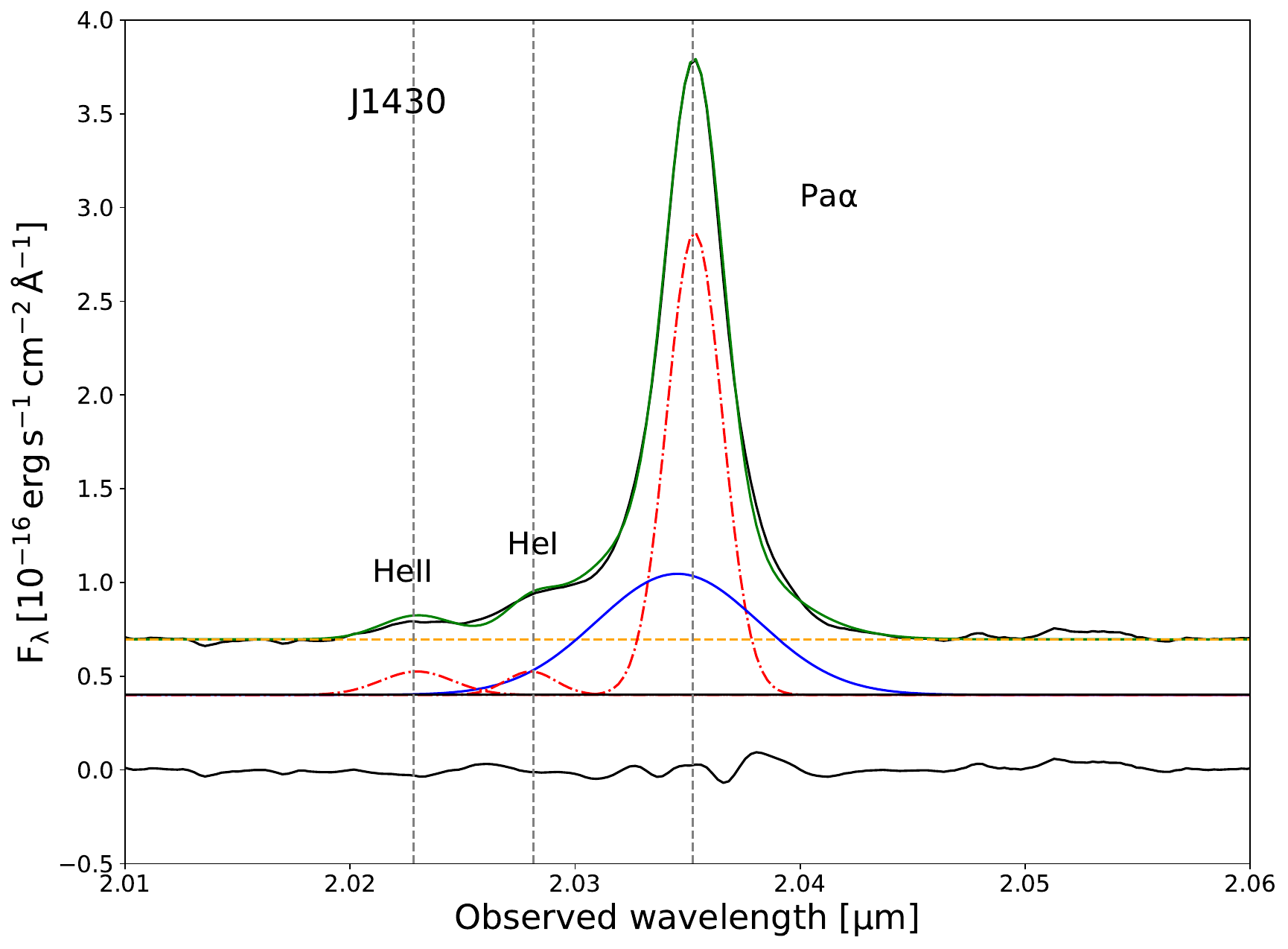}
\includegraphics{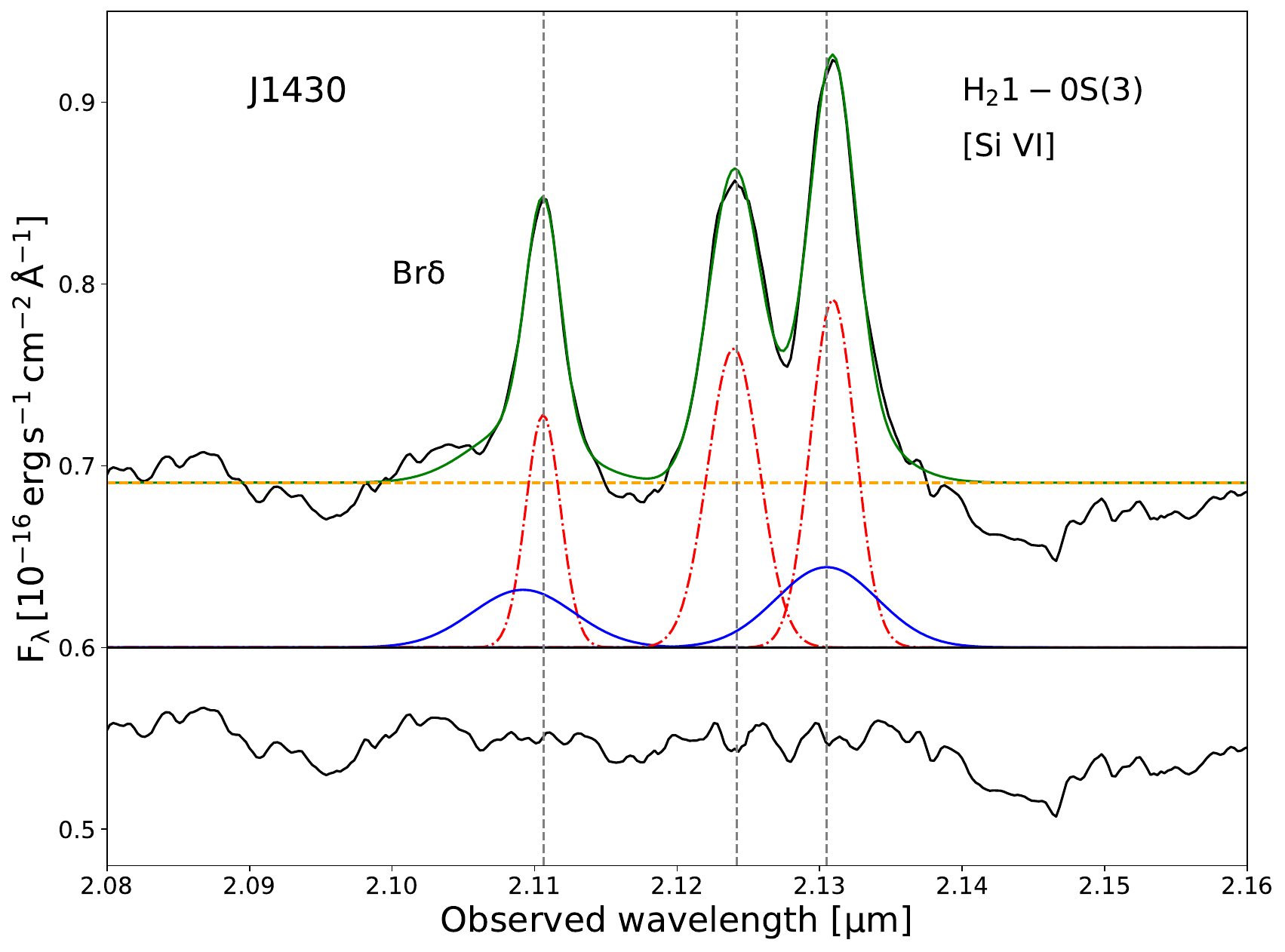}
\includegraphics{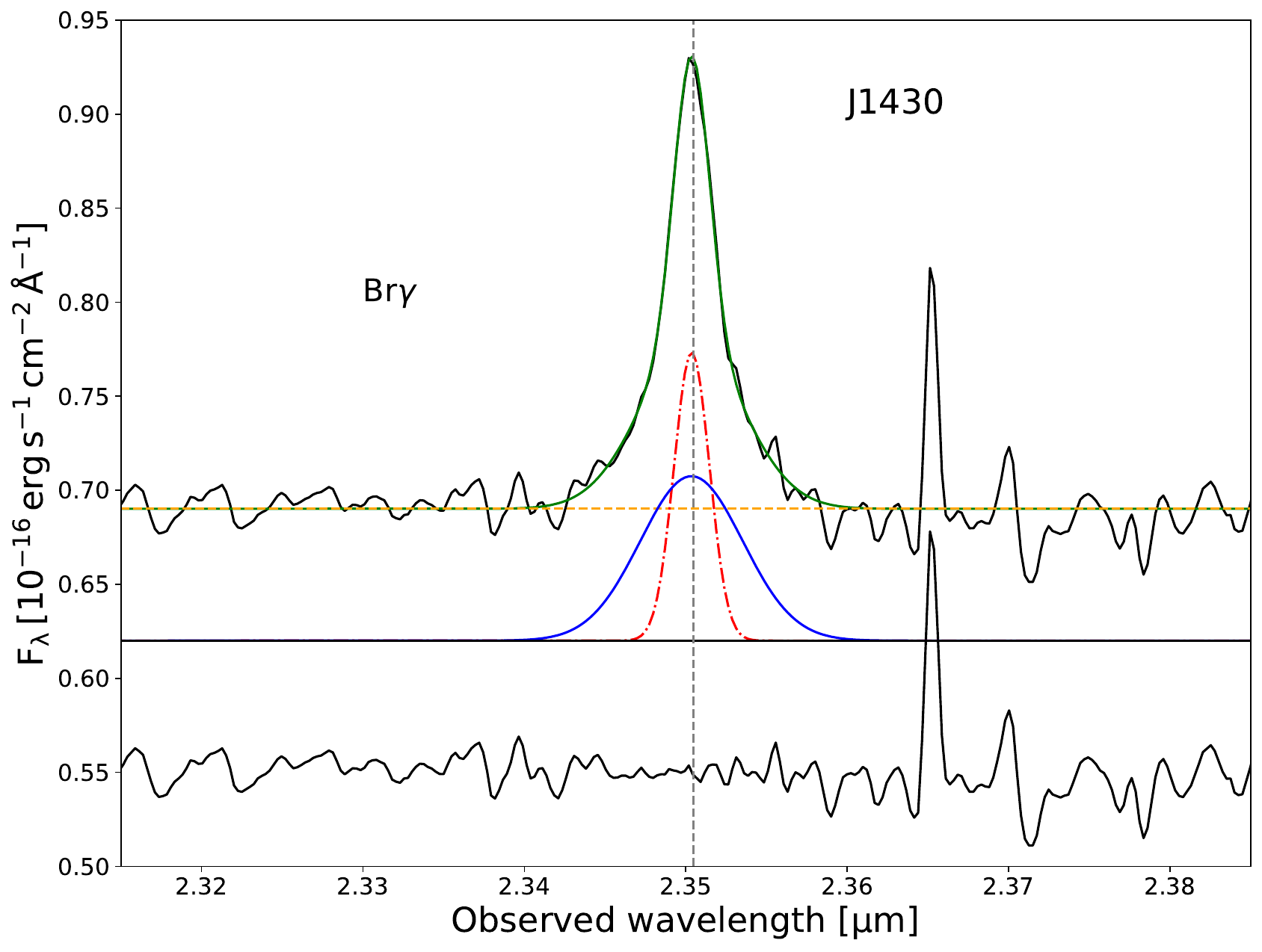}}
\caption{Line profiles showing the nuclear spectra of J1356N (top panels) and the Teacup (bottom panels), extracted in a circular aperture of \SI{0.8}{\arcsecond} diameter and smoothed using a 3 pixel boxcar.
The corresponding fits are shown as solid green lines. Solid blue and dot–dashed red Gaussians are the broad- and narrow- line components, respectively, and the orange dashed line is the continuum. The Gaussians have been vertically shifted for displaying purposes. The insets at the bottom of each panel are the residuals. The gray dotted vertical lines mark the expected position of the line peaks according to the redshift of the QSO2s.}
\label{fig:nuclear_fits}
\end{figure*}

\begin{figure*}[ht]
\resizebox{\hsize}{!}{
\includegraphics{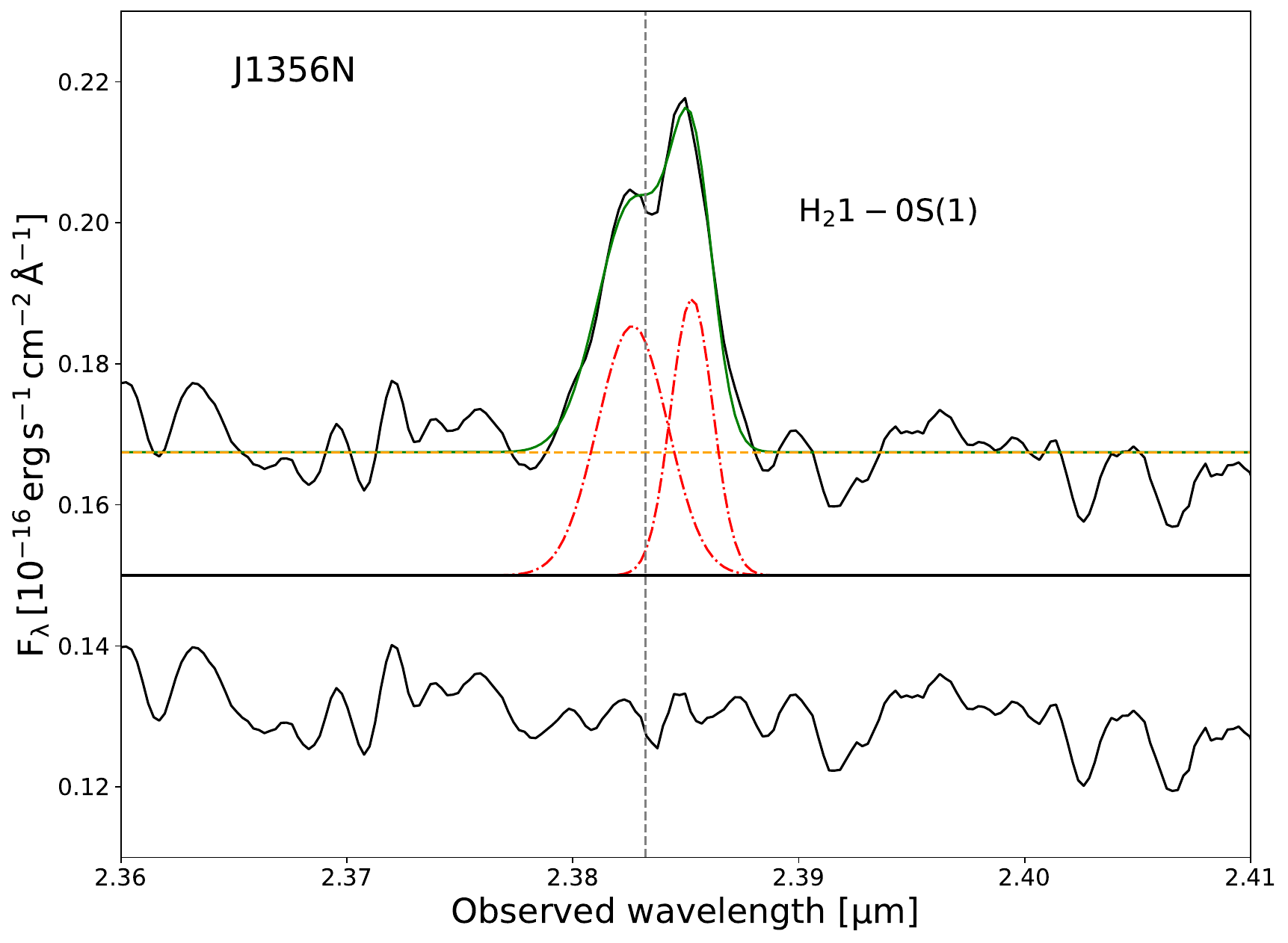}
\includegraphics{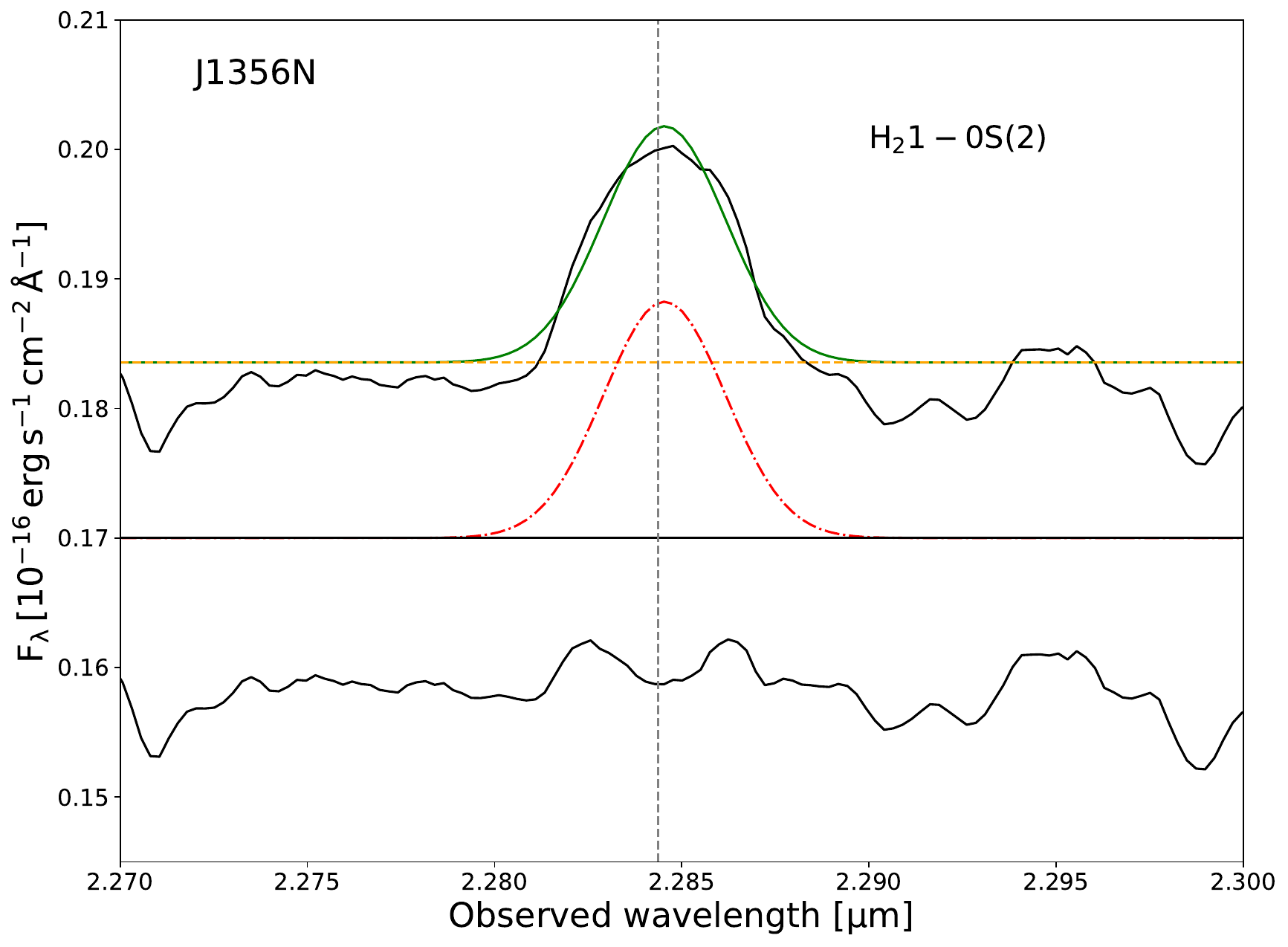}}
\resizebox{\hsize}{!}{
\includegraphics{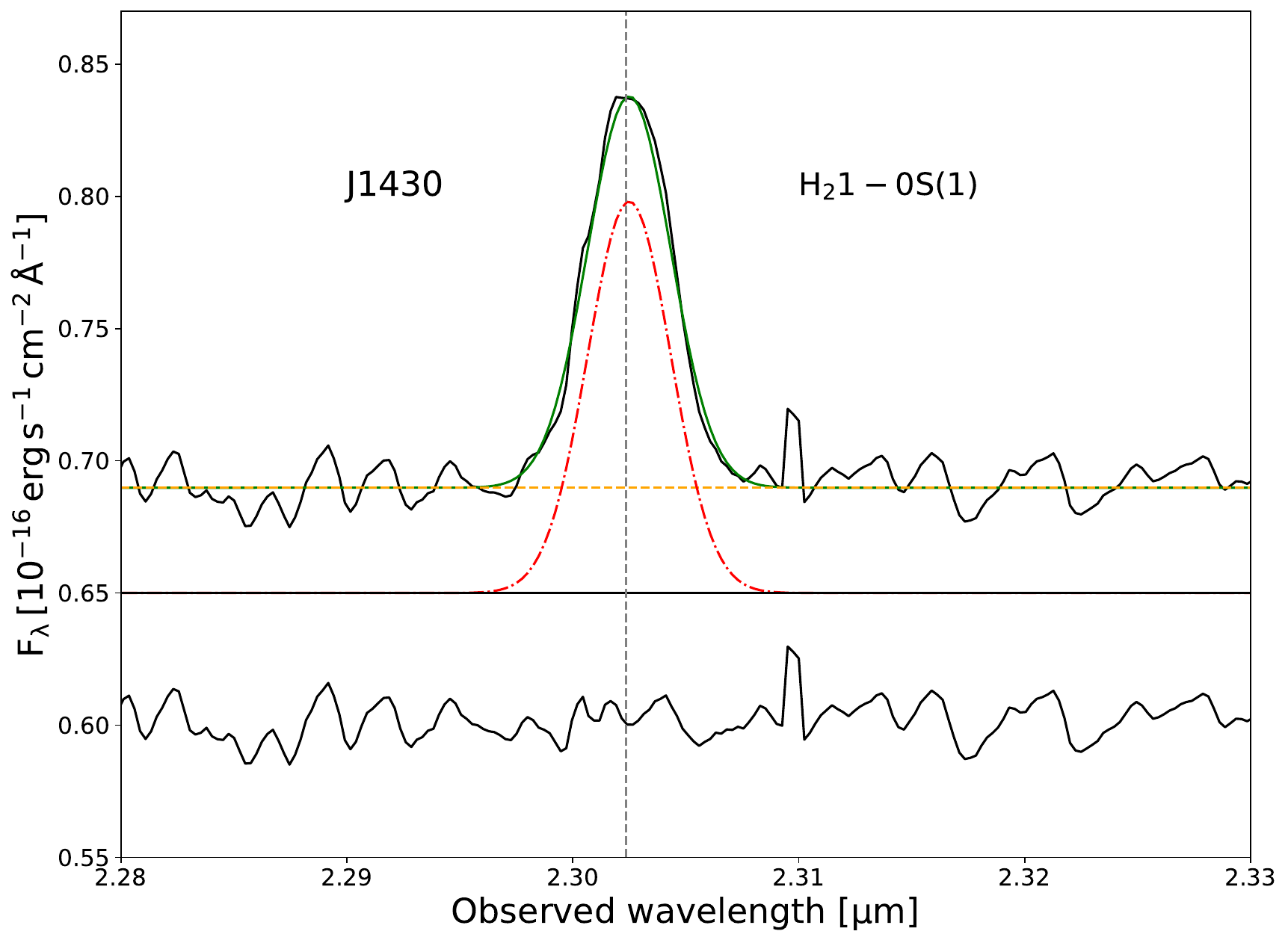}
\includegraphics{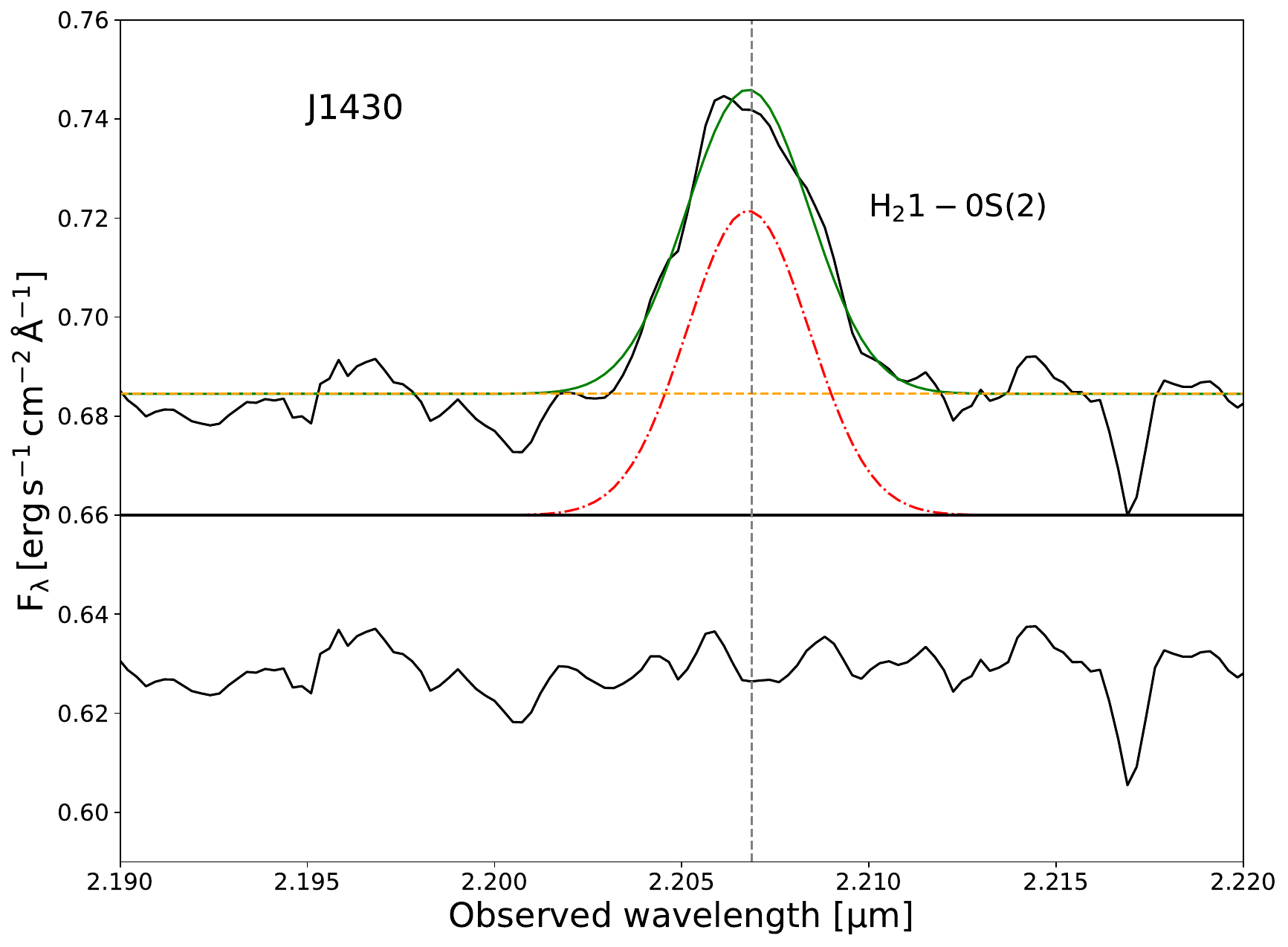}}
\caption{Same as Figure \ref{fig:nuclear_fits} but for the \hdone~and \hdtwo~line profiles.}
\label{fig:H2_nuclear_fits}
\end{figure*}

We extracted the K-band spectra of the nuclear region of the Teacup, J1356N, and J1356S nuclei, using circular apertures of \SI{0.8}{\arcsecond} diameter ($\sim$1.2 and 1.8 kpc for the Teacup and J1356, respectively), centered at the maximum of the \Pa~emission. The aperture was chosen to match the spatial resolution set by the seeing ($<$\SI{0.8}{\arcsecond}). In the following, we will refer to the spectra extracted with this aperture as nuclear spectra, which are shown in Figure \ref{fig:nuclear-spectra} with the emission lines labeled.

The most prominent emission features in the nuclear spectra of both galaxies are \Pa, \Brd, HeI$\rm \lambdaup$2.060 \textmu m, \Brg, and the coronal line $\rm[SiVI]\lambdaup 1.963$ \textmu m. We also detect several \hd emission lines which are indicative of the presence of a nuclear warm molecular gas reservoir.
We perform the fit of these emission lines adopting a combination of Gaussian profiles using the \texttt{astropy} Python library. 
In Figures \ref{fig:nuclear_fits}, \ref{fig:H2_nuclear_fits}, and \ref{fig:nuclear-fit_app}, we show the profiles and corresponding fits of %the most prominent emission lines in the nuclear spectrum, namely 
\Pa, \Brd, \Brg, \sisix, \hdone, \hdtwo~and \hdthree.
In Tables \ref{tab:J1356N_fit} and \ref{tab:nuclear_fit} we report the FWHMs, velocity shifts ($\rm V_s$), and fluxes resulting from our fits with their corresponding errors. 
Velocity shifts are calculated with respect to the red-shifted central wavelength, assuming z = 0.1232 and 0.0851 for J1356 and the Teacup (\SP).
The uncertainties in $\rm V_s$ include the wavelength calibration error ($\sim$8 \kms, as measured from the sky spectrum) and the individual fit uncertainties. In the case of fluxes, the errors have been determined by adding quadratically the flux calibration error \citep[15\%,][]{ramosalmeida2017} and the fit uncertainties. The FWHMs reported in Tables \ref{tab:J1356N_fit} and \ref{tab:nuclear_fit}  are corrected for instrumental broadening ($\sim$75 \kms).

We require three Gaussians to reproduce the \Pa~line profile in J1356N nuclear spectrum.
This includes two narrow blue- and red-shifted components of FWHM $\sim$ 400 \kms~and a broad red-shifted component of FWHM $\sim$ 1280 \kms. 
We also notice that the He I and He II lines, which are blended with \Pa, require only one Gaussian with FWHM $\sim$ 600-750 \kms. 
Similarly to \Pa, we were able to fit the \hdone~line profile with two narrow components (FWHM $\sim$ 450 and $\sim$ 265 \kms), blue- and red-shifted with respect the systemic velocity calculated using the QSO2 redshift.
Being a merging system, it is not surprising that this QSO2 shows complex and irregular gas kinematics. In fact, \SP~also reported evidence of two narrow, two intermediate, and one broad components in the [O III] doublet line profile traced by MEGARA, which has even higher spectral resolution than SINFONI. We identify the broad \Pa~component with the nuclear counterpart of the ionized outflow detected in [O III] for which \SP~reported a FWHM of $\sim$ 1300 \kms.
In the case of \sisix, we fit a single red-shifted component of FWHM $\sim$ 890 \kms~with $\rm V_s$ $\sim$ 230 \kms.
In the case of \Brd~and the \hdtwo, S(3), S(4), and S(5) molecular lines, single Gaussians with FWHM $\sim$ 500-740 \kms~were sufficient to reproduce the profiles (see Table \ref{tab:J1356N_fit}).

In the case of J1356S, we require only one Gaussian to reproduce the line profiles in the nuclear spectrum (see Table \ref{tab:nuclear_fit}). 
In particular, the \Pa, He I, and He II lines, which are blended together, are fitted with one Gaussian component of FWHM $\sim$ 700-940 \kms.
The \sisix~line profile requires only one Gaussian with FWHM $\sim$ 940 \kms~and $\rm V_s$ $\sim$ 50 \kms. In the case of \Brd, a single Gaussian with FWHM $\sim$ 470 \kms~and $\rm V_s$ $\sim$ -10 \kms~was sufficient to reproduce the line profile.
The \hdone, S(2), S(3), S(4), and S(5) molecular lines are fitted with single Gaussians with FWHM $\sim$ 620-1175 \kms~(see Table \ref{tab:nuclear_fit}). 
Finally, the \Brg~emission line shows a low signal-to-noise ratio in the spectra of J1356, being at the edge of the SINFONI wavelength coverage. % in a noisy region of the spectra. 
We fitted the line profile with a narrow Gaussian of FWHM $\sim$ 465  \kms~and a broad one of $\sim$1410 \kms~in J1356N spectrum,  while in the case of J1356S the line appears undetected. % in J1356S nuclear spectrum.

For what concerns the Teacup nuclear spectrum, we require two Gaussians to reproduce the \sisix~line profile. This includes a narrow and broad component of FWHM $\sim$ 510 \kms~and $\sim$ 1175 \kms~respectively. \citet{ramosalmeida2017} already detected a broad component of \sisix~line at nucleus, which they identified with the coronal gas counterpart of the nuclear outflow detected with optical spectroscopic data \citep{villarmartin2014,harrison2015}. 
In the case of \Pa, we fitted a narrow component of FWHM $\sim$ 420 \kms~and a broad component of FWHM $\sim$ 1245 \kms~blue-shifted by about 90 \kms~with respect to the narrow component. This fit is not consistent with what was reported by \citet{ramosalmeida2017}, who measured a FWHM of the broad \Pa~of $\sim$1795 \kms, likely due to the fact that the latter authors did not fit the He I and He II line profiles.
To reproduce the profiles of He I and He II lines, we used only one Gaussian with FWHM $\sim$ 375 \kms~and 550 \kms, respectively. 
If we neglect the contributions of He I and He II and fit the \Pa~line profile with only a narrow and a broad component, we find FWHM $\sim$ 1655 \kms~for the broad \Pa~component (see Appendix \ref{app:nuclear_spectra} and Figure \ref{fig:teacup_Pa}), which is consistent within 2$\sigma$ with the FWHM reported by \citet{ramosalmeida2017}. 
In the case of \Brd, we fitted a broad component of FWHM $\sim$ 1180 \kms~and $\rm V_s$ $\sim$ -200 \kms~relative to the central wavelength of the narrow component. The \Brg~line profile is well reproduced by a narrow and a broad component with FWHM $\sim$ 320 \kms~and 950\kms, respectively. 
Finally, in the case of the molecular lines, as earlier, the line profiles are fitted by a single Gaussian with FWHM $\sim$ 480-600 \kms~(Table \ref{tab:nuclear_fit} and bottom panels of Figure \ref{fig:H2_nuclear_fits}).

\subsection{The warm molecular gas}\label{sec:warmgas}

\begin{figure*}
\resizebox{\hsize}{!}{\includegraphics{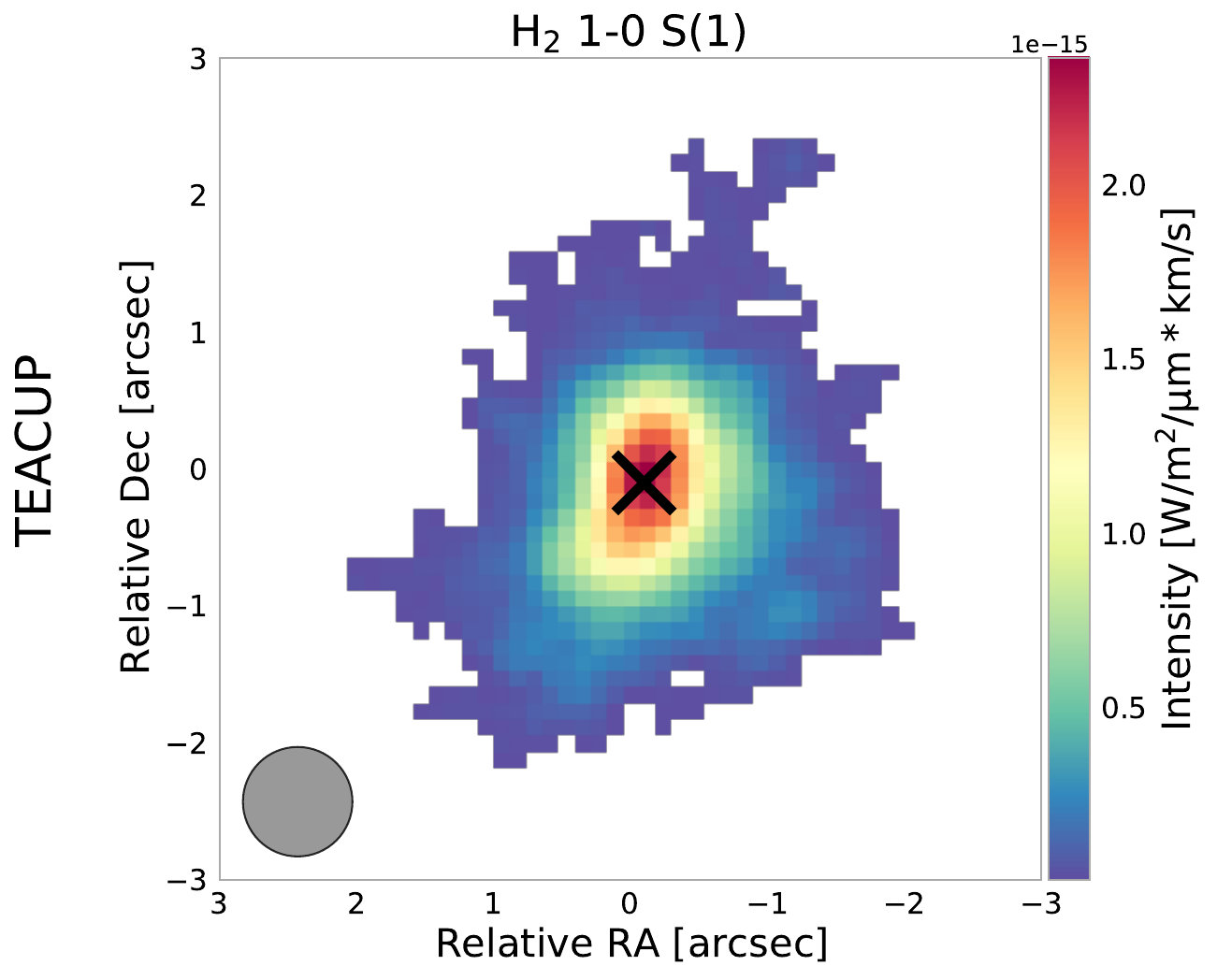}
\includegraphics{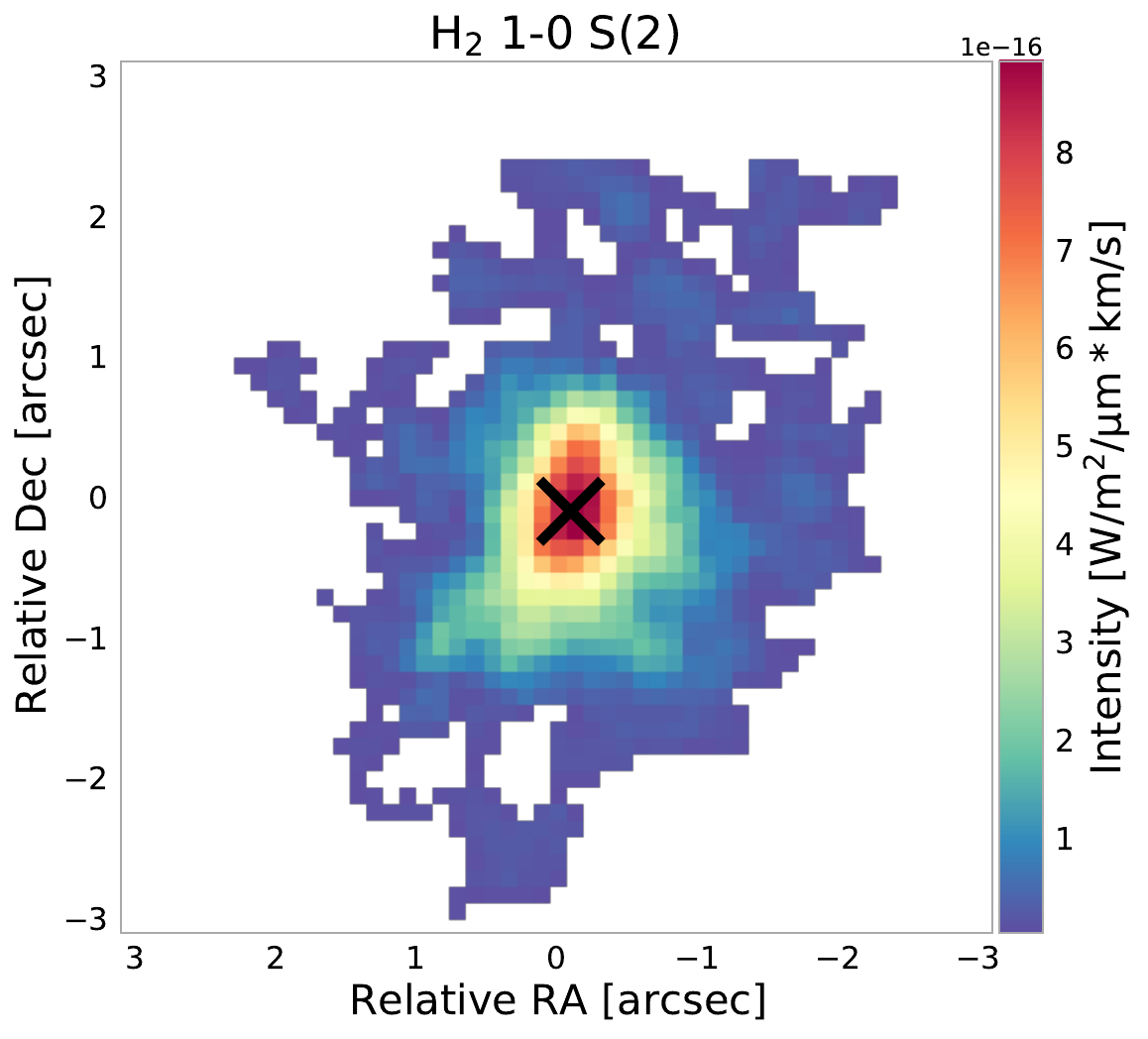}
\includegraphics{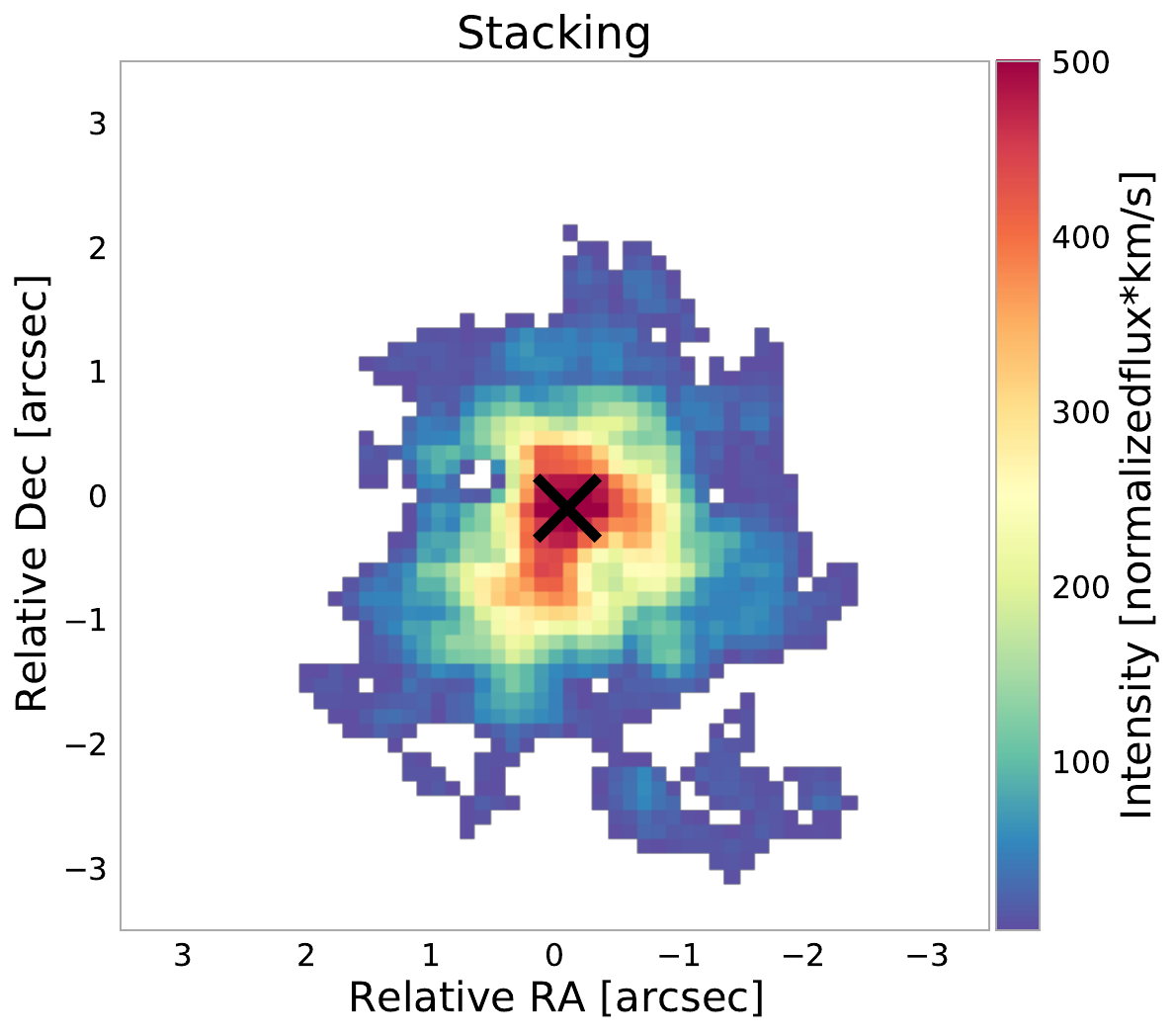}}
\resizebox{\hsize}{!}{\includegraphics{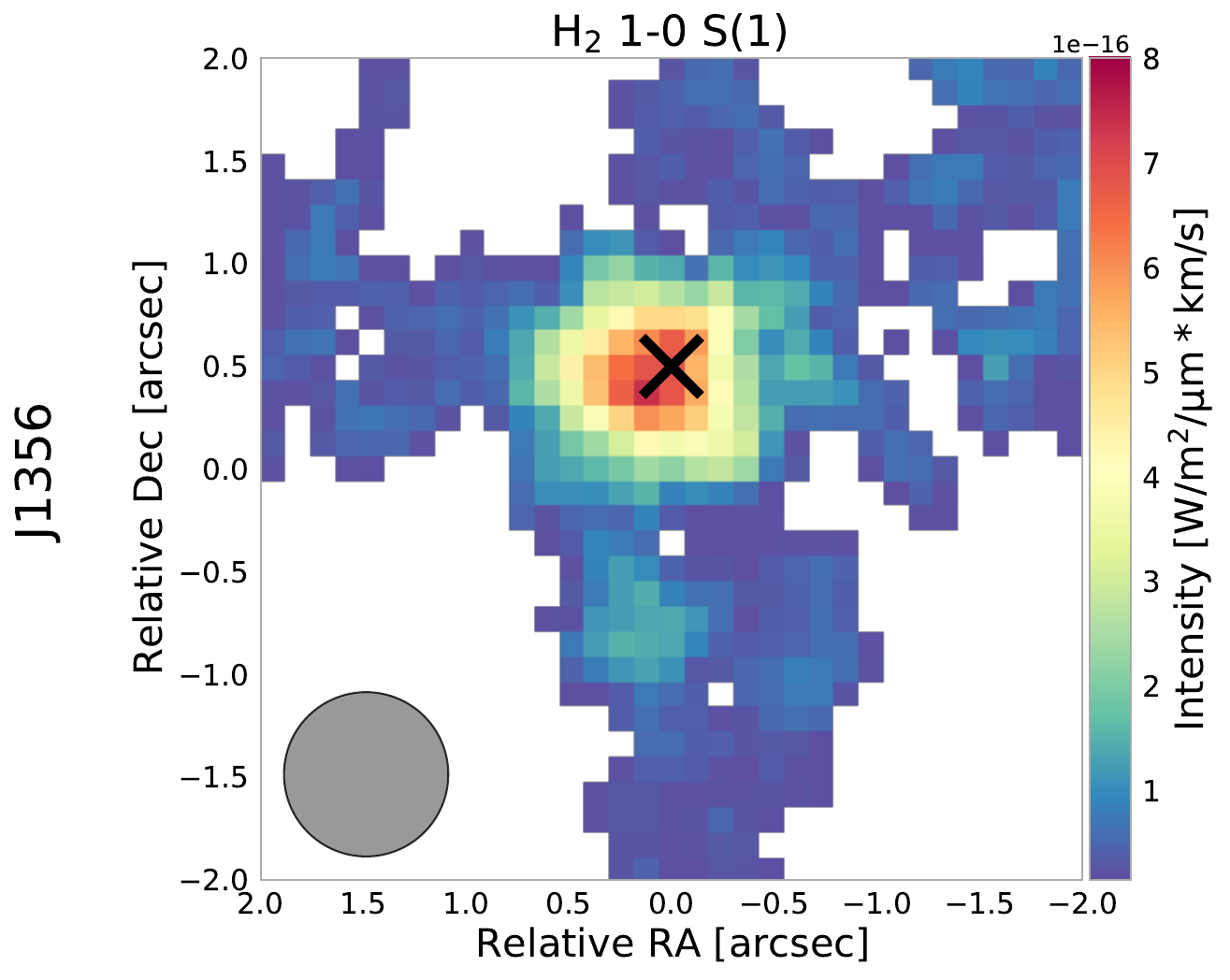}
\includegraphics{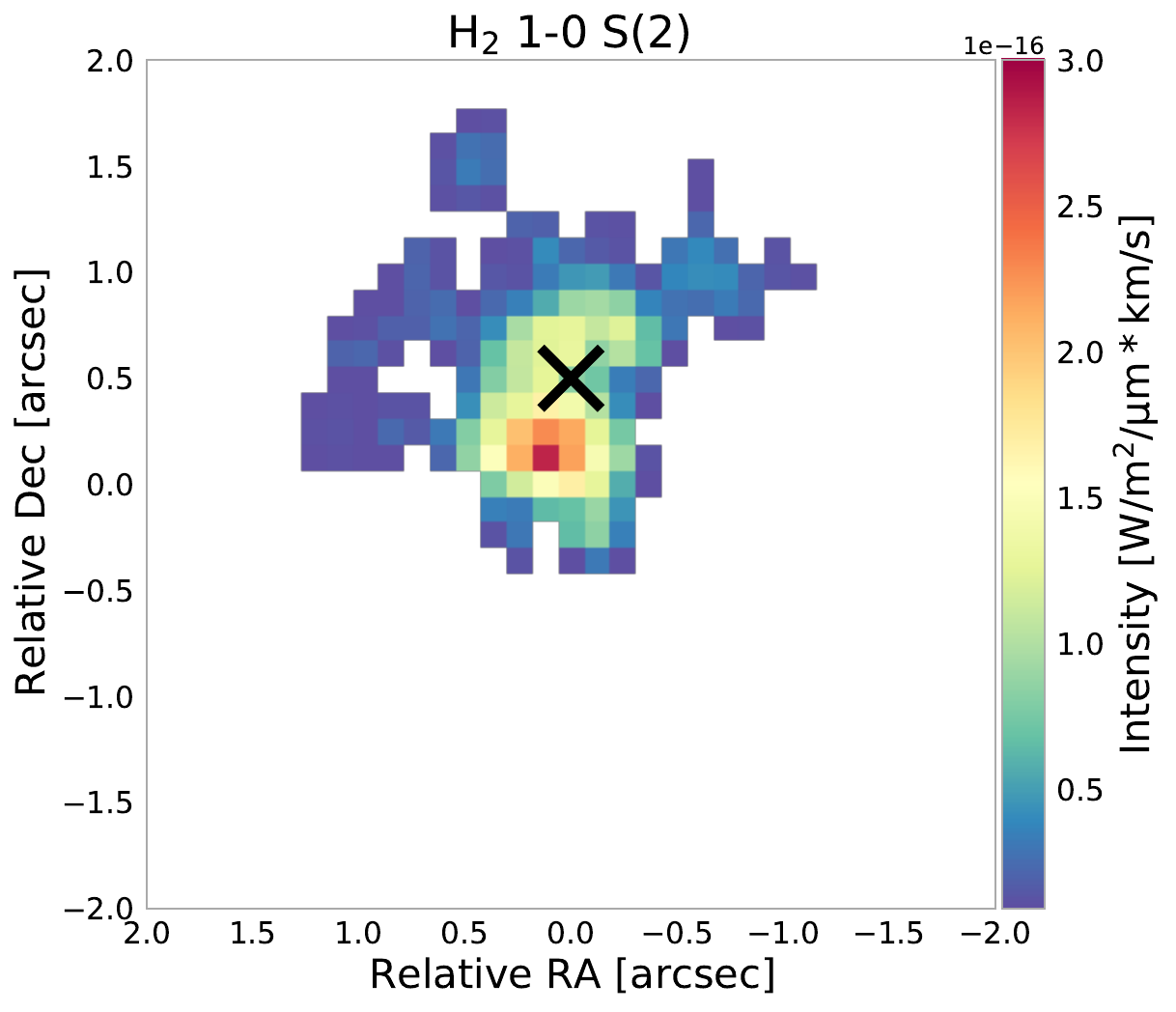}
\includegraphics{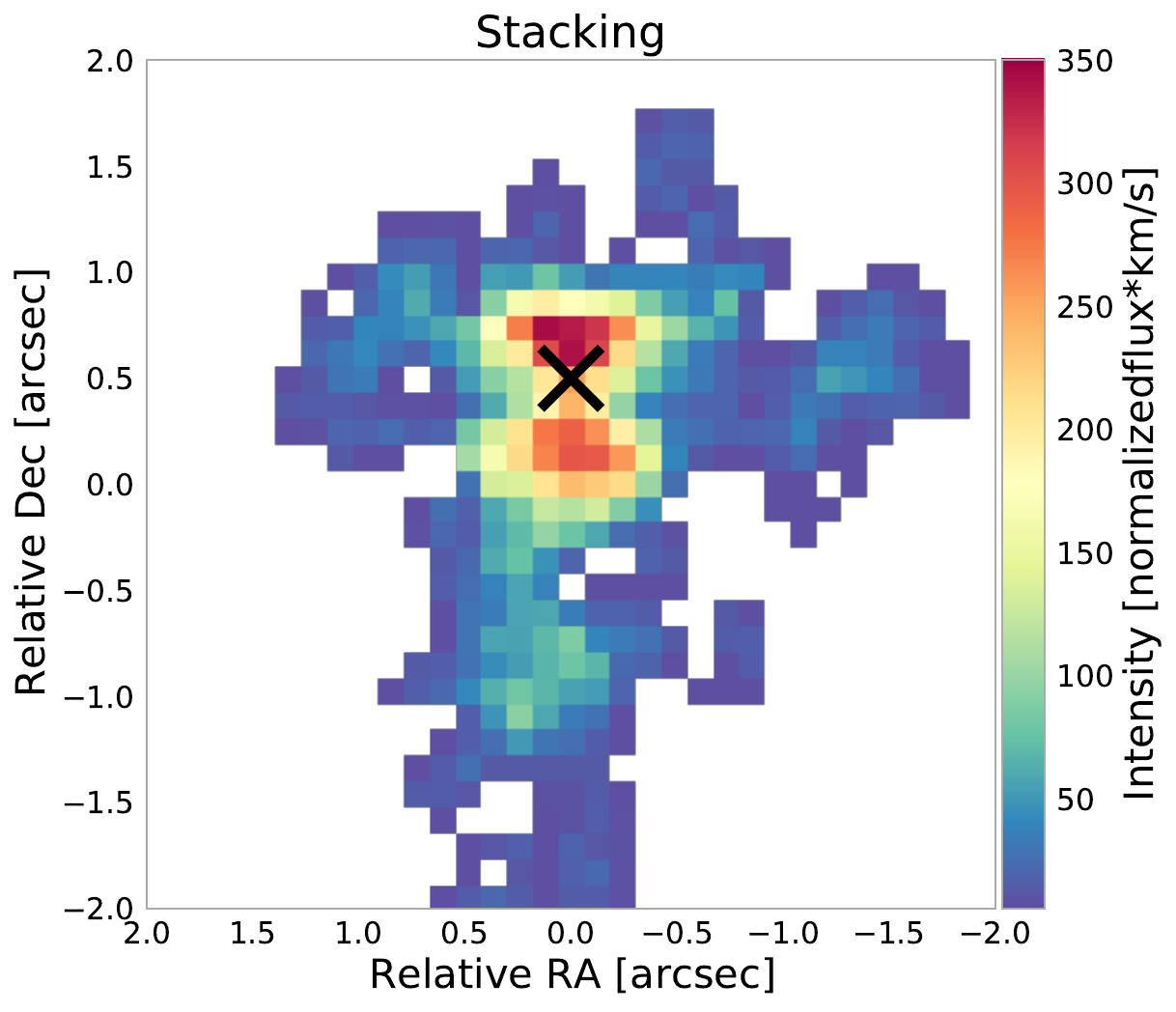}}
 \caption{Integrated intensity (i.e. moment 0) maps of the \hdone, S(2), and stacked \hd of J1430 (top panels) and of J1356 (bottom panels). The black cross marks the peak of the near-infrared continuum. The maps have been smoothed using a two-spaxel boxcar for presentation purposes. Regions below 2$\sigma$ are masked out.}
  \label{fig:mom0}
\end{figure*} 

In the following, we will focus on the molecular \hd lines, which trace the warm molecular gas phase of the ISM. In fact, our goal is to characterize the properties and kinematics of warm molecular gas and to compare with the cold molecular gas phase traced with CO line and presented in \RA~and \citet{audibert2023}. Since the \hdthree~line is blended with \sisix, we will focus the analysis on the \hdone~and \hdtwo~emission, being the other two most prominent \hd lines in the spectra.

We determine the mass of warm molecular gas measured  from the nuclear spectra of the two QSO2s from the \hdone~line luminosity and adopting the following relation from \citet{mazzalay2013}:
\begin{equation}\label{eq:mass}
    M_{H_2} \simeq 5.0875 \times 10^{13} ~\Biggl( \frac{D}{Mpc}\Biggr) ^2 
    \Biggl( \frac{F_{H_2 1-0S(1)}^{corr}}{erg ~s^{-1} cm^{-2}} \Biggr),
\end{equation}
\noindent where $D$ is the luminosity distance and $F_{H_2 1-0S(1)}^{corr}$ is the extinction corrected \hdone~line flux. 
The \hdone~line fluxes are corrected for extinction using the infrared extinction $A_K$, derived from the optical extinction $A_V$ assuming that $A_K \approx 0.1 \times A_V$. In the case of the Teacup we adopted the $A_V = 1.2$ mag, which is the value measured in the same region from the VLT/MUSE data reported by \citet{venturi2023}. For J1356 we adopted $A_V = 0.9$ mag from VLT/MUSE data of this galaxy \citep{bianchininprep}.
Therefore, using Equation \ref{eq:mass} we measure $\rm M_{H_2} \, = (5.85 \pm 0.90) \, \times 10^3 $ \msun~in the Teacup for the central \SI{0.8}{\arcsecond} diameter.
\citet{ramosalmeida2017} reported $\rm M_{H_2} \sim 3 \times 10^3 $ \msun~in an aperture of \SI{0.5}{\arcsecond} diameter and $\rm M_{H_2} \sim 10 \times 10^3 $ \msun~in an aperture of \SI{1.25}{\arcsecond}.
We find that the north nucleus is the brightest in \hd, as it is also the case for CO and [O III] (\RA; \SP).
For J1356N and J1356S, we measured $\rm M_{H_2} \, = \, (4.05 \pm 0.69) \, \times 10^3 $ \msun~and $\rm (1.47 \pm 0.37) \, \times 10^3 $ \msun~using the same aperture of \SI{0.8}{\arcsecond} diameter. According to our findings, J1356N contains the 75$\%$ of the warm molecular gas mass contained in the system. Therefore, in the following, the kinematic analysis will be focused on J1356N only, and we omit the N/North wording from this point onward. 

We produced \hdone~and S(2) integrated intensity (moment 0) maps of the Teacup and J1356, which are shown in the left and middle panels of Figure \ref{fig:mom0}. In order to compare with the the cold phase of the molecular gas, we also measured the total warm \hd masses. Thus, we extracted the \hdone~emission from the line emitting regions above $2\sigma$ (see Figure \ref{fig:mom0}), and we followed the procedure described above. We measured total warm molecular gas masses of $\rm M_{H_2}^{warm} \sim 4.5~and~1.3 \times 10^4$ \msun~for the Teacup and the J1356 system, respectively, with maximum spatial extents of the \hdone~emission at 2$\sigma$, measured from the AGN position, of 4.8 kpc and 5.8 kpc.

Furthermore, in order to increase the S/N of the molecular emission lines, we performed stacking of the two emission lines.
%The stacking analysis was performed using a standard stacking procedure, which is summarized below.
Firstly, we cut the SINFONI data cubes around \hdone~and S(2) emission lines, creating separate cubes for each line.  
We then modeled and subtracted the continuum for each spaxel individually. We removed the noise spikes by interpolating the spectrum of each spaxel. Then, we rebinned the spectra onto a common velocity grid centered on the observed wavelengths of each spectral line.
We obtained continuun subtracted cubes for each line. %The Teacup \hdone~and S(2) datacubes have root mean square (rms) noise ($\sigma$) of 3.23 $\rm 10^{-19}$ \noise~and 2.36 $\rm 10^{-19}$ \noise~per channel of $\sim$10 \kms, respectively. While, J1356 \hdone~and S(2) datacubes have $\sigma$ = 3.26 $\rm 10^{-19}$ \noise~and $\sigma$ = 2.56 $\rm 10^{-19}$ \noise~per channel of $\sim$30 \kms, respectively. 
We then created a weight map for each line by dividing the maximum intensity of the spaxel spectrum by the sum of the maxima of the two spectra. Finally, we normalized and combined the cubes using these weight maps.
%The Teacup and J1356 stacked \hd datacubes have $\sigma$ = 0.15 and $\sigma$ = 0.21 per channel of $\sim$20 \kms, respectively.
The right panels of Figure \ref{fig:mom0} show the moment 0 map of the stacked cube of the \hd lines of each QSO2.

The Teacup intensity map shows an elongated distribution of \hd at the nucleus, consistent with the S(1) intensity map shown in \citet{ramosalmeida2017}.
The optical emission, both in continuum and in [O III], was reported to show a single-peaked compact morphology by several observations at different spatial resolutions \citep[HST data at \SI{0.1}{\arcsecond}, VLT/MUSE at \SI{0.6}{\arcsecond}, MEGARA at \SI{1.1}{\arcsecond};][\SP]{keel2012,harrison2015,venturi2023}
On the contrary the morphology of the cold molecular gas, traced by CO(2-1) at \SI{0.2}{\arcsecond}, shows a double peaked morphology, with the two peaks separated by $\sim$ \SI{0.8}{\arcsecond} (i.e., 1.3 kpc) with PA $\sim$ -\ang{10} \citep[\RA;][]{audibert2023}. 
Due to the limited angular resolution of the SINFONI data, we cannot resolve the two peaks.
\citet{ramosalmeida2017} reported that the \hdone~emission is elongated by approximately 1.4 kpc along the N-S direction, roughly perpendicular to the \Pa~and \sisix~emissions. 
In our \hd maps we observe the same, and thanks to the combination with new SINFONI observations, that is, increasing the exposure time by two times with respect to \citet{ramosalmeida2017}, we recover more extended \hdone~diffuse emission. The \hdtwo~emission shows a similar morphology, but is characterized by a lower S/N relative to \hdone. 
Thanks to the stacking procedure, we can further investigate the diffuse emission of \hd by combining the two \hd lines. With respect to \citet{ramosalmeida2017}, we were able to recover emission up to \SI{2}{\arcsecond} (i.e. 3.2 kpc) along the E-W direction, and up to \SI{3}{\arcsecond} (i.e. 4.8 kpc) toward the south in the stacked moment map.
\begin{figure*}[ht]
\resizebox{\hsize}{!}{\includegraphics{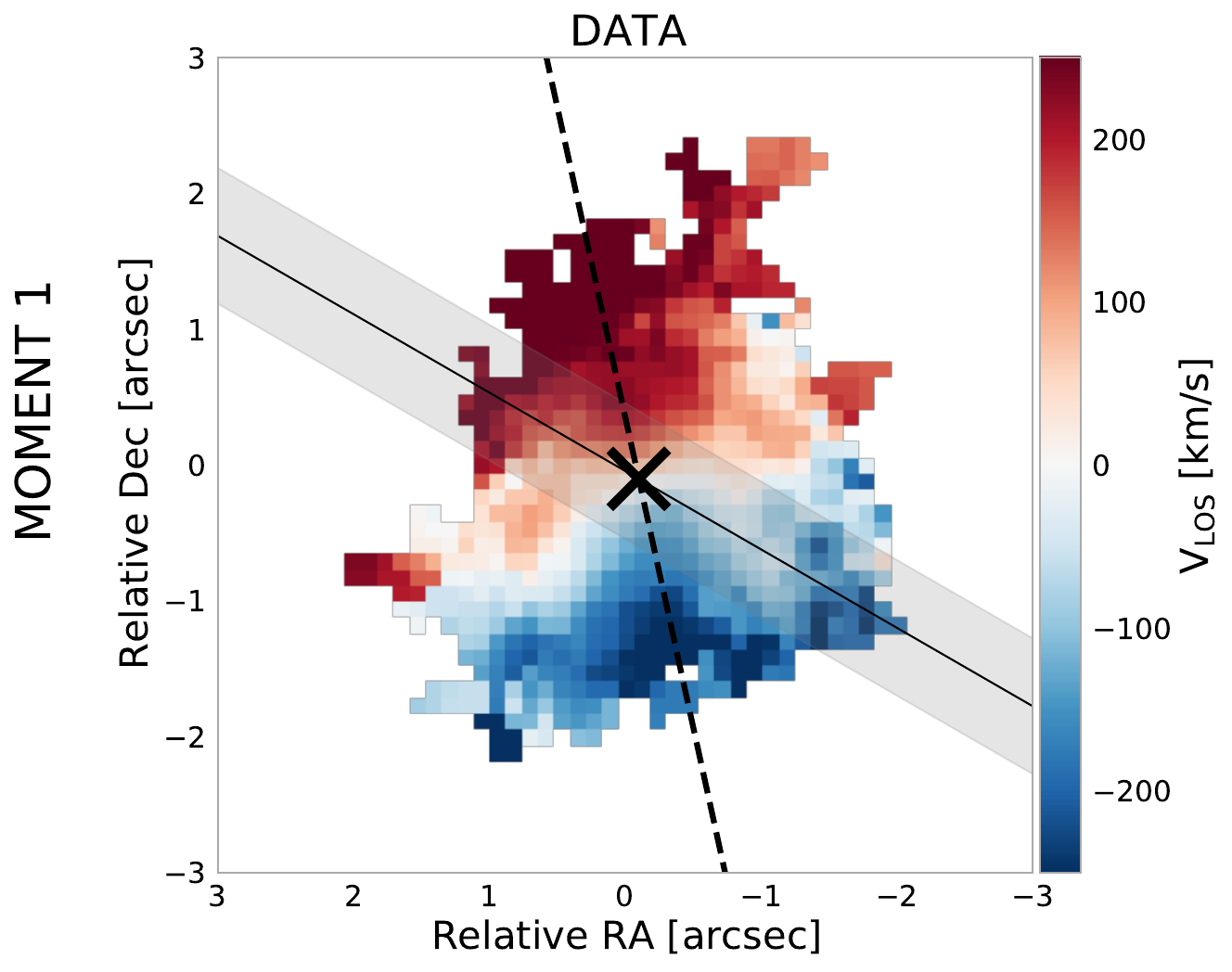}
\includegraphics{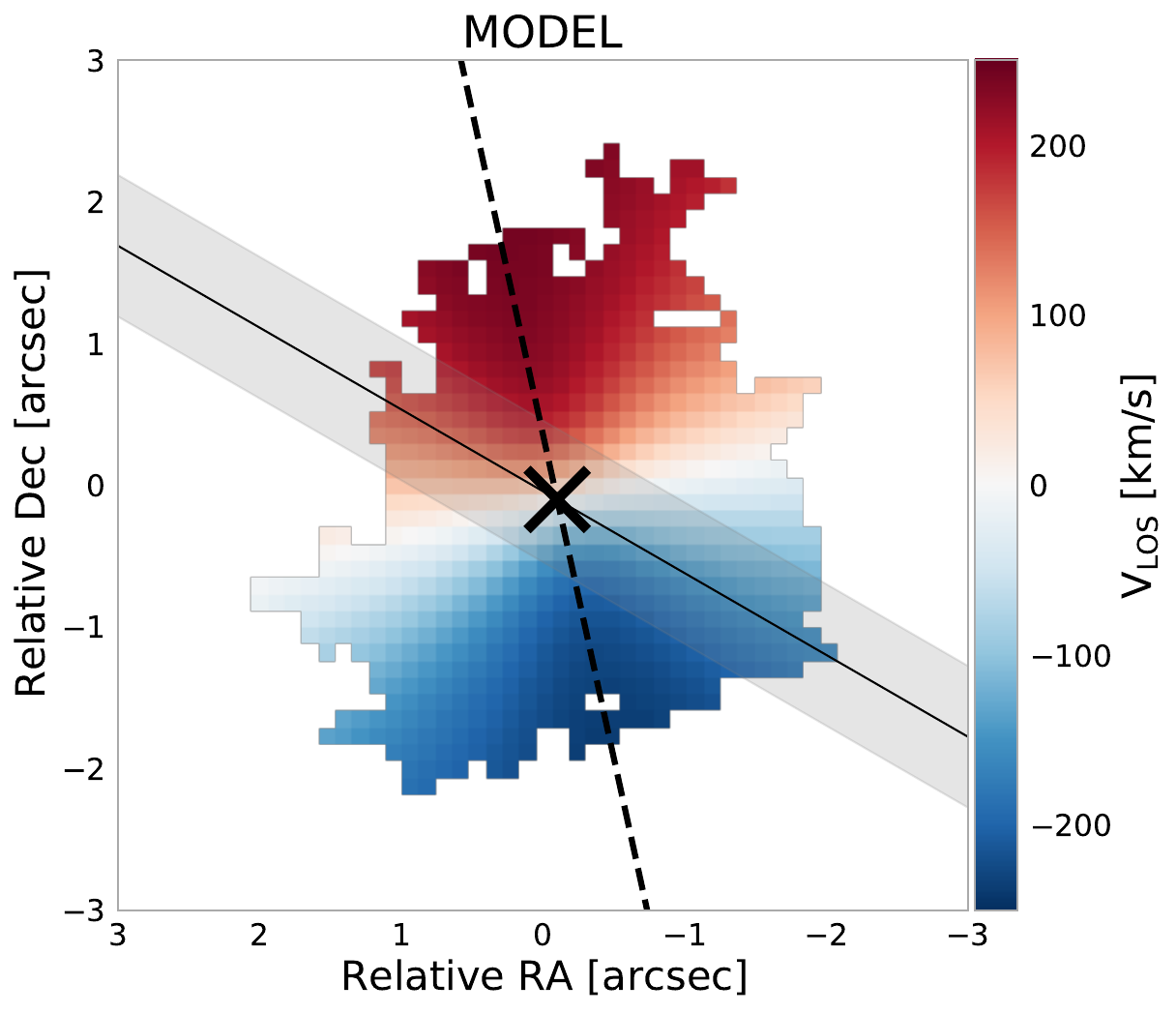}
\includegraphics{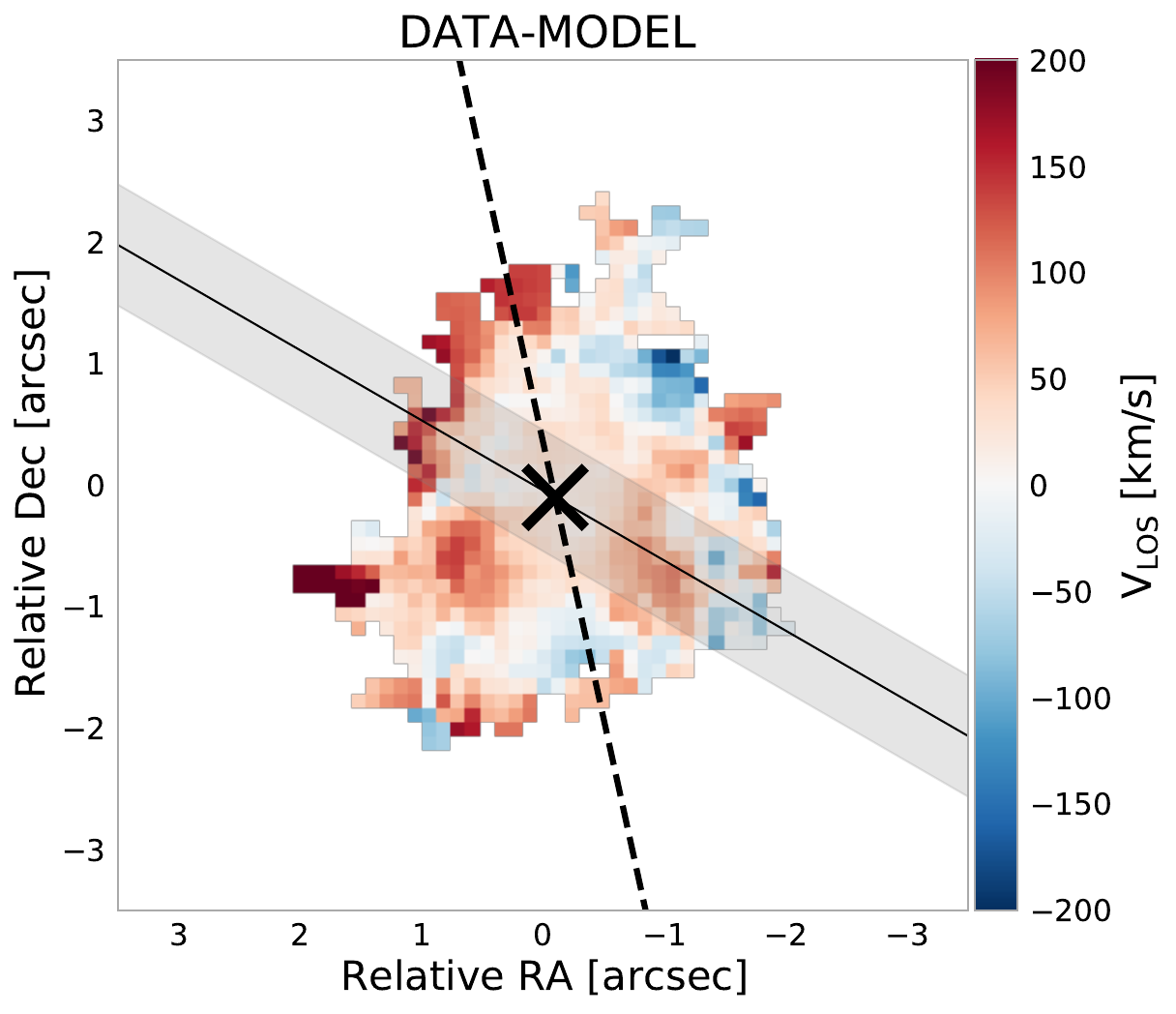}}
\resizebox{\hsize}{!}{\includegraphics{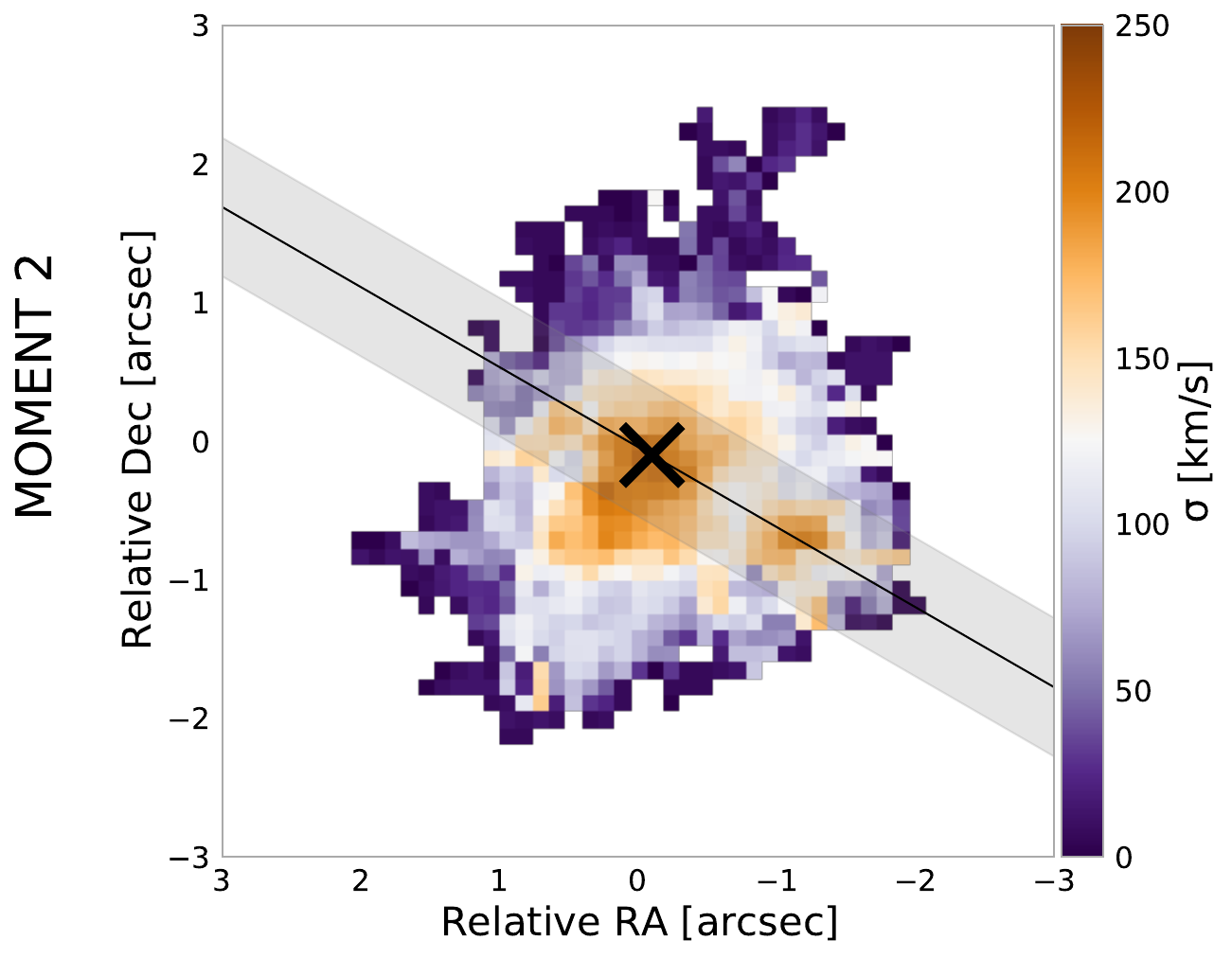}
\includegraphics{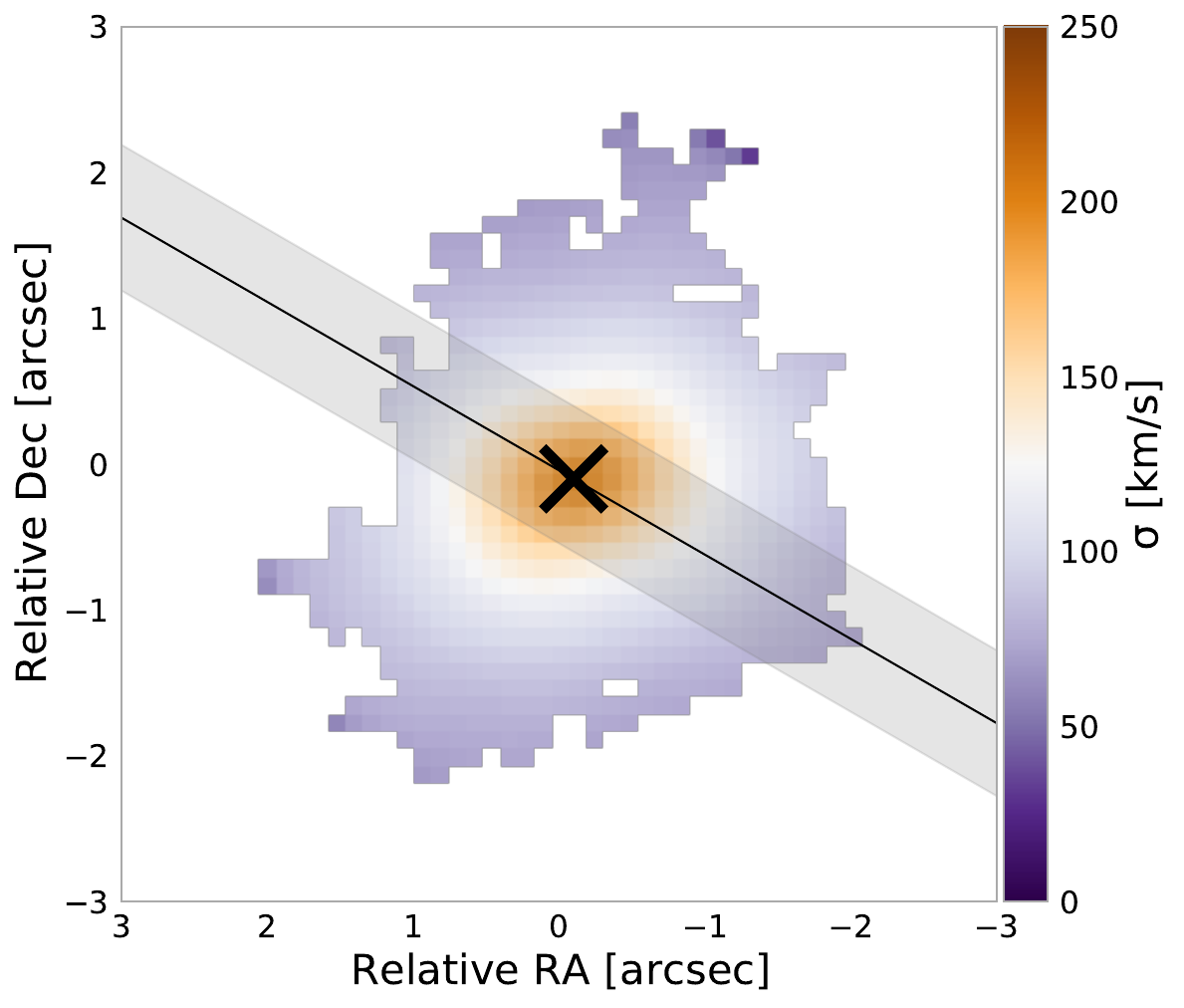}
\includegraphics{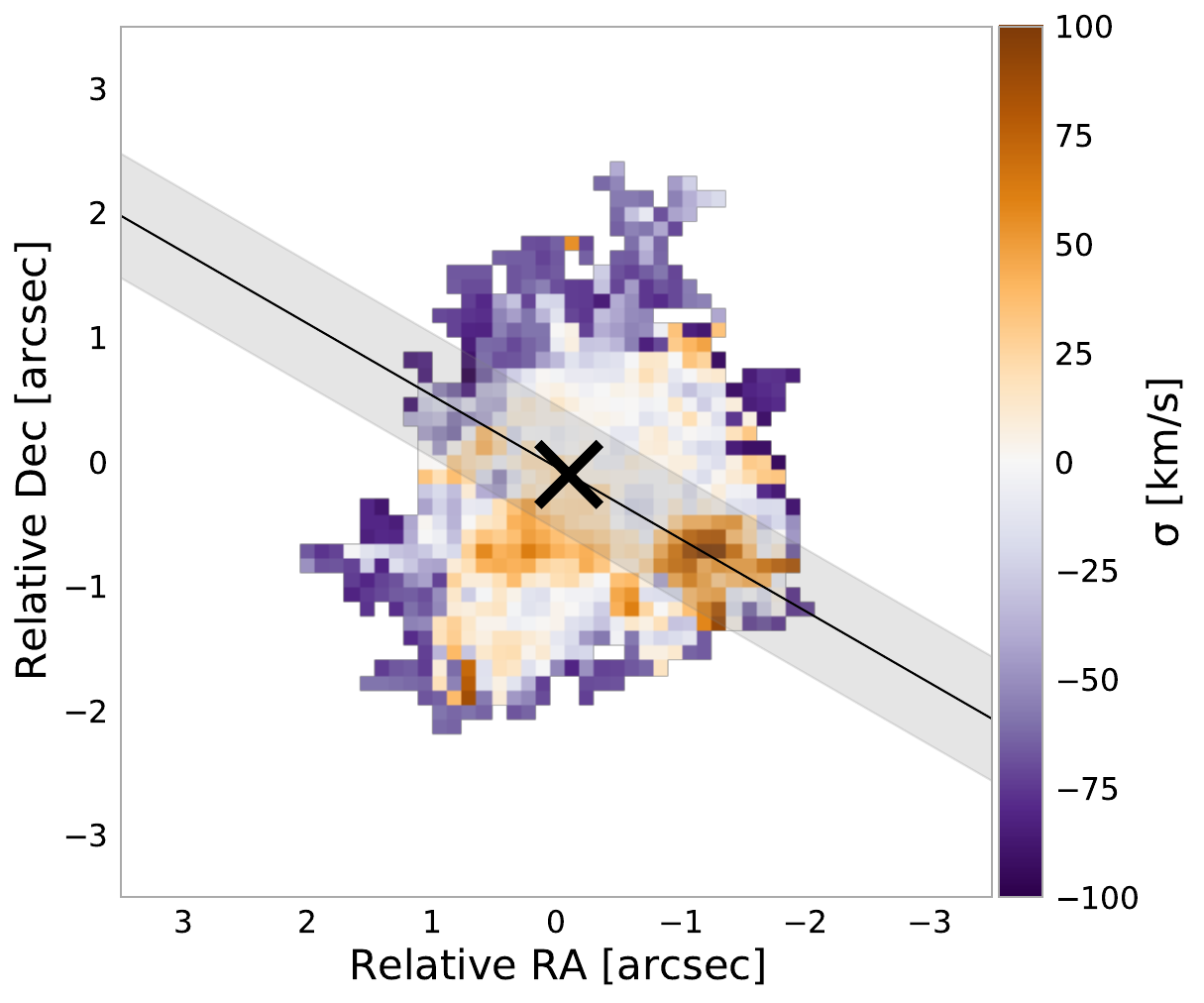}}
 \caption{\hdone~moment 1 and 2 maps (i.e. velocity and velocity dispersion) of the Teacup. The \barolo~models of the moment 1 and 2 maps and the corresponding residuals are also shown. Regions below 2$\sigma$ are masked out. The black cross mark the peak of the near-infrared continuum as in Figure \ref{fig:mom0}. The kinematic major axis is shown as a black dashed line in the moment 1 maps. The $\rm PA_{jet}$ = \ang{60} is indicated as a black solid line and gray shaded area. The maps have been smoothed using a two-spaxel boxcar for presentation purposes. North is up and east to the left.}
  \label{fig:j1430_barolo}
\end{figure*} 

Focusing on J1356, we notice that the \hdtwo~intensity peak is shifted by about \SI{0.5}{\arcsecond} in the south direction with respect to the \hdone~intensity peak. In fact, the resulting map, obtained by stacking the two lines, shows a double peaked feature due to this mismatch between the S(1) and S(2) intensity peaks. J1356 shows an irregular and complex gas morphology, with the \hdone~line showing elongated emission extending southward, exceeding \SI{2}{\arcsecond} ($\sim$4.2 kpc) and also diffuse emission to the northeast and west regions of the map. The [O III] emission also appears elongated towards the south (\SP). 
In contrast, the S(2) emission is concentrated toward the center, with a maximum extension of $\sim$ \SI{1.5}{\arcsecond} (3.3 kpc), mainly in the northeast region of the map. The CO(2-1) emission is also quite compact, concentrated toward the central \SI{0.5}{\arcsecond} ($\sim$1 kpc) radius region (\RA).
Instead, the intensity map obtained from the stacked cube clearly shows that the diffuse emission is dominated by the \hdone~line, exhibiting a similar morphology to the S(1) intensity map. 

\begin{figure*}
\resizebox{\hsize}{!}{\includegraphics{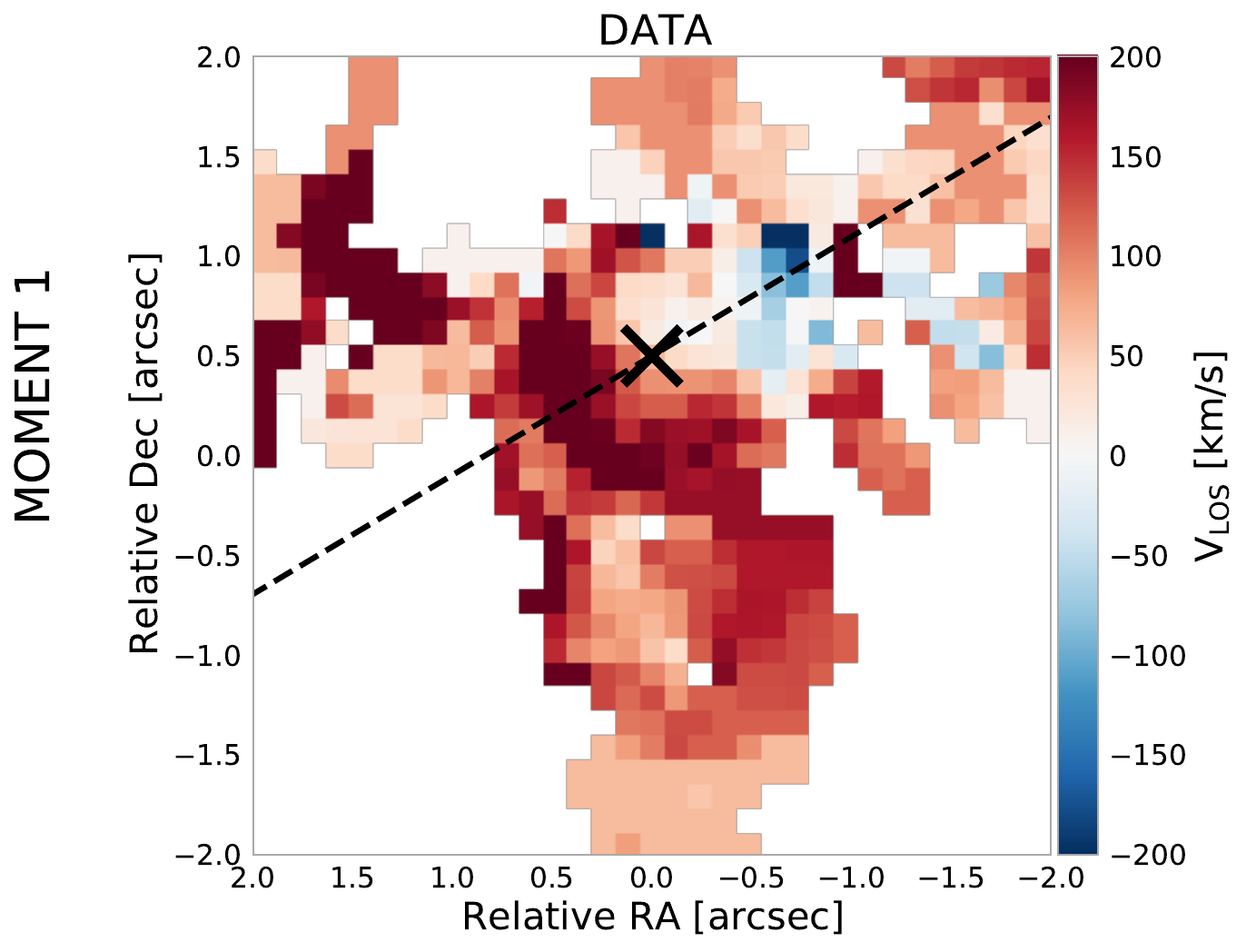}
\includegraphics{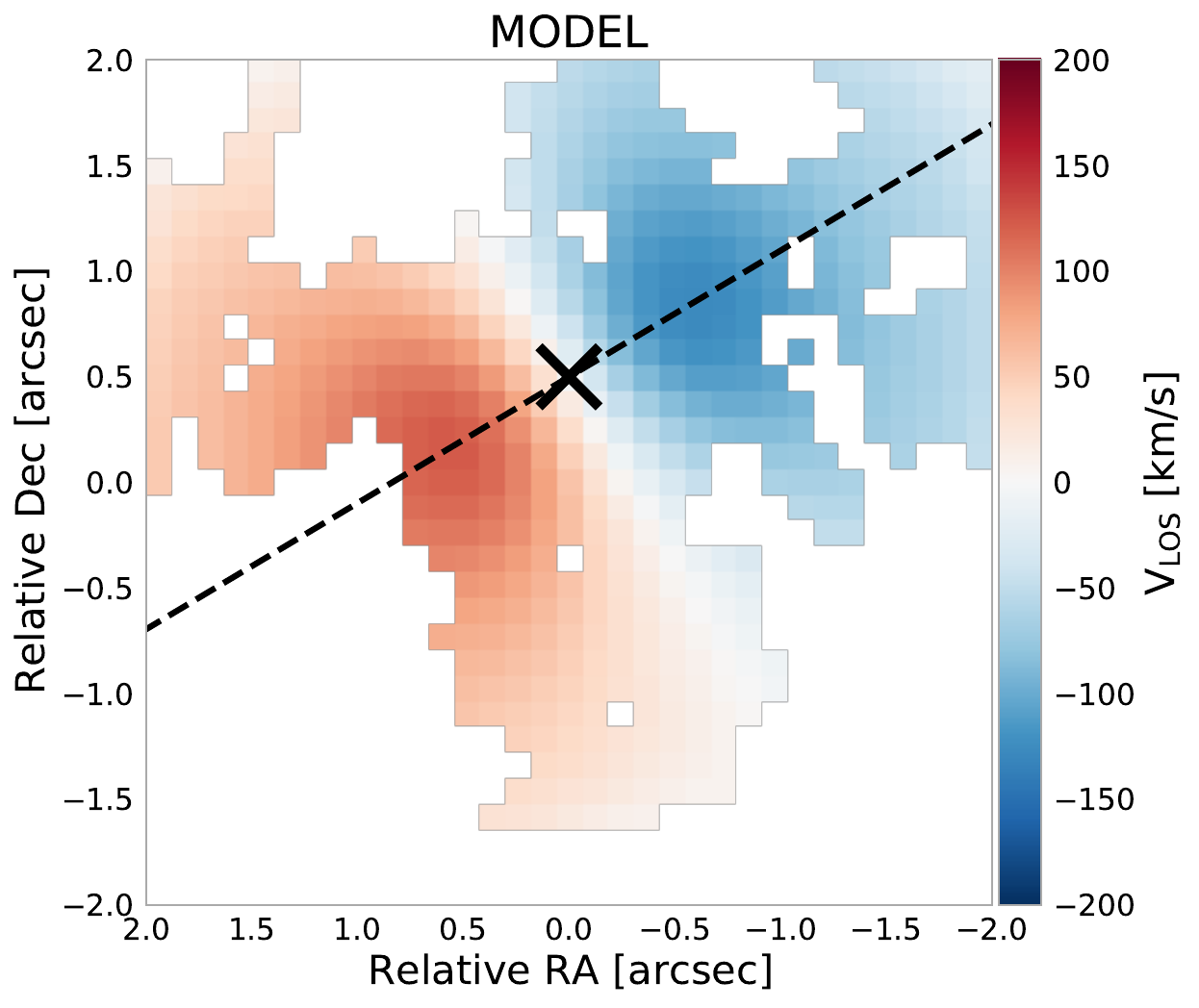}
\includegraphics{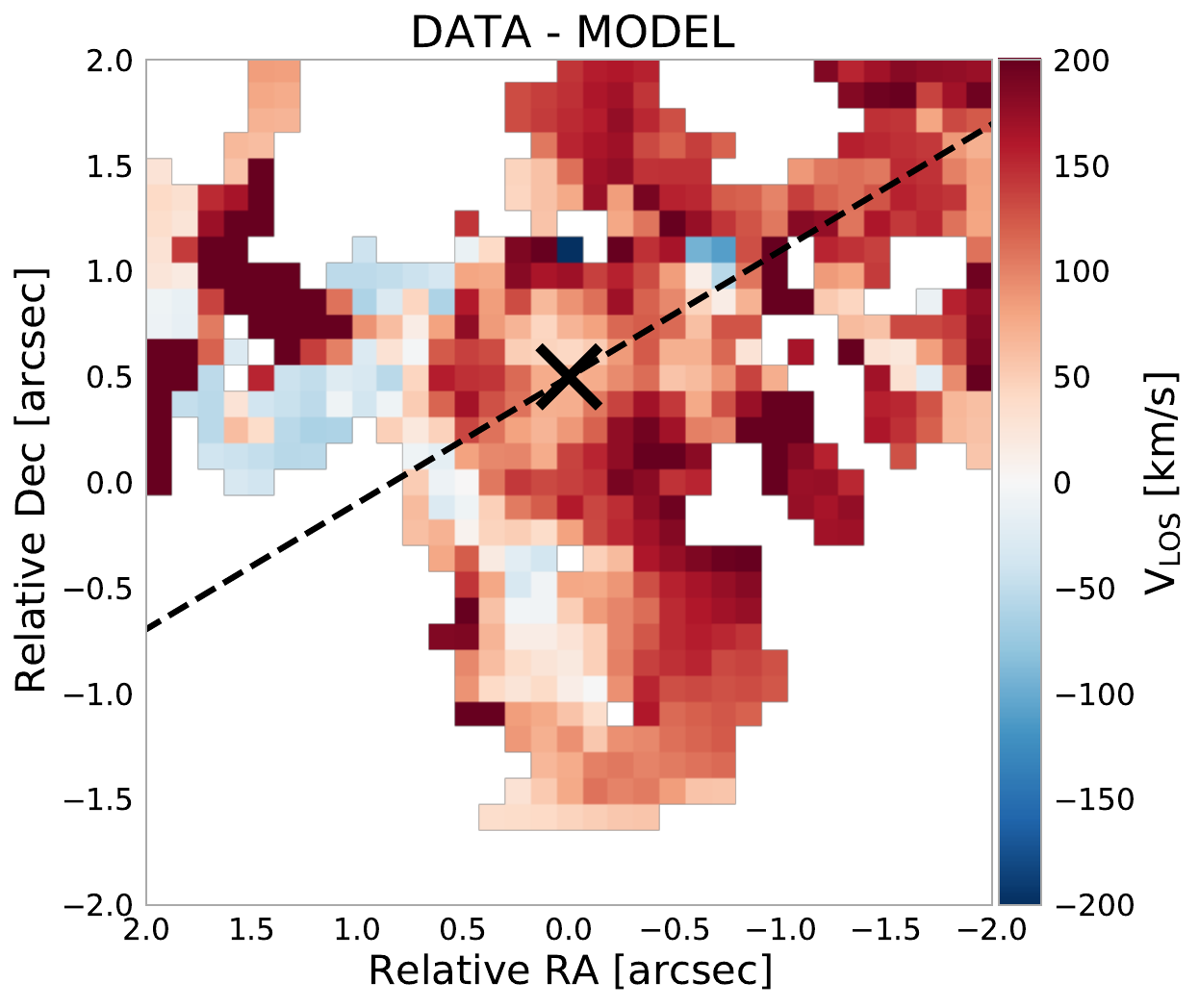}}
\resizebox{\hsize}{!}{\includegraphics{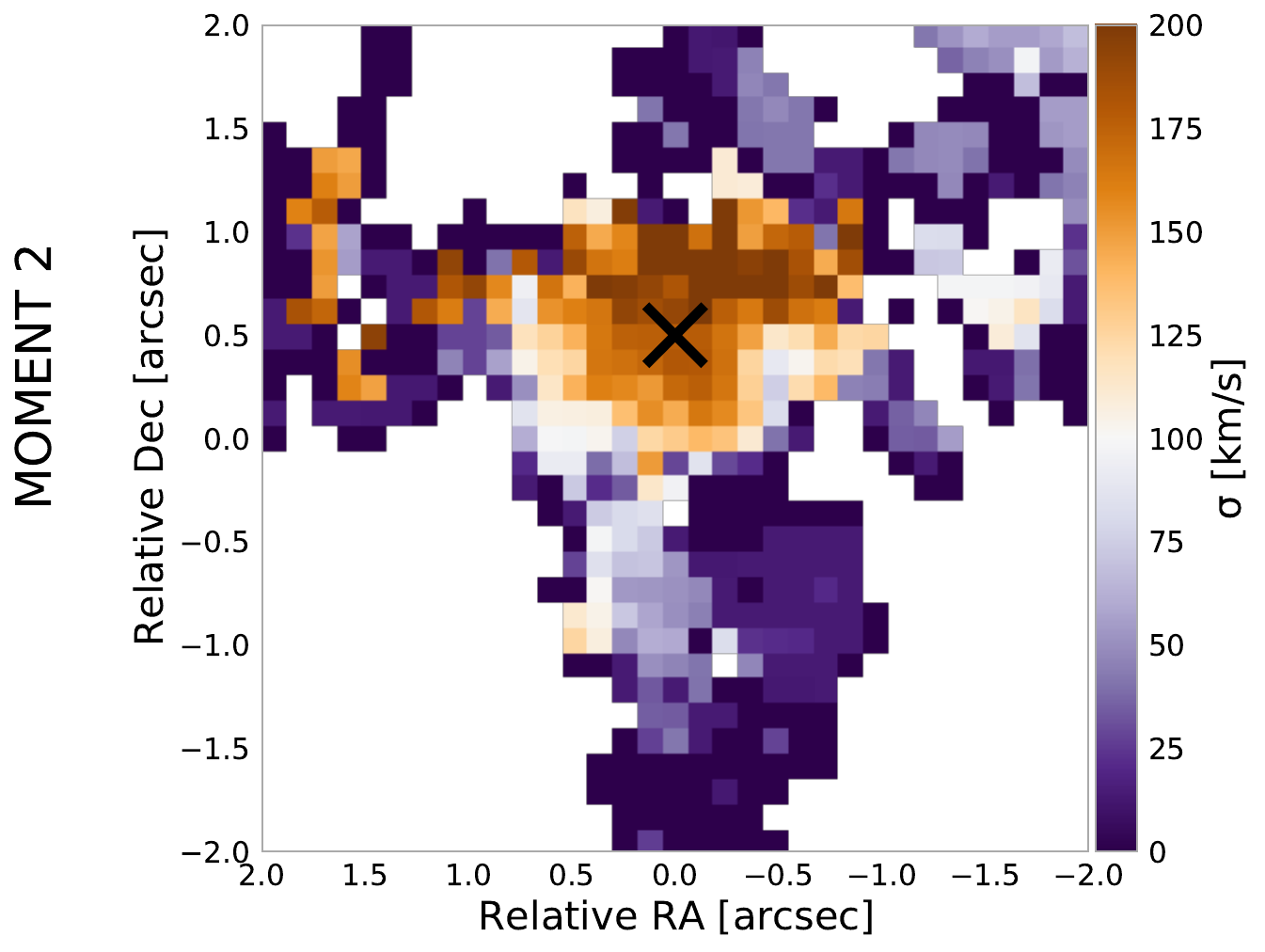}
\includegraphics{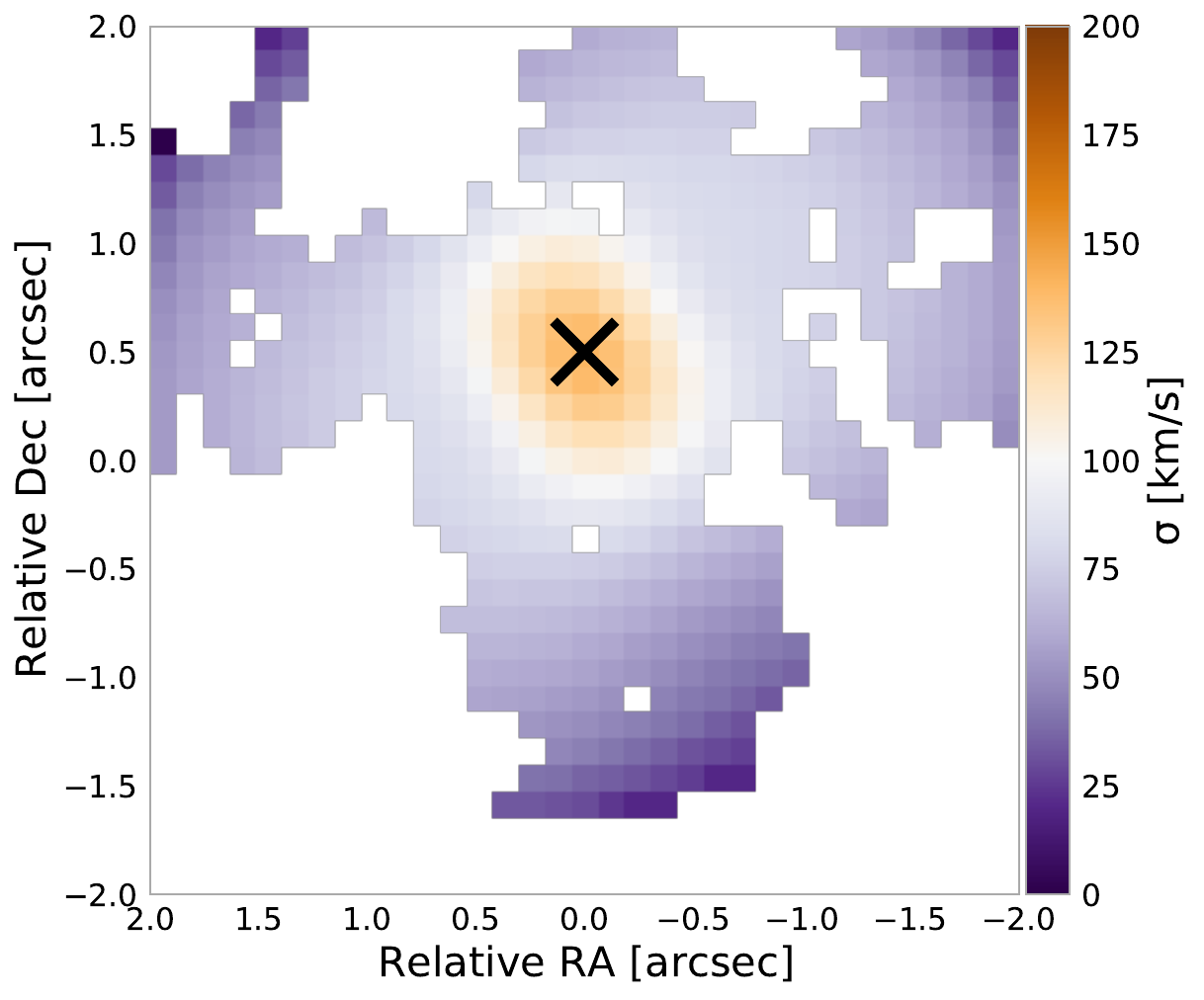}
\includegraphics{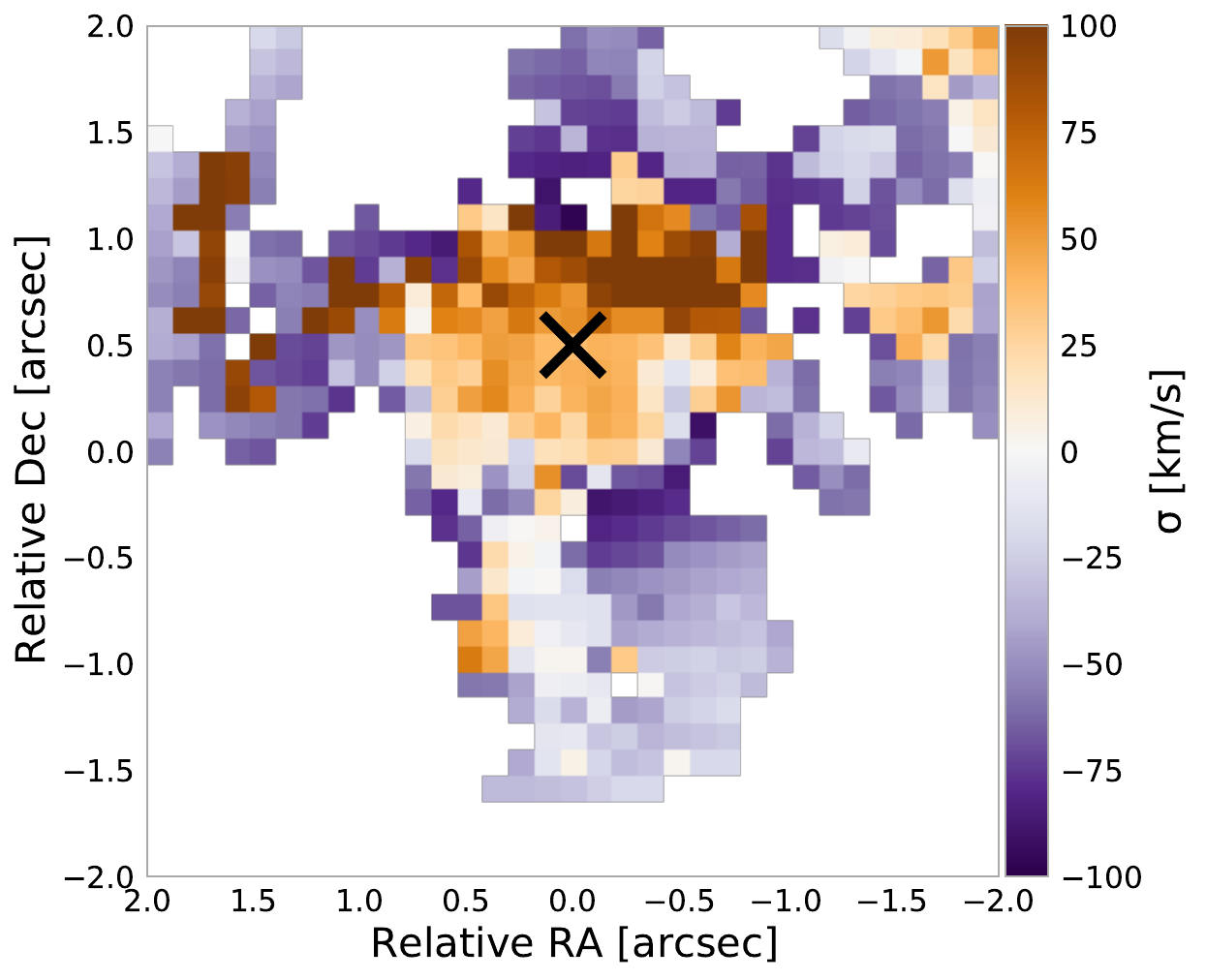}}
 \caption{Same as in Figure \ref{fig:j1430_barolo} but for J1356.}
  \label{fig:j1356_barolo}
\end{figure*}

\subsection{\hd kinematics}\label{sec:kiematics}

In order to interpret the warm molecular gas kinematics, we built a dynamical model of the systems fitting the observed \hdone~data cubes with the 3D-Based Analysis of Rotating Objects from Line Observations \citep[\barolo,][]{diteodoro}. 
We fit a 3D tilted-ring model to J1430 and J1356 \hdone~data cubes using \SI{0.4}{\arcsecond} and \SI{0.3}{\arcsecond} wide annuli up to a maximum radial distance from the AGN of \SI{1.6}{\arcsecond} and \SI{1.2}{\arcsecond}, respectively. 
In the first run, we allowed four parameters to vary: rotation velocity, velocity dispersion, disc inclination, and position angle, fixing the kinematic center to the peak position of the near-infrared continuum emission. 
For J1356, we ran the model again fixing the inclination to \ang{52}, i.e., equal to the best fit disc inclination found by \RA, so that in this case only PA, rotation velocity, and velocity dispersion can vary. We follow this approach in this case because the inclination and position angle are strongly degenerate due to the lower S/N.

By subtracting the model velocity map from the observed one, we obtained mean-velocity residual maps that we used to investigate deviations from circular motions. We did the same with the velocity dispersion maps (see Figures \ref{fig:j1430_barolo} and \ref{fig:j1356_barolo}).
Finally, we produced position-velocity diagrams (PVDs) along the kinematics minor and major axis, adopting a slit width of \SI{0.8}{\arcsecond}, i.e., equivalent to the spatial resolution of the SINFONI data cubes (see Figures \ref{fig:PVD} and \ref{fig:PVD_stacked}).

\subsubsection{The Teacup}

Figure \ref{fig:j1430_barolo} shows the \hdone~velocity (moment 1) and velocity dispersion (moment 2) maps of the Teacup. 
As already reported by \citet{ramosalmeida2017}, the Teacup \hdone~velocity field is dominated by rotation with a gradient approximately along the north-south direction and maximum velocities of $\pm$ 250 \kms.
The rotation pattern is also clearly visible from the S-shaped structure in the PVD taken along PA = \ang{13} $\pm$ \ang{6}, which is the kinematic major axis according to our modeling with \barolo~(see Figure \ref{fig:PVD}, top left panel).
The \hdone~moment 2 map shows high values of velocity dispersion up to $\sim$200 \kms~in the central region and orthogonal to the radio jet orientation ($\rm PA_{jet}$=\ang{60}), that decrease towards the outer regions. The kinematics maps shown in this work are overall consistent with those presented in \citet{ramosalmeida2017} using VLT/SINFONI data and in \RA~and \citet{audibert2023} using ALMA CO(2-1) and CO(3-2) data, respectively. 

The moment 1 and 2 maps of J1430 \barolo~model and the corresponding residual moment maps (data-model) are shown in the middle and right panels of Figure \ref{fig:j1430_barolo}. The \barolo~disc model provides a position angle PA=\ang{13} and inclination i=\ang{42} as best fit parameters.
Therefore, the kinematic major axis of the warm \hd gas disc differs from the CO major axis \citep[$\rm PA_{CO}$ = \ang{4}, \RA;][]{audibert2023}, from the [O III] major axis ($\rm PA_{[O III]}$ = \ang{27}, \citealt{harrison2014,venturi2023}; \SP), and from the galaxy major axis ($\rm PA_{gal}$ =\ang{-19}). 
This could be related with the past merger event that Teacup undergone and/or with the influence of the jet on the ionized and molecular gas.
The velocity residual moment map, shown in the top right panel of Figure \ref{fig:j1430_barolo}, does not show strong residuals.
The largest residuals are detected in the residual velocity dispersion map, shown in the bottom right panel of Figure \ref{fig:j1430_barolo}, and they are located towards the southwest, along PA = \ang{240}, which is the direction of the jet (PA = \ang{60} = \ang{240}).

Figure \ref{fig:PVD} shows the PVDs extracted from the \hdone~cube: the left panels correspond to the PVDs extracted along the J1430 kinematic major (top) and minor (bottom) axis, and the middle panels to the PVDs extracted along (top) and perpendicular (bottom) to the jet direction. The blue and red contours trace the \hdone~emission above 2$\sigma$ in SINFONI data cube and \barolo~model cube, respectively. Overall, the \barolo~model reproduces well the kinematics of J1430. We consider as potential non circular motions only the gas emission outside \barolo~contours at 2$\sigma$. In the case of the Teacup, these non circular motions are not very extended and for this reason we also considered the PVDs extracted from the stacked cube (see Figure \ref{fig:PVD_stacked}).
The PVDs obtained along the kinematic major and minor axis show high velocity gas emission ($v > |300|$ \kms) within the central \SI{1}{\arcsecond} radius region that corresponds to non circular motions. In particular, red-shifted high velocity gas ($v > 300$ \kms) is detected to the north along the kinematic major axis and to the west along the kinematic minor axis. In the PVDs extracted perpendicular to the jet direction from the stacked cube (bottom central panel of Figure \ref{fig:PVD_stacked}), we also to detect blue-shifted high-velocity gas ($v < -300 $ \kms) not accounted for by rotation.
Furthermore, red-shifted gas at $v \sim 250$ \kms~outside \barolo~contours is detected along the kinematic minor axis to the west, along the jet direction to the south-west, and perpendicular to the jet direction to the south-east. 

\subsubsection{J1356}

The \hdone~moment 1 and 2 maps of J1356 are shown in Figure \ref{fig:j1356_barolo}, as well as the corresponding model and residual (data-model) moment maps.
The \hdone~moment 1 map shows positive velocities to the southeast and negative velocities to the northwest.
The moment 2 data map shows high velocity dispersion values of up to 200 \kms~in the center. Overall, the kinematics is quite complex and disturbed because of the ongoing merger.

The \barolo~best-fit model is a rotating disc with a position angle PA = \ang{121} $\pm$ \ang{8} and inclination fixed to \ang{52}, although as can be seen from the large residuals, the gas kinematics cannot be well reproduced by rotation. 
In this case the kinematic major axis is consistent with the CO major axis ($\rm PA_{CO}$ = \ang{110}, \RA), therefore the \hdone~and CO emission are tracing gas in the same disc. This warm and cold molecular gas is coplanar with the galaxy disc but has a different PA ($\rm i_{gal}$ = \ang{55}, $\rm PA_{gal}$ = \ang{156}, \RA, from r-band SDSS DR6 photometry). 
In contrast, the [O III] kinematics is consistent with a rotation disc of kinematic major axis equal to \ang{45}  (\citealt{harrison2014}; \SP). 
However, a misalignment between the galaxy and the nuclear molecular discs is not surprising in an ongoing major merger.

\begin{figure*}
\resizebox{\hsize}{!}{
\includegraphics{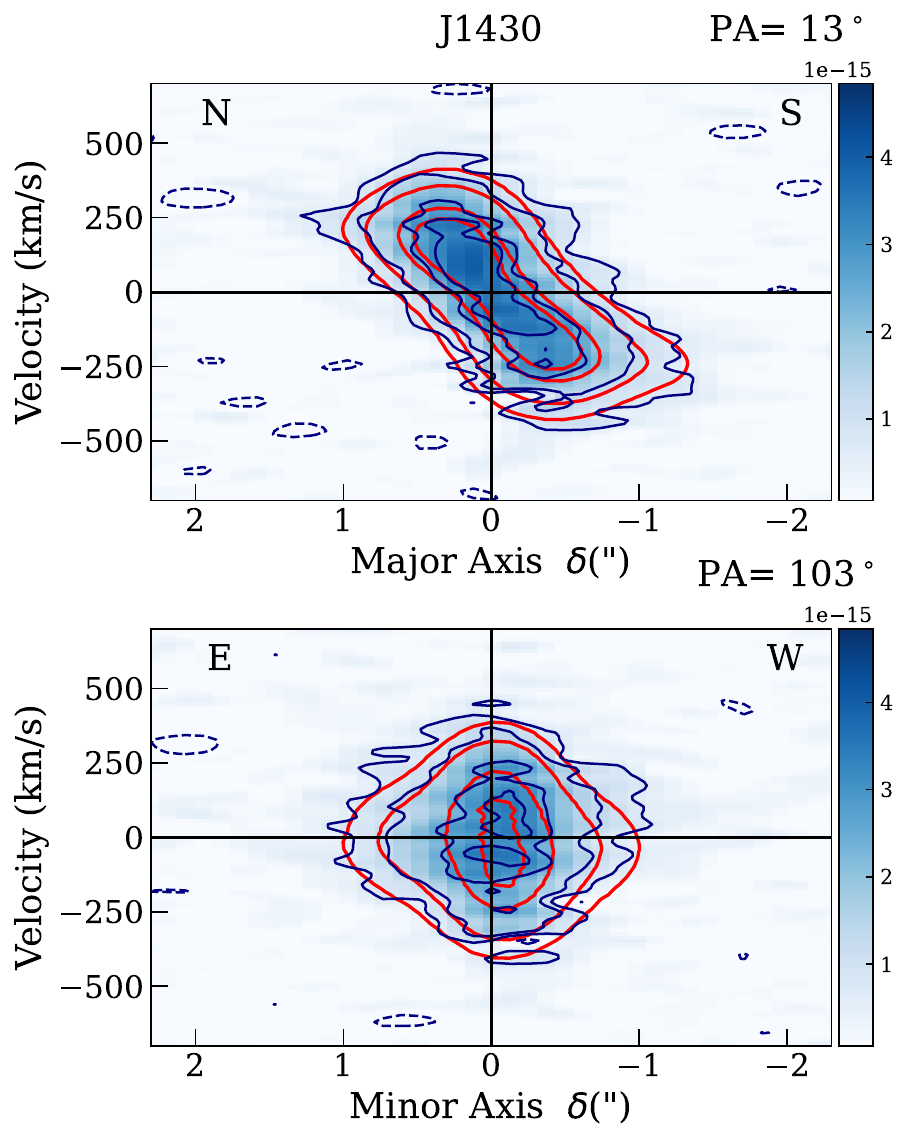}
\includegraphics{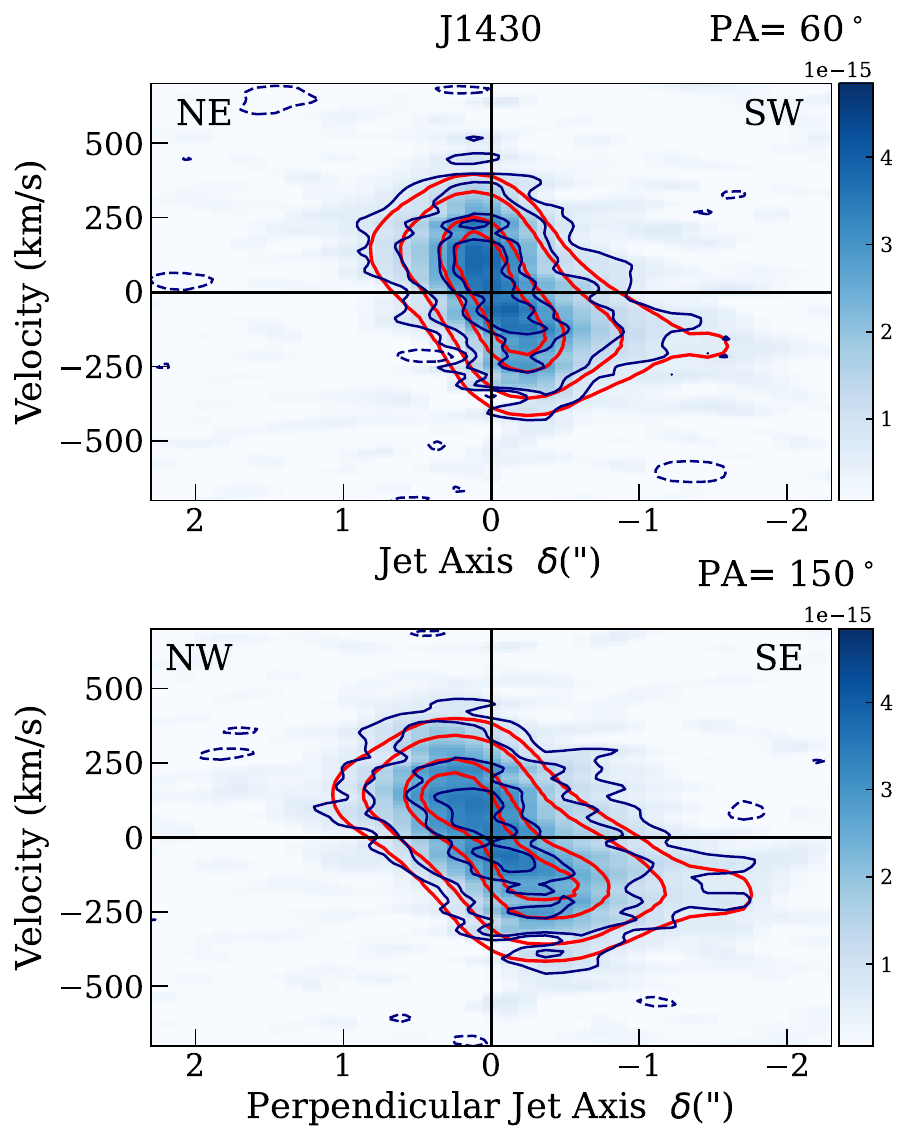}
\includegraphics{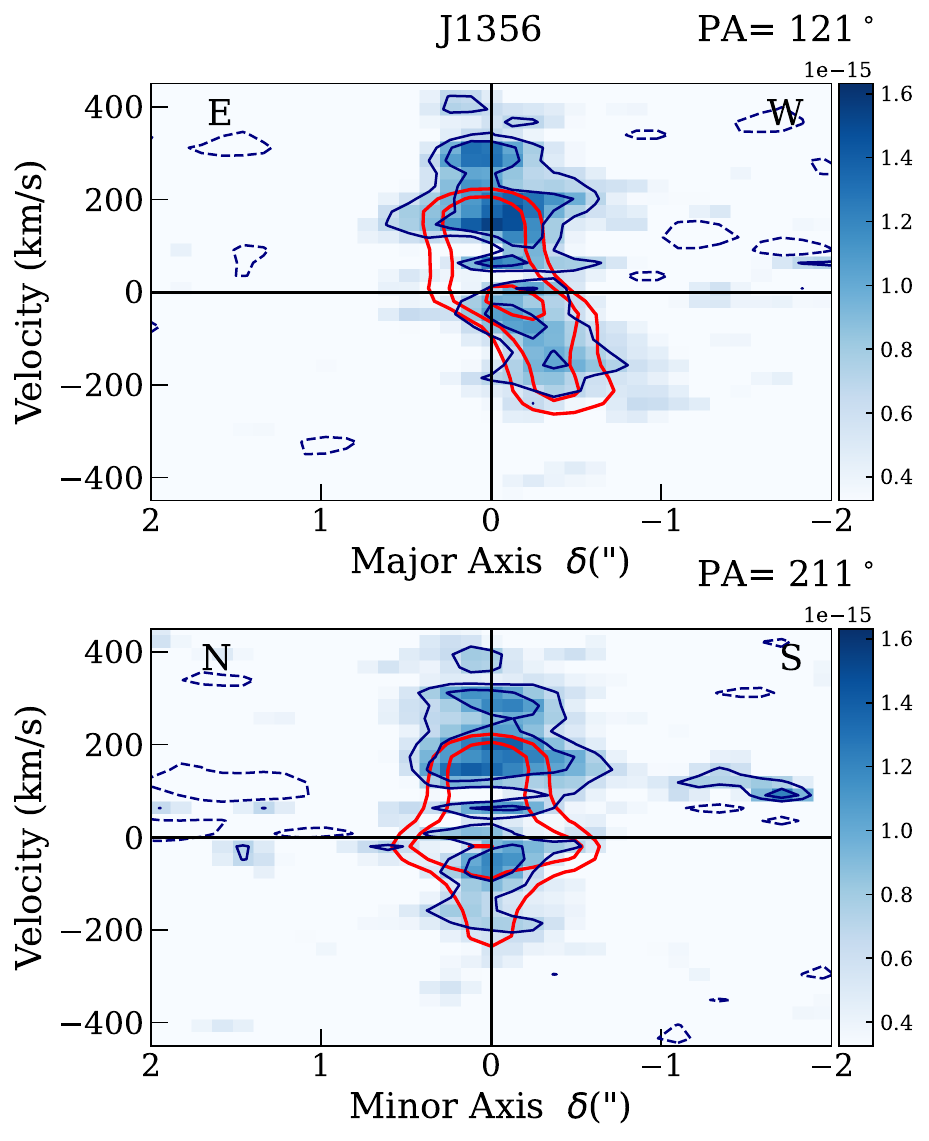}
}
\caption{PVDs extracted from the \hdone~emission line data cubes. The blue contours trace the \hdone~emission in SINFONI data cube, while the red contours trace the corresponding \barolo~model. Left panels show the PVDs along the kinematic major (PA=\ang{13}, top) and minor axis (PA=\ang{103}, bottom) in J1430. Middle panels show the PVDs along (PA=\ang{60}, top) and perpendicular (PA=\ang{150}, bottom) to the radio jet direction in J1430. In the left and middle panels the blue and red contours are drawn at (2, 4, 8, 10)$\sigma$ with $\sigma$ = 3.23 $\rm 10^{-19}$ \noise. Right panels: PVDs along the kinematic major (PA=\ang{121}, top) and minor axis (PA=\ang{211}, bottom) in J1356. The blue and red contours are drawn at (2, 3, 5)$\sigma$ with $\sigma$ = 3.26 $\rm 10^{-19}$ \noise. All the PVDs were extracted using a slit of width \SI{0.8}{\arcsecond}.}
\label{fig:PVD}
\end{figure*}

Large residual emission is shown in both moment 1 and 2 residual maps.
The residual mean velocity map shows mainly red-shifted velocities concentrated toward the center. Similarly, the residual velocity dispersion map exhibits the strongest perturbations in the inner \SI{1}{\arcsecond} ($\sim$2.2 kpc) radius region, reaching values of up to 100 \kms. In fact, only the red-shifted side of the CO outflow was detected \citep[\RA; ][]{sun2014,audibertprep}. 
The right panels of Figure \ref{fig:PVD} show the \hdone~PVDs obtained along the kinematic major (top) and minor (bottom) axis, and the same PVDs, but extracted from the stacked cube are shown in Figure \ref{fig:PVD_stacked}.
As in the Teacup, we identify gas emission outside \barolo~at the 2$\sigma$ level as potential non circular motions.
We detect high-velocity gas 
%detected the largest residuals outside \barolo~ contours at 2$\sigma$, 
in the inner \SI{0.8}{\arcsecond} radius region, both along the kinematic major and minor axis. These high-velocity non circular motions have positive velocities, $\rm v_{max} \sim$ 400 \kms, as it was also the case for CO (\RA).
Along the kinematics major axis to the west we barely resolve red-shifted low velocity gas, with v$\sim$ 100-200 \kms, outside the \barolo~model contours.   
However, we identify as outflow only the high-velocity non circular motions (see Section \ref{sec:outflow}). 
%The blue-shifted side of the warm \hd outflow in J1356 is detected only the PVDs extracted from the stacked cube at 1$\sigma$, see Figure \ref{fig:PVD_stacked}.

\subsection{Warm molecular outflows}\label{sec:outflow}

In both QSO2s we detect high-velocity molecular gas that cannot be accounted for by rotation. This is evident from the PVDs reported in Figures \ref{fig:PVD} and \ref{fig:PVD_stacked}, where the SINFONI data (blue contours) exceeds \barolo~disc model (red contours) at the same $\sigma$. In undisturbed spiral galaxies, such as the other two QSO2s studied by \RA, cold molecular outflows are usually coplanar with the CO discs, and non circular motions consistent with outflowing gas are usually found along the kinematic minor axis (e.g. \RA). However, as discussed in \citet{audibert2023}, in spheroidal galaxies which are undergoing or have undergone a galaxy interaction/merger, the molecular outflows might have a more 3D geometry and their preferential direction might not necessarily be the minor axis (see, e.g. the case of the Teacup in CO). Therefore, here we follow the prescription from Scenario \textsc{II} described in \citet{audibert2023} for measuring outflow masses.

We assume that only the high-velocity gas participates in the outflow, i.e., only the gas faster than the maximum velocity of the \barolo~models at 2$\sigma$, shown in Figure \ref{fig:PVD}. In the case of Teacup this corresponds to velocities larger than $\pm$370 km~s$^{-1}$, and in J1356, to gas faster than 270 km~s$^{-1}$, without blueshifted counterpart. Although we cannot discard that part of this high-velocity gas corresponds to complex motions associated with the mergers, specially in the case of J1356 (see \RA~and \citealt{audibertprep}), here we assume that it corresponds to outflowing gas.

Then, to derive the outflow flux, we created integrated intensity maps by selecting only channels above these velocities and emission above 2$\sigma$.
%Then, we selected the emission above 2$\sigma$ in the velocity ranges [-660 \kms , -370 \kms] and [370 \kms , 650 \kms] in J1430, and in the velocity ranges [-465 \kms , -270 \kms] and [270 \kms , 510 \kms] in J1356. 
The velocity channel maps are shown in Figures \ref{fig:J1430_chanmap} and \ref{fig:J1356_chanmap}, and the integrated intensity maps of the blue- and red-shifted high-velocity components are shown in Figure \ref{fig:outflowmaps}. We note that the integrated intensity outflow maps shown in Figure \ref{fig:outflowmaps} and corresponding fluxes measured from them do not include the emission coming from spikes of noise that we spotted in the channel maps, which are probably residuals of sky subtraction. These noise spikes can be seen in the channel map at -491 \kms~in the case of Teacup (to the south of the AGN; see Figure \ref{fig:J1430_chanmap}) and in the channel maps at -463 and -324 \kms~in J1356 (see Figure \ref{fig:J1356_chanmap}). 
We computed the outflow masses from the extinction-corrected, integrated outflow flux, converting the \hdone~line luminosity to warm molecular gas mass using Equation \ref{eq:mass} and the infrared extinctions reported in Section \ref{sec:warmgas}. The corresponding values are shown in Table \ref{tab:outflow}.
For the outflows of Teacup and J1356 we measured total warm molecular gas masses of $\rm M_{H_2} \sim 2.6 \times 10^3 $ and $\sim 1.5 \times 10^3 $ \msun, which correspond to 44 and 37\% of the mass of warm molecular gas measured in the central \SI{0.8}{\arcsecond} diameter of the galaxies, respectively.  

In order to derive the outflow mass rates we assume a time-averaged thin expelled shell geometry and we adopted the following equation:

\begin{equation}\label{eq:outflow}
    \dot{M}_{out} = v_{out} \times \frac{M_{out}}{r_{out}}.
\end{equation}

\noindent In this equation $\rm M_{out}$, $\rm v_{out}$, and $\rm r_{out}$ are the outflow mass, velocity, and radius.
We adopt as the outflow radius the maximum extent of the high-velocity gas contours at 3$\sigma$ level shown in Figure \ref{fig:outflowmaps},
while the outflow velocity is the luminosity-weighted average of the velocity channels adopted to derive the integrated outflow flux as in \RA~and \citet{audibert2023}. 
For the Teacup, from Figure \ref{fig:outflowmaps} we measure outflow radii of \SI{1.2}{\arcsecond} (1.9 kpc) for the blue- and red-shifted sides of the outflow, and outflow velocities of -470 \kms~and 430 \kms. From this we measure warm molecular outflow rates of 4.2 $\times 10^{-4}$ \sfr~and 2.0 $\times 10^{-4}$ \sfr, for the blue- and red-shifted sides of the outflow.

For J1356, we measure an outflow velocity of 370 \kms~for the red-shifted side of the outflow. From the red contours in Figure \ref{fig:outflowmaps} we measure an outflow radius of \SI{0.9}{\arcsecond} (2.0 kpc). From this we measure a warm molecular outflow rate of 2.9 $\times 10^{-4}$ \sfr~for the red-shifted side of the outflow.
Finally, we calculate the kinetic power of the warm molecular outflow as follows:

\begin{equation}\label{eq:Ekin}
    \dot{E}_{kin} = \frac{\dot{M}_{out}}{2} ~v_{out}^2. 
\end{equation}

\noindent With the above assumptions, we derive kinetic powers of $\rm log ~\dot{E}_{kin}$/(\ergs) $\sim$ 35 for both the two QSO2s.
The summary of the outflow properties is reported in Table \ref{tab:outflow}. Taking into account the total warm mass outflow rate of the Teacup of 6.26 \sfr, we obtain a total kinetic power $\rm log ~\dot{E}_{kin}$/(\ergs) = 35.6 for the warm molecular outflow.

\begingroup
\setlength{\tabcolsep}{8pt} % Default value: 6pt
\renewcommand{\arraystretch}{1.3} % Default value: 1
\begin{table*}
     \caption[]{Warm molecular outflow properties.}
         \label{tab:outflow}
\centering                          
\begin{tabular}{l c c c c c c c c c c c c}        
\hline  
QSO2 & \multicolumn{2}{c}{$\rm F_{H_2 1-0S(1)}$} & \multicolumn{2}{c}{$\rm M_{out}$} & \multicolumn{2}{c}{$\rm v_{out}$} & \multicolumn{2}{c}{$\rm r_{out}$} & \multicolumn{2}{c}{$\rm \dot{M}_{out}$}  & \multicolumn{2}{c}{$\rm log ~\dot{E}_{kin}$} \\  
 & \multicolumn{2}{c}{($\rm 10^{-16} ~erg ~s^{-1} cm^{-2}$)} & \multicolumn{2}{c}{($\rm 10^3$ \msun)} & \multicolumn{2}{c}{(\kms)} & \SI{}{(\arcsecond)} & (kpc)& \multicolumn{2}{c}{($\rm 10^{-4}$ \sfr)}  & \multicolumn{2}{c}{(\ergs)} \\ 
 & blue & red & blue & red & blue & red & & & blue & red  & blue & red\\ 
\hline
J1430 & 2.22 & 1.17  & 1.70 & 0.89 & -470 & 430 & 1.2 & 1.9 & 4.23 & 2.03 & 35.5 & 35.1\\
%J1356 & 0.28 & 0.88 & 0.47 & 1.49 & -390 & 370 & 0.8 & 1.8 & 1.06 & 3.20 & 34.7 &  35.1\\
J1356 & - & 0.88 & - & 1.49 & - & 370 & 0.9 & 2.0 & - & 2.86 & - &  35.1\\
\hline
\end{tabular}
\flushleft 
\tablefoot{The columns correspond to the integrated, extinction corrected \hdone~outflow flux, \hd mass, velocity, radius, mass ouflow rate, and kinetic energy. The outflow properties are reported for the blue- and red-shifted sides of the outflow in the case of the Teacup, and only for the red-shifted side in the case of J1356.}
\end{table*}

\section{Discussion}\label{sec:discussion}

In the present work, our aim is to complete the picture of the multiphase quasar-driven outflows for our QSO2s by adding the warm molecular measurements obtained in the near-infrared domain with SINFONI. We note that these warm molecular outflows are not detected in the nuclear spectra of the QSO2s shown in Figures \ref{fig:nuclear_fits} and \ref{fig:H2_nuclear_fits}. This is because  the amount of molecular gas participating in the outflow is small, the velocities are not high, and the signal-to-noise of the nuclear spectra at the reddest wavelengths is relatively low. Thanks to the analysis of the integral field data that we have done, including the fits with \barolo~and inspection of the PVDs, the warm molecular outflows were revealed.

Hereafter we compare the warm molecular gas properties with cold molecular ones, and then we discuss the contribution of the warm molecular phase to the multiphase outflow, its impact on the galaxy host, and the interplay of the compact radio jets with the gas of the ISM.

\begin{figure*}
\resizebox{\hsize}{!}{
\includegraphics{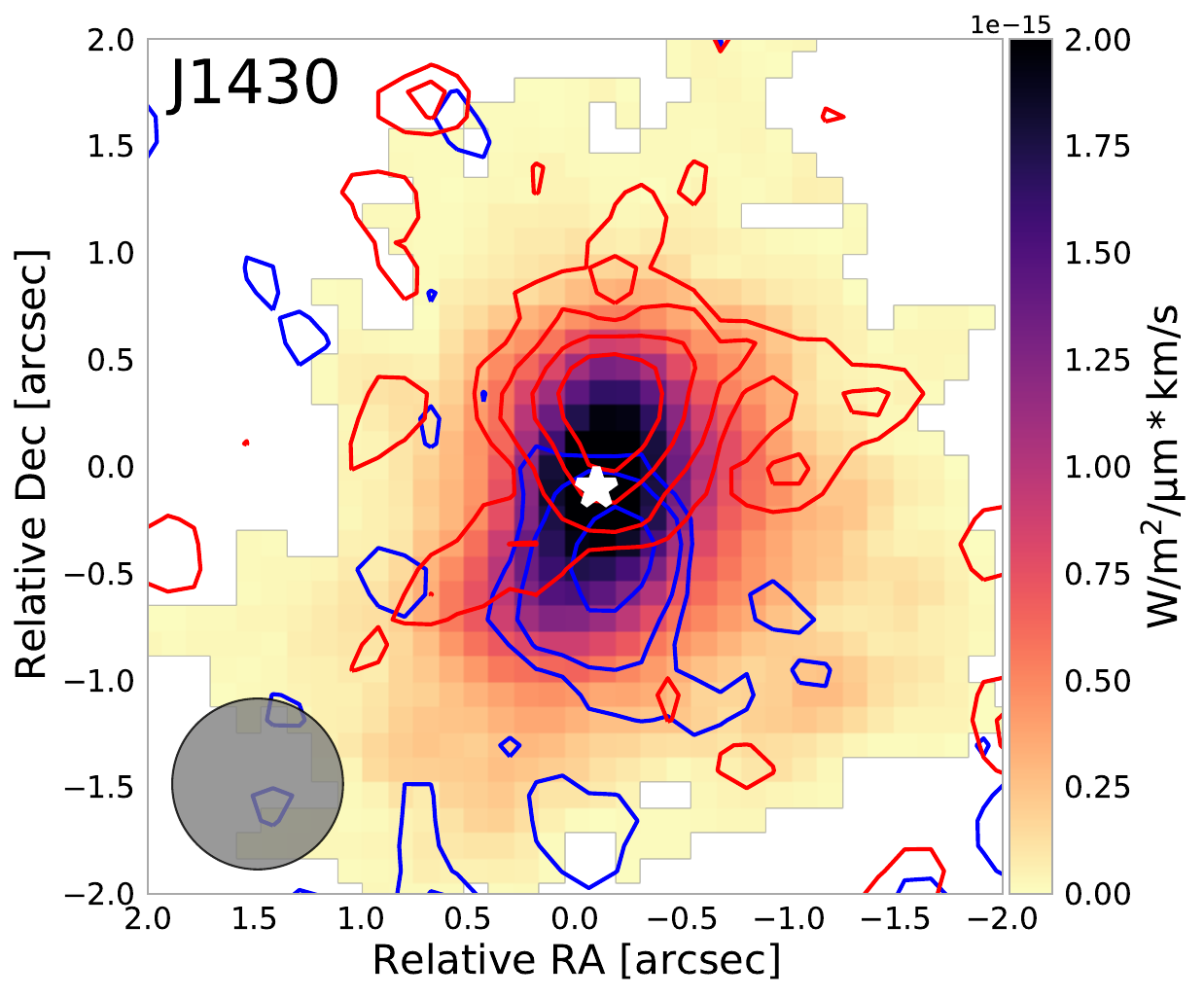}
\includegraphics{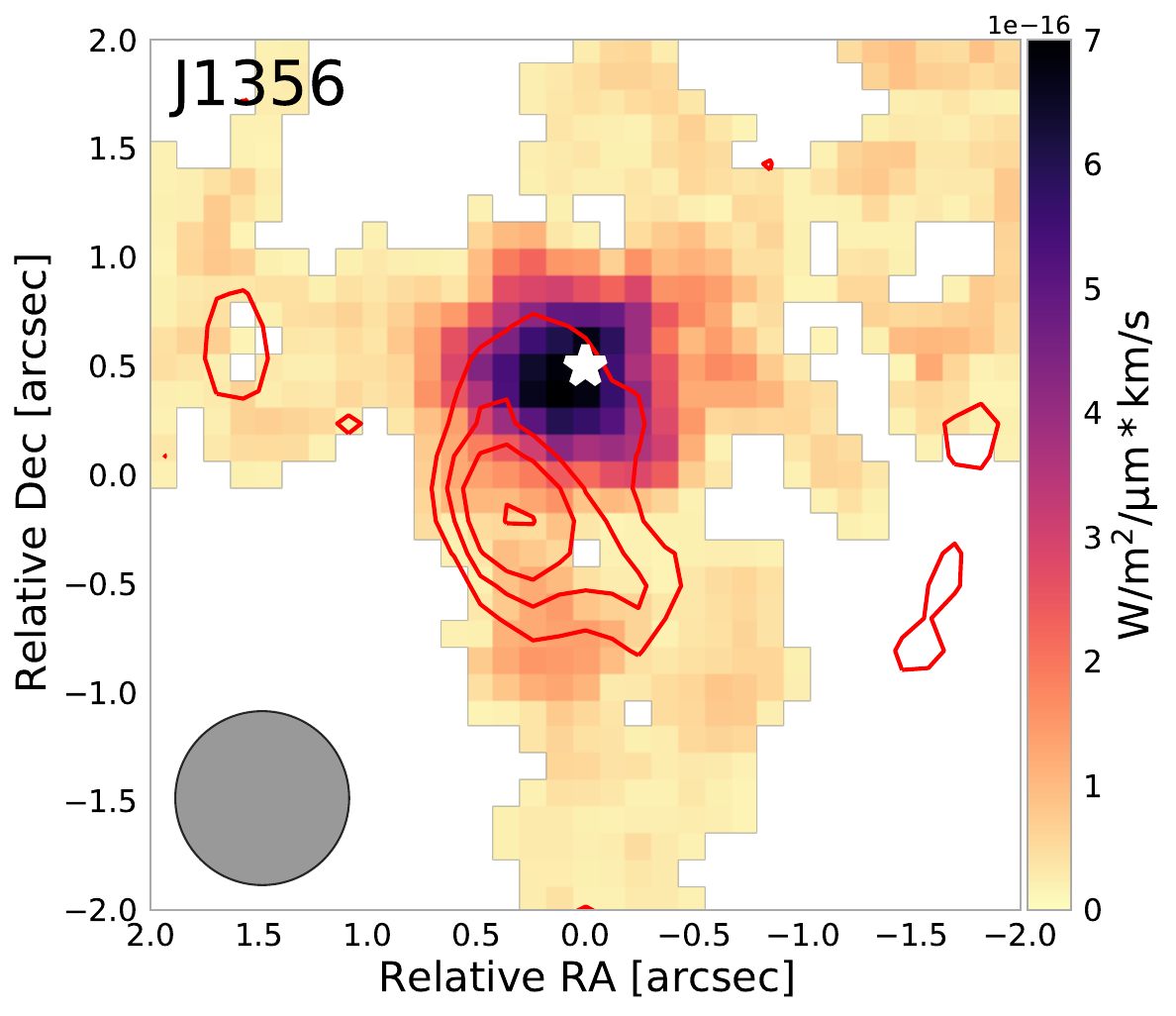}
}
\caption{Intensity maps of the high velocity components of the \hdone~emission. The color map corresponds to the moment 0 of the total \hdone~emission, while the contributions from high velocities are shown as red and blue contours. The spatial resolution of the SINFONI data is indicated with a gray ellipse in the bottom left corner. The white stars mark the peak of the near-infrared continuum as in Figure \ref{fig:mom0}. Left panel: the red and blue contours of J1430 refer to velocities above $\sim$370 \kms~with $\sigma_{red}$ = 2.5 $\rm 10^{-17}$ \flux and $\sigma_{blue}$ = 2.7 $\rm 10^{-17}$ \flux~. The red and blue contours are drawn at (2 ,3 ,5 ,7)$\sigma$. Right panel: the red contours of J1356 refer to velocities above $\sim$270 \kms~with $\sigma_{red}$ = 4.4 $\rm 10^{-17}$ \flux. The red contours are drawn at (2 ,3 ,4 ,6)$\sigma$.}
\label{fig:outflowmaps}
\end{figure*}

\subsection{\hd gas properties and kinematics}

In Section \ref{sec:warmgas} we analyzed the nuclear spectra of the Teacup and of the two nuclei of J1356, finding that warm molecular gas masses of $\rm M_{H_2}^{warm} \sim$ 5.9, 4.1, and 1.5 $\times 10^3$ \msun~in the central \SI{0.8}{\arcsecond} regions of J1430, J1356N, and J1356S, respectively.
In the same Section, we also measured the total warm \hd masses that we can compare with the the cold phase of the molecular gas. \RA~reported %total cold molecular gas masses of
$\rm M_{H_2}^{cold} \sim 6.2~and~11.5 \times 10^{9}$ \msun~from the CO(2-1) emission of J1430 and J1356, respectively, with maximum radii of 0.8 and 2.3 kpc. These findings result in total warm-to-cold molecular gas ratios of 7 $\times 10^{-6}$ and 1 $\times 10^{-6}$ for the Teacup and J1356, respectively. 
Indeed, only a small fraction of the molecular gas reservoir in galaxies is expected to be found in the warm molecular phase of the ISM \citep[e.g.][]{dale2005, mazzalay2013}. 
A warm-to-cold mass ratio of $6-7 \times 10^{-5}$ was measured in two nearby Luminous Infrared Galaxies (LIRGs) with \citep{pereirasantaella2016} and without \citep{Emonts2014} nuclear activity, combining NIR and sub-mm observations.
Recently, \citet{garcia-burillo2024} reported warm-to-cold molecular gas mass fractions $\leq 10^{-4}$, measured on nuclear scales in a sample of 45 nearby AGN ($D_L$ = 7 - 45 Mpc).

In Section \ref{sec:kiematics} we modeled the \hdone~kinematics with \barolo~tilted disc models. 
In the Teacup we find that the CO and the warm \hd disc have the same inclination but different PA. Indeed, our best fit \barolo~model provides a PA of the warm \hd disc of \ang{13}, which lies in between the findings for the CO \citep[$\rm PA_{CO}$ = \ang{4}, \RA;][]{audibert2023},  and the [O III] disc ($\rm PA_{[O III]}$ = \ang{27}, \citealt{harrison2014,venturi2023}; \SP).
The small difference between CO and warm \hd PA could be of low significance, as high uncertainties affect PA and inclination values, which can be highly degenerate. However, the major axis of the galaxy, $\rm PA_{gal}$ = \ang{-19}, is highly inconsistent with the above values. These differences could be due to a warp in the discs and/or to disturbed motions which could be related to the past merger event and/or to the influence of the radio jet on the gas morphology and kinematics (\citealt{audibert2023}; \SP).

J1356 is hosted in an ongoing merger system and thus, its kinematics is more disturbed. We found that the CO and the warm \hd major axis have consistent PA, therefore the two gas phases seem to trace the same disc. These warm and cold \hd discs are coplanar with the galaxy but have a different PA, which is not surprising in a merger.

In both QSO2s, the \hdone~velocity gradient ranges are consistent with the CO velocity gradients (\RA, shown in their Figures 9 and 11), with maximum values of $\pm$250 \kms~in the Teacup and of $\pm$200 \kms~in J1356 (see Figures \ref{fig:j1430_barolo} and \ref{fig:j1356_barolo}). On the contrary the \hdone~moment 2 maps show velocity dispersion values of up to 200-250 \kms~at the center, which are higher than the 120-160 \kms~values shown in CO moment 2 maps (\RA, Figure 9 and 11). This difference could be related to the different spatial resolution of the two observations, but is more likely due to the higher temperature and lower density of the warm gas, which is usually characterized by a higher velocity dispersion \citep[see][and references therein]{garcia-burillo2024}.

Finally, we detect a double-peaked emission line profile in both \Pa~and \hdone~in the J1356N nuclear spectra shown in Figures \ref{fig:nuclear_fits} and \ref{fig:H2_nuclear_fits}. This is not detected in the CO(2-1) line profile shown in Figure 15 of \RA.
%and could be due to an inner warm molecular gas disc not resolved in our data, but mergers can also show double peaked spectra \citep{Maschmann2023}.
Double peaked profiles can arise from the approaching and receding sides
of a rotating gas disc or correspond to non-circular motions, sometimes associated with mergers \citep{Maschmann2023}. 
Furthermore, as can be seen from Figure \ref{fig:mom0}, we also detect a shift in the \hdone~and S(2) emission peaks that could be related to the ongoing merger and/or to the radio jet interaction with the ISM. However, higher spatial resolution data for both radio emission and molecular \hd lines are needed to distinguish between the different scenarios. 

\subsection{The multiphase outflow}

\begin{table*}
     \caption[]{Multiphase outflows properties}
         \label{tab:multiphase_outflow}
\centering                          
\begin{tabular}{l c c c c c c c c c c}        
\hline  
\hline
QSO2 & Data & Emission & $\rm r_{out}$ & $\rm v_{out}$ & $\rm t^{out}_{dyn}$ & $\rm M_{out}$ & $\rm \dot{M}_{out}$ &  Ref. \\
 &  & line & kpc & \kms & Myr & $\rm 10^6$ \msun & \sfr & \\
\hline
J1430 & ALMA & CO(2-1) & 0.5 & -180|250 & 2.5 & 31 & 15.8 & a \\
& ALMA & CO(2-1)$^*$ & 1.2 & 300 & 3.9 & 160 & 41.0  & b \\
& GTC/MEGARA & [O III] & 3.7|3.1 & -760|529 & 4.8|5.7 & 5 & 1.1 &  c\\
& VLT/SINFONI & \hd & 1.9 & -470|430 & 4.0|4.3 & 2.6$\times 10^{-3}$ & 6.2$\times 10^{-4}$ & d \\
\hline
J1356 & ALMA & CO(2-1) & 0.4 & 310 & 1.4 & 14 & 7.8  & a \\
& ALMA & CO(2-1)$^*$ & 1.3 & 300 & 4.3 & 129 & 26.2  & b \\
& GTC/MEGARA & [O III]& 12.6|6.8 & -631|483 & 19.5|13.8 & 35 & 2.0 &  c \\
& VLT/SINFONI & \hd & 2.0 & 370 & 5.3 & 1.5$\times 10^{-3}$ &  2.9$\times 10^{-4}$ &  d \\
\hline
\end{tabular}\\
\flushleft 
\footnotesize{\textbf{Notes}. The table reports the multiphase outflows properties derived from this work and from literature for J1430 and J1356. Two values are reported when both the blue- and red-shifted sides of the outflows are detected. The [O III] masses are derived using the [S II]-based electron density (\SP) and the corresponding mass outflow rates are divided by 3 to match our outflow geometry. (*) Outflow properties are computed following Scenario II from \citet{audibert2023}. Column 6 are the dynamical times of the outflows $\rm t^{out}_{dyn} = r_{out}/v_{out}$.}
\flushleft 
\footnotesize{\textbf{References}. (a) \RA; (b) \citet{audibertprep}; (c) \SP; (d) this work. }
\end{table*}

We detected blue- and red-shifted high-velocity warm \hd gas in the nuclear region of the Teacup up to a radius of 1.9 kpc. In J1356 only the receding side of the warm \hd outflow is detected up to 2.0 kpc from the AGN.
We identified this gas as the warm component of the multiphase outflows, whose ionized (\citealt{greene2012,ramosalmeida2017,venturi2023}; \SP) and cold molecular \citep[\RA;][]{audibert2023, girdhar2024} phases have been discussed in previous works. 
In particular, using the CO(2–1) transition observed with ALMA at \SI{0.2}{\arcsecond} resolution, \RA~detected spatially resolved cold molecular outflows of 0.5 and 0.4 kpc in J1430 and J1356, respectively. 
For the same objects \SP~detected, using MEGARA observations of the [O III] line at \SI{1.2}{\arcsecond} resolution, ionized outflows which extend on galactic scales, from 3 kpc up to 13 kpc. 
Table \ref{tab:multiphase_outflow} reports the cold molecular and warm ionized outflow properties of the two QSO2s (\RA; \citealt{audibertprep}; \SP).
%Taking into account the mass budget of the central kiloparsec from the AGN, where the molecular phase is dominant, 
Considering the warm and cold molecular outflow masses reported in Table \ref{tab:multiphase_outflow}, we derive warm-to-cold molecular gas ratios of $\sim$1 $\times 10^{-4}$ if we use the outflow masses from \RA~and of $\sim$1 $\times 10^{-5}$ if we use the values corresponding to Scenario II from \citet{audibertprep}.  
The cold molecular outflow masses were computed adopting a CO-to-\hd conversion factor $\rm \alpha_{CO} = 0.8$ \msun(K \kms)$^{-1}$, while the total cold molecular masses $\rm M_{H_2}^{cold}$ were estimated using $\rm \alpha_{CO} = 4.35$ \msun(K \kms)$^{-1}$ (\RA).
The ionized outflows of J1430 and J1356 have masses of 5 and 35 $\times 10^{6}$ \msun, respectively (see Table \ref{tab:multiphase_outflow} and \SP) therefore, this gas phase is the fastest, but the cold molecular phase dominates the mass budget of the outflow.
Indeed, in the local Universe and at AGN luminosities below $10^{47}$ \ergs, it has been widely reported that molecular outflows carry more mass than their ionized counterparts \citep[e.g.][]{veilleux2020,feruglio2010,rupke2013,cicone2014,carniani2015, fiore2017, fluetsch2021}.

In this work we measured warm molecular outflow rates of 6.2 and 2.9 $\times 10^{-4}$ \sfr~in the Teacup and J1356, which are consistent with the estimates available in the literature for other AGN of different luminosities, which range between $10^{-5}$ and $10^{-2}$ \sfr~\citep{diniz2015,ramosalmeida2019,riffel2020,bianchin2022,riffel2023}. 
In \citet{riffel2023}, whose warm molecular outflow compilation is the largest available in the local Universe to date, they characterized the \hdone~emission and kinematics using data obtained with Gemini/NIFS of a sample of 33 AGN hosts with $ \rm 0.001 < z < 0.056$ and hard X-ray luminosities $\rm 41 < log L_X/(erg~s^{-1}) < 45$. They reported warm outflow masses of $\rm 10^{0}-10^{4}$ \msun~and warm molecular outflow rates of $10^{-5}-10^{-2}$ \sfr.

The warm molecular outflow mass rates reported in this work and the value measured for the QSOFEED QSO2 J1509+0434 \citep[0.001 \sfr,][]{ramosalmeida2019} can be compared with the cold molecular and warm ionized gas outflows reported for them by \RA, \citet{audibertprep}, and \SP. These are $\rm \dot{M}_{CO}$ = 7.8-41.0 \sfr~and $\rm \dot{M}_{[O III]}$ = 1.1-2.0 \sfr~(see Table \ref{tab:multiphase_outflow}).
In the comparison between the different outflow phases, it should be noted that the observations were carried out at different angular resolutions (\SI{0.2}{\arcsecond}, \SI{1.2}{\arcsecond}, and \SI{0.8}{\arcsecond} in the case of ALMA, MEGARA, and SINFONI, respectively) and that different methodologies were employed to derive the outflow properties.
\RA~identified as outflow only the high-velocity gas detected along the kinematic minor axis showing deviations from the circular motions modeled with \barolo. 
\citet{audibert2023}, considering four different scenarios to calculate the outflow mass rates of the Teacup using the same CO(2–1) data analyzed by \RA, reported values of $\rm \dot{M}_{CO}$ from 6.7 to 44 \sfr. 
Table \ref{tab:multiphase_outflow} reports the CO outflow properties derived by \citet{audibertprep} for the two QSO2s studied here. The authors used the same method of this work, Scenario II from \citet{audibert2023}, and the same CO(2–1) data analyzed by \RA.
%They reported outflow mass rates of 41 \sfr for Teacup and 26 \sfr for J1356. 
We note that in the case J1356N, only the red-shifted side of the cold molecular outflow was detected \citep[\RA;][]{sun2014,audibertprep}.

\SP~derived the ionized outflow properties by integrating all the high-velocity [OIII] emission above 3$\sigma$, and reported the detection of both the blue-shifted and red-shifted sides of the ionized outflows for Teacup and J1356.
Taking into account the caveats mentioned above (see also \citealt{hervella2023}), we can compare our values with those of the ionized and cold molecular gas, and we find that the relative contribution of the warm molecular phase to the total (i.e., ionized + warm molecular + cold molecular) mass outflow rate is approximately 0.001\%.
Figure \ref{fig:mdot} shows the relative contribution of the ionized, cold, and warm molecular phases to the total mass outflow rate in J1356 and the Teacup and, for the sake of completeness, in J1509 and J1100, which are also part of the QSOFEED sample. We note that the relative contribution of the ionized and cold molecular phases varies among the sources, while the contribution of the warm molecular phase is in the range 0.001-0.005\%. The higher relative contribution of warm molecular outflowing gas is the one found in J1509 by \citet{ramosalmeida2019}, who detected outflows in both the ionized and warm molecular gas phases, analyzing near-infrared spectroscopic data from the EMIR instrument on the GTC.  
%(Espectrógrafo Multiobjeto Infra-Rojo)
From the \hdone~line luminosity they obtained an outflow mass of $\sim 10^4$ \msun, a factor 100-400 lower than the mass of the ionized outflow, and a mass outflow rate of 0.001 \sfr~with an outflow radial size (FWHM) of $\approx$ 1.5 kpc. 
%However, we do not compare with the he warm-to-cold molecular gas ratio of this QSO2, since its cold molecular outflow rate is an upper limit \citep{ramosalmeida2022}.

In this work, we aimed to provide a more comprehensive view of quasar-driven outflows by putting together the cold and warm molecular and warm ionized outflow measurements available for these QSOFEED quasars. Indeed, having a large sample with multiphase outflow measurements for the same targets is the ultimate goal of the QSOFEED project (\RA). 
It is worth mentioning that measurements of warm molecular gas outflows are scarce in the literature and that, at least in nearby AGN (z$<$0.002), the warm molecular gas emission seems to be dominated by galaxy rotation \citep{davies2014, davies2024, riffel2018, storchi-bergmann2019, esparza2024}. 
In fact, \citet{ramosalmeida2017} did not report evidence of broad \hd components in the nuclear spectrum of the Teacup, with the velocity map showing a dominant rotation pattern. They addressed as possible signatures of outflowing molecular gas the nuclear narrow components blue-shifted with respect to the systemic velocity. Thanks to the deeper
SINFONI observations analyzed here, we increased the exposure time by two times compared to the data analyzed in \citet{ramosalmeida2017}, and we were able to resolve the warm molecular outflow, which extends up to 1.9 kpc. The higher-sensitivity of JWST and its near- and mid-infrared instruments will surely contribute to increase number of warm molecular outflows detected in AGN. 

\begin{figure}
\resizebox{\hsize}{!}{
\includegraphics{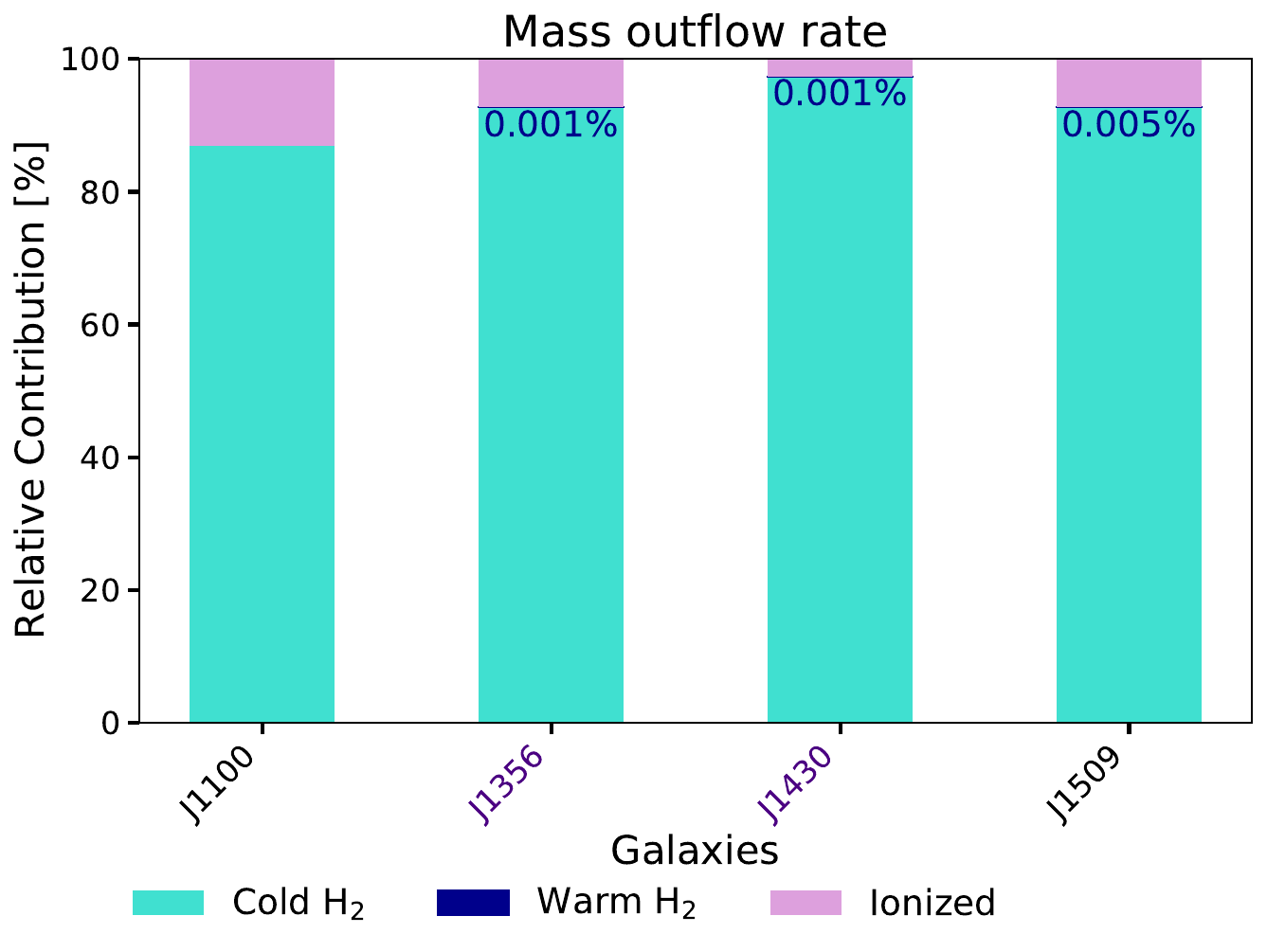}}
\caption{Relative contribution of the cold molecular (turquoise), ionized (pink),
and warm molecular (blue) phases to the total mass outflow rate. The blue labels refer to the relative contribution of the warm molecular phase in \%. The galaxies whose names are in violet are the targets of this work. The ionized mass outflow rates are from \SP, adopting their [S II]-based electron densities and divided by us by a factor of 3 to match our outflow geometry.
The cold molecular mass outflow rates are from Scenario II in \citet{audibertprep}, and the warm molecular mass outflow rate of J1509 is from \citet{ramosalmeida2019}. 
No warm molecular outfflow detections are yet reported for J1100.} %The black arrow indicates that the CO outflow in J1509 is an upper limit. }
\label{fig:mdot}
\end{figure}

\subsection{The interaction of the outflow with the galaxy host}

Using the SFRs measured for our two QSO2s reported by \SP, of 69 and 12 \sfr, we can calculate the mass loading factor, $\eta = \rm \dot{M}_{\text{out}}/\text{SFR}$, for the warm molecular gas. We find $\eta = 5.2 \times 10^{-5}$ for the Teacup and $\eta = 4.2 \times 10^{-6}$ for J1356, which are rather small values. By combining our loading factors with those of the cold molecular phase and ionized gas, we obtain $\eta_{\text{tot}} > 1$ only for the Teacup, as already reported by \RA~and \SP. This suggests that in J1356, the star formation process is more effective in removing gas than the outflows.
As discussed in \SP, it would be more accurate to compare the multiphase mass outflow rates with the SFRs calculated in the same regions. Here we are using SFRs derived from total far-infrared luminosity, representing a galaxy-wide SFR. The outflows, on the other hand, are more compact, especially in the case of the molecular outflows, which have radii of 0.4-0.5  kpc (\RA) and of 1.2-1.3 kpc \citep{audibertprep} for the cold molecular phase and of 1.9-2.0 kpc for the warm molecular phase.
Another point to consider are the different timescales of star formation and outflows. Given that QSO2 outflows have dynamical timescales of approximately 1–20 Myr \citep[see Table \ref{tab:multiphase_outflow} and \RA; \SP;][]{bessiere2022,audibertprep}, their characteristics should be compared to those of recent star formation, which can be probed through resolved stellar population analysis \citep{bessiere2022} or through polycyclic aromatic hydrocarbon (PAH) emission features detected in the mid-infrared, which are related to young and massive stars. %On the other hand, the PAH properties can vary with distance from the AGN 
\citep{garciabernete2022,garciabernete2024,lai2022,lai2023,Zhang2024,ramosalmeida2023,zanchettin2024}. 
Indeed, the spatial distribution of PAHs in the two QSO2s will be studied thanks to available Cycle 2 JWST observations (PI: Ramos Almeida, proposal 3655).
This approach will allow us to determine whether the outflows inhibit or stimulate recent star formation, or if both processes can occur simultaneously \citep{cresci2015, carniani2016,bessiere2022}.

The two QSO2s analyzed in this work are radio-quiet objects with AGN luminosities of $10^{45.83}$ \ergs~(The Teacup) and of $10^{45.54}$ \ergs~(J1356), derived by \SP~from the extinction correct [O III] luminosities reported by \citet{kongho2018}. Therefore, the multiphase gas outflows should in principle be driven by radiation pressure and/or wind angle winds, typical of the quasar-mode feedback \citep{fabian2012}. However, it has been widely suggested that low-power compact jets in radio-quiet AGN can compress and shock the surrounding gas as they expand, contributing to the driving of both ionized and molecular outflows \citep[e.g.][]{venturi2018, jarvis2019,audibert2023}. 
The two QSO2s have 6 GHz VLA observations at both high ($\sim$ \SI{0.25}{\arcsecond} beam) and low ($\sim$ \SI{1}{\arcsecond} beam) angular resolutions, which were analyzed by \citet{jarvis2019}. The authors inferred that only a small fraction, i.e. 3-6\%, of the radio emission is due to the star formation process. They identified the extended radio emission with a jet or a lobe in J1430, while the radio emission identification as a jet or a lobe is more tenuous in J1356, where the extended radio feature is only visible in the low-resolution map.
\SP~noted that the red-shifted counterpart of J1356 ionized outflow, which extends up to $\sim$6.8 kpc is perfectly aligned with the 6 GHz radio contours at \SI{1}{\arcsecond} resolution, constituting a possible evidence for a jet-like structure accelerating the ionized gas through the galaxy ISM or to shocks induced by the outflow itself, which are seen in the radio \citep{Fischer2023}. However, due to the compactness of the warm molecular outflow detected with the SINFONI data, it is quite challenging to link the properties of the warm molecular outflow with the extended radio emission reported by \citet{jarvis2019}. 

As for the Teacup, the high-angular resolution 6 GHz VLA observations reveal the presence of a compact radio jet along a PA = \ang{60}, extending up to $\sim$0.8 kpc \citep{harrison2015, jarvis2019}, with a jet power of $\rm P_{\text{jet}} \sim 10^{43}$ \ergs~\citep{audibert2023}. This compact jet-like structure seems to subtend a small angle relative to the cold molecular gas disc, and consequently to the warm molecular gas as well.
Our \hdone~residual dispersion map, shown in Figure \ref{fig:j1430_barolo}, shows enhanced velocity dispersion in the south and southwest direction, both along and perpendicular to the jet direction. 
Additionally, PVDs along and perpendicular to the jet direction (see middle panels of Figures \ref{fig:PVD} and \ref{fig:PVD_stacked}) show non circular red-shifted emission at modest velocities, i.e. 50-250 \kms~to the southwest and southeast.%south and west.
According to simulations \citep{mukherjee2018, meenakshi2022}, radio jets can efficiently inject kinetic energy into the surrounding gas, inducing local turbulence and shocks when the jets have a low inclination relative to the gas disc, as in the case of the Teacup. 
\citet{audibert2023} performed a comparison between ALMA CO(2-1) and CO(3-2) observations and the simulations in \citet{mukherjee2018}, finding that the jet compresses and accelerates the molecular gas, driving a lateral outflow characterized by increased velocity dispersion and higher gas excitation. 
%Due to the low resolution of our SINFONI data, we cannot either confirm or discard these results. However, we note that the PVDs along and perpendicular to the jet direction closely resemble the ALMA CO(2-1) and CO(3-2) ones.

The results presented here, along with those presented in previous studies of the same targets, might indicate that compact, low power jets could also be an efficient outflow driving mechanism even in radio-quiet AGN. This might contribute to explaining the dispersion in the $\rm \dot{M}_{out}-L_{bol}$ plane found in recent studies \citep[e.g. \RA;][]{lamperti2022}. In the particular case of warm molecular outflows, kiloparsec scale radio/sub-mm continuum emission is detected in the three quasars with outflow measurements (J1430, J1356, and J1509), indicating a potential causal connection between the two.
However, another quasar belonging to the QSOFEED sample, J0945 \citep{speranza2022} also shows extended ($\sim$1 kpc) radio emission and no warm molecular outflow was found from the analysis of high angular resolution Gemini/NIFS observations. The orientation between the jets and the molecular gas discs could be a determinant factor here \citep{harrisonramos2024}.

\section{Conclusions}\label{sec:conclusions}

In this paper we investigate deep near-infrared IFS data obtained with VLT/SINFONI of the type-2 quasars J1356 and J1430. We analyze the gas dynamics in the two systems, by using the \hdone~and S(2) emission lines detected in the QSO2 spectra.
We detect warm molecular outflows with radii of up to 2.0 and 1.9 kpc, respectively, but contributing very little to the total mass budget.
We summarize our main findings as follows.
\begin{enumerate}
    \item The warm molecular gas phase in the Teacup and in J1356 represents a small fraction of the total molecular gas.  We measure masses of $\rm (5.85 \pm 0.90) \, \times 10^3 $ \msun, $\rm (4.05 \pm 0.69) \, \times 10^3 $ \msun, and $\rm(1.47 \pm 0.37) \, \times 10^3 $ \msun~in the inner \SI{0.8}{\arcsecond} diameter region of the Teacup, J1356 north and south nuclei, respectively.  The total warm \hd masses are $\sim 4.5 \times 10^4$ \msun~and $\sim 1.3 \times 10^4$ \msun~in the Teacup and in J1356, with maximum radii of 4.8 and 5.8 kpc. Considering the CO-derived masses of cold molecular gas of these QSO2s, this implies warm-to-cold molecular gas ratios of 7 and 1 $\times 10^{-6}$, respectively.
    \item In the Teacup, the kinematics of the warm molecular gas is consistent with rotation, but the warm \hd PA differs from the galaxy PA, probably related to the past merger and/or to the influence of the jet on the ionized and molecular gas.  In J1356 only a fraction of the warm molecular gas is rotating and the kinematics is disturbed, but the CO and warm \hd PAs are consistent, suggesting that are tracing the same disc. In both QSO2s, the \hdone~emission shows higher velocity dispersion at the nucleus compared to CO. 
    \item In both QSO2s, high-velocity gas is detected, indicating the presence of warm molecular outflows, with velocities of 370 and 450 \kms~and outflow masses of 1.5 and 2.6 $\times 10^3$ \msun~for J1356 and the Teacup, respectively. The warm-to-cold gas ratios that we measure in the outflow regions are $\sim$ 1 $\times 10^{-5}$, which are significantly higher than the value measured in the central \SI{0.8}{\arcsecond} radius (1.8 and 1.3 kpc).
    \item We measure warm molecular mass outflow rates of 6.2 $\times 10^{-4}$ \sfr~and of 2.9 $\times 10^{-4}$ \sfr~in J1430 and J1356, respectively, which are approximately 0.001\% of the total (i.e., ionized + cold molecular + warm molecular) mass outflow rate. Adopting the SFRs derived from the total far-infrared luminosity, we measure mass loading factors of $\eta = 5.2 \times 10^{-5}$ for the Teacup and of $\eta = 4.2 \times 10^{-6}$ for J1356. 
    \item  We detect an enhancement of velocity dispersion in the \hdone~residual dispersion map of the Teacup, both along and perpendicular to the direction of the compact radio jet, approximately \SI{1}{\arcsecond} ($\sim$1.6 kpc) from the AGN. This local turbulence could be due to the injection of kinetic energy from the radio jet, as already suggested by the comparison of the CO emission with simulations.
\end{enumerate}

This study aimed to expand our understanding of multiphase quasar-driven outflows in our QSO2 by including warm molecular gas observations obtained in the near-infrared with VLT/SINFONI.
Our findings reveal that the warm molecular gas, both in the disc and outflows, represents only a minor fraction of the total gas reservoir (considering warm ionized, cold molecular, and warm molecular phases and excluding the neutral gas phase). 
%Interestingly, all three QSO2s with warm molecular outflow detections from the QSOFEED sample (J1430, J1356, and J1509) exhibit kiloparsec-scale radio/sub-mm emission, potentially indicating a causal connection.Thanks to newly approved JWST observations (PI: Ramos Almeida, proposal 3655) targeting PAH emission, it will be possible to investigate whether these outflows suppress or stimulate recent star formation, or if both processes might occur simultaneously.

\begin{acknowledgements}
      Based on observations made with ESO Telescopes at the Paranal Observatory under programs ID 097.B-0923(A) and 094.B-0189(A).
      Maria Vittoria Zanchettin acknowledges financial support from the PRIN MIUR 2017PH3WAT contract, and PRIN MAIN STREAM INAF "Black hole winds and the baryon cycle".
      Cristina Ramos Almeida, Jose Acosta-Pulido, Anelise Audibert and Pedro Cezar acknowledge support from the Agencia Estatal de Investigaci\'on of the Ministerio de Ciencia, Innovaci\'on y Universidades (MCIU/AEI) under the grant ``Tracking active galactic nuclei feedback from parsec to kiloparsec scales'', with reference PID2022$-$141105NB$-$I00 and the European Regional Development Fund (ERDF).
\end{acknowledgements}

\bibliographystyle{aa} % style aa.bst
\bibliography{bibliografia} % your references Yourfile.bib


\begin{appendix} 

\section{Nuclear Spectra}\label{app:nuclear_spectra}

Figure \ref{fig:teacup_Pa} shows the fit of the \Pa~line profile adopting a narrow and a broad component and not considering the contribution of He I and He II lines. We find that the narrow component has a FWHM = 445 $\pm$ 3 \kms, $\rm V_s$ = 10 $\pm$ 8 \kms~and Line Flux $\rm (8.59 \pm 1.29) \times 10^{15}~erg~cm^{-2}~s^{-1}$. While the broad component is well fitted by a Gaussian of FWHM = 1655 $\pm$ 26 \kms, $\rm V_s$ = -197 $\pm$ 16 \kms~and Line Flux $\rm (5.59 \pm 0.86) \times 10^{15}~erg~cm^{-2}~s^{-1}$. Thus our findings are consistent with the best fit values found by \citet{ramosalmeida2017} with the same procedure.

\begin{figure}[h]
\resizebox{\hsize}{!}{
\includegraphics{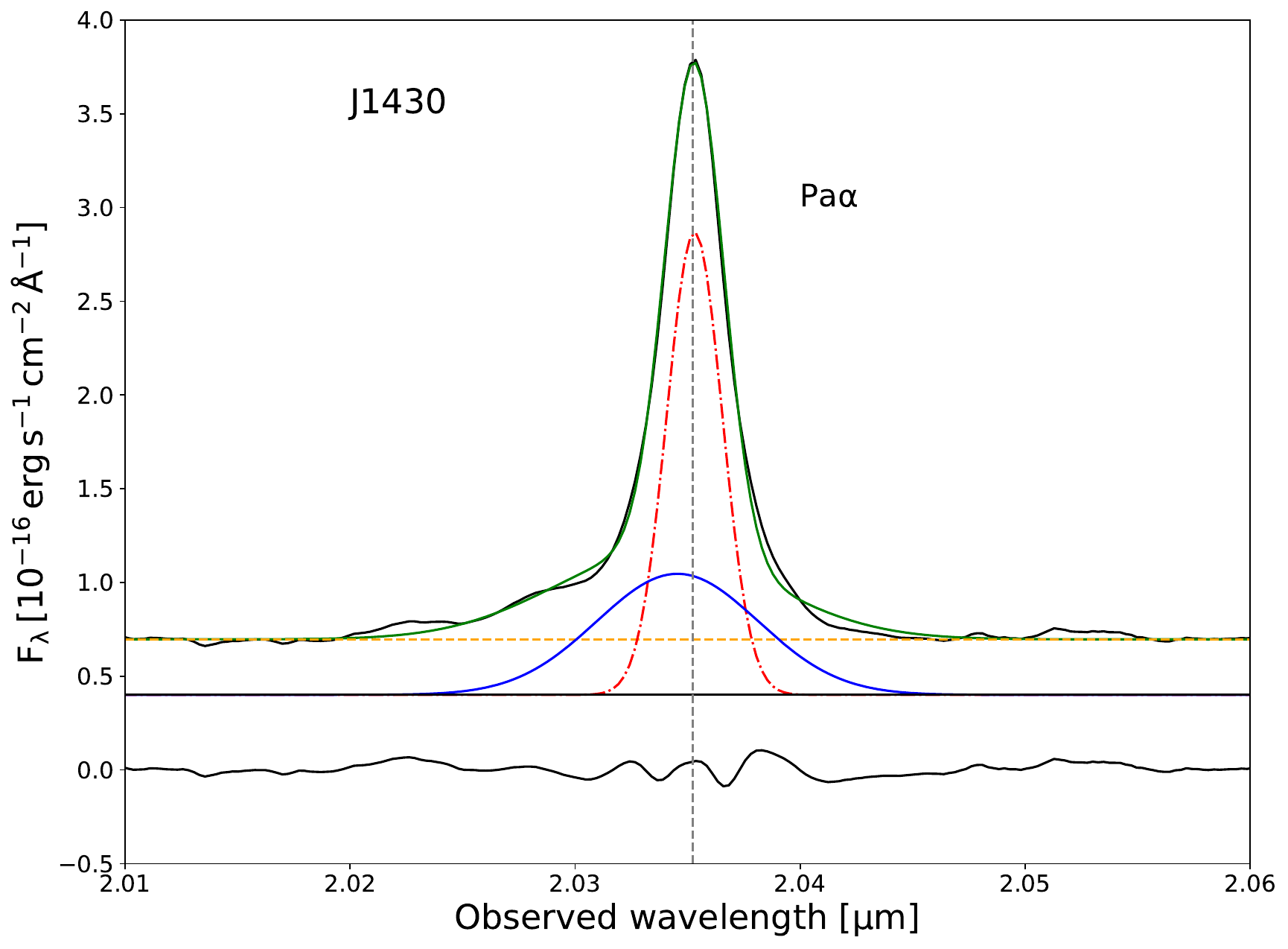}}
\caption{Same as Figure \ref{fig:nuclear_fits} and \ref{fig:H2_nuclear_fits} but for \Pa~in the Teacup.}
\label{fig:teacup_Pa}
\end{figure} 

In the following, we report the profiles of the most prominent lines in the K-band nuclear spectrum of the south nucleus of J1356 (J1356S). We extracted the spectrum by adopting a circular aperture of \SI{0.8}{\arcsecond} diameter (i.e., 1.8 kpc) and centered on the maximum emission of \Pa, which is the most prominent line of the spectrum. As described in Section \ref{sec:nuclearspectra}, we perform the fits of the emission lines adopting a combination of Gaussians. In Table \ref{tab:nuclear_fit} we report the FWHMs, the velocity shift and the fluxes along with their corresponding errors, obtained from our fits of J1356S and J1430 nuclear spectra. Figure \ref{fig:nuclear-fit_app} shows the line profiles with the corresponding fits in J1356S. In this case, we adopted only one Gaussian for each line, due to the low S/N of the spectrum. Indeed, \hdone~and \hdtwo~are detected with a S/N of 4 and 3, respectively. In Table \ref{tab:nuclear_fit} we report the 3$\sigma$ upper limit of the \Brg~emission line that remains undetected.

\begingroup
\setlength{\tabcolsep}{8pt} % Default value: 6pt
\renewcommand{\arraystretch}{1.3} % Default value: 1
\begin{table*}
\caption[]{Same as Table \ref{tab:J1356N_fit} but for the Teacup and J1356S. In the case of J1356S, the tabulated flux for \Brg~corresponds to 3$\sigma$ upper limit, as the line is not detected there.}
\label{tab:nuclear_fit}
\centering                          
\begin{tabular}{l |c c c | c c c}  
\hline
\hline  
 \multicolumn{7}{c}{Nuclear Spectrum} \\
\hline
&\multicolumn{3}{c}{J1430} & \multicolumn{3}{c}{J356S}\\
\hline
Line & FWHM & $\rm V_{s}$ & Line Flux $\times$ $\rm 10^{15}$ & FWHM & $\rm V_{s}$ & Line Flux $\times$ $\rm 10^{15}$ \\ 
 & (\kms) & (\kms) & ($\rm erg~cm^{-2}~s^{-1}$) & (\kms) & (\kms) & ($\rm erg~cm^{-2}~s^{-1}$)\\
\hline
\Pa~& 421 $\pm$ 4 & 12 $\pm$ 8 & 7.63 $\pm$ 1.15 & 702 $\pm$ 23 & -52 $\pm$ 26 & 0.57 $\pm$ 0.10 \\ 
\Pa~(b) & 1245 $\pm$ 41 & -100 $\pm$ 20 & 5.82 $\pm$ 0.97 & - & - & - \\
\Brd~& 395 $\pm$ 27 & -2 $\pm$ 14 & 0.39 $\pm$ 0.07 & 472 $\pm$ 239 & -10 $\pm$ 144 & 0.01 $\pm$ 0.01 \\
\Brd~(b) & 1180 $\pm$ 155 & -200 $\pm$ 105 & 0.28 $\pm$ 0.11 & - & - & - \\
\Brg~& 321 $\pm$ 26 & -11 $\pm$ 12 & 0.42 $\pm$ 0.09 & - & - & 0.04$^*$ \\
\Brg~(b) & 949 $\pm$ 79 & -13 $\pm$ 29 & 0.69 $\pm$ 0.19 & - & - & - \\
He I  & 376 $\pm$ 47 & -20 $\pm$ 23 & 0.35 $\pm$ 0.89 & 939 $\pm$ 371 & 200 $\pm$ 182 & 0.13 $\pm$ 0.06 \\
He II  & 549 $\pm$ 43 & 20 $\pm$ 25 & 0.50 $\pm$ 0.09 & 767 $\pm$ 258 & -119 $\pm$ 196 & 0.04 $\pm$ 0.02 \\
$\rm[Si VI]$ & 511 $\pm$ 41 & 61 $\pm$ 18 & 0.75 $\pm$ 0.20 & 939 $\pm$ 663 & 50 $\pm$ 753 & 0.04 $\pm$ 0.05 \\
$\rm[Si VI]$ (b) & 1175 $\pm$ 405 & 0 $\pm$ 237 & 0.39 $\pm$ 0.38 & - & - & - \\
\hdone~& 560 $\pm$ 9 & 18 $\pm$ 10 & 0.68 $\pm$ 0.10 & 619 $\pm$ 68 & 20 $\pm$ 41 & 0.09 $\pm$ 0.02 \\
\hdtwo~&  518 $\pm$ 15 & -14 $\pm$ 12 & 0.25 $\pm$ 0.04 & 502 $\pm$ 82 & 29 $\pm$ 50 & 0.03 $\pm$ 0.01 \\
\hdthree & 596 $\pm$ 32 & -32 $\pm$ 24 & 0.75 $\pm$ 0.14 & 793 $\pm$ 143 & 204 $\pm$ 176  & 0.1 $\pm$ 0.06 \\
\hdfour & 564 $\pm$ 123 & -95 $\pm$ 74 & 0.20 $\pm$ 0.08 & 1175 $\pm$ 181 & -79 $\pm$ 98 & 0.05 $\pm$ 0.01 \\
\hdfive & 484 $\pm$ 48 & -9 $\pm$ 30 & 0.32 $\pm$ 0.07 & 733 $\pm$ 100 & -82 $\pm$ 60 & 0.06 $\pm$ 0.02 \\
\hline
\end{tabular}\\
\end{table*}
\endgroup

\begin{figure*}[ht]
\resizebox{\hsize}{!}{
\includegraphics{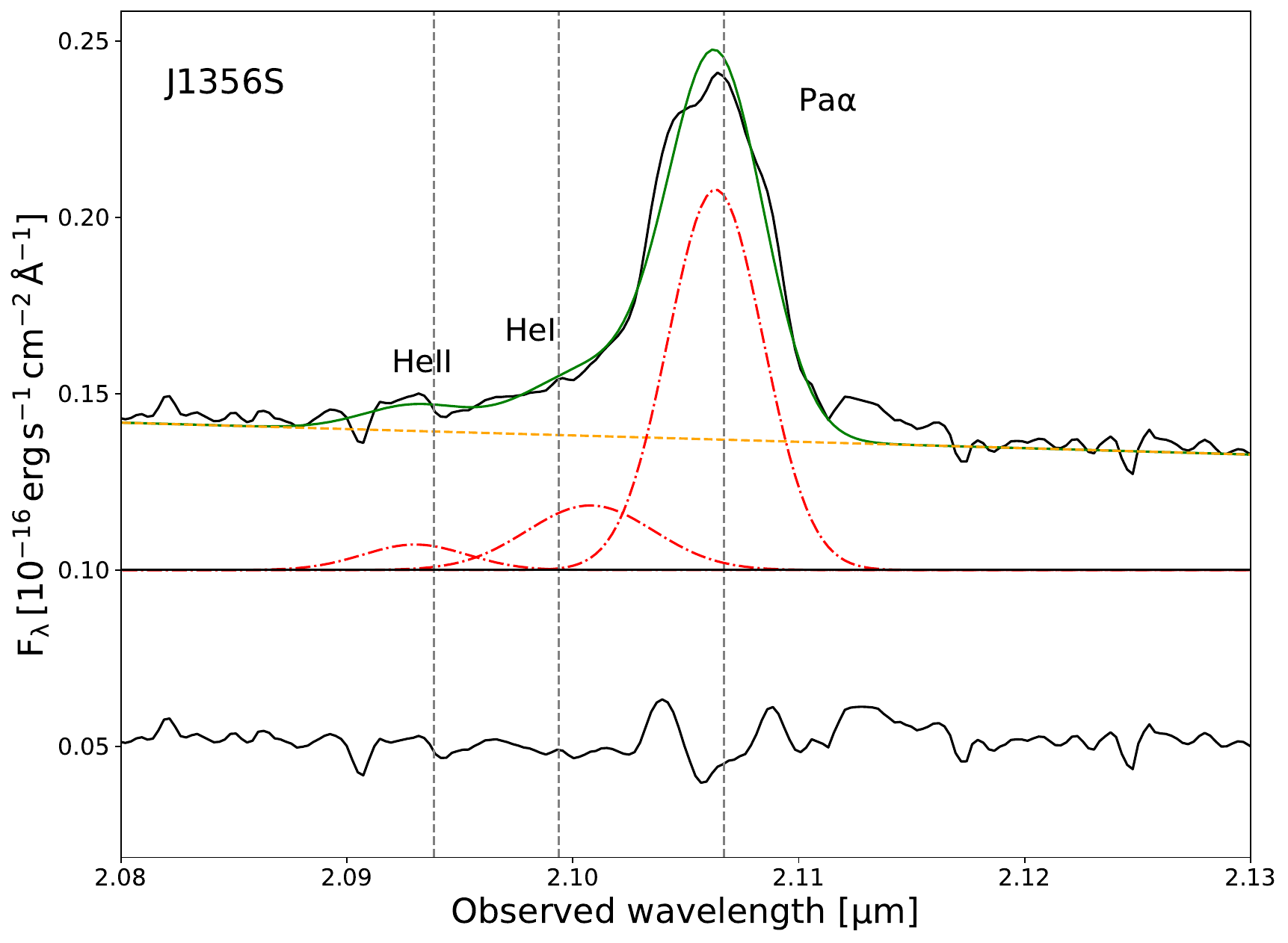}
\includegraphics{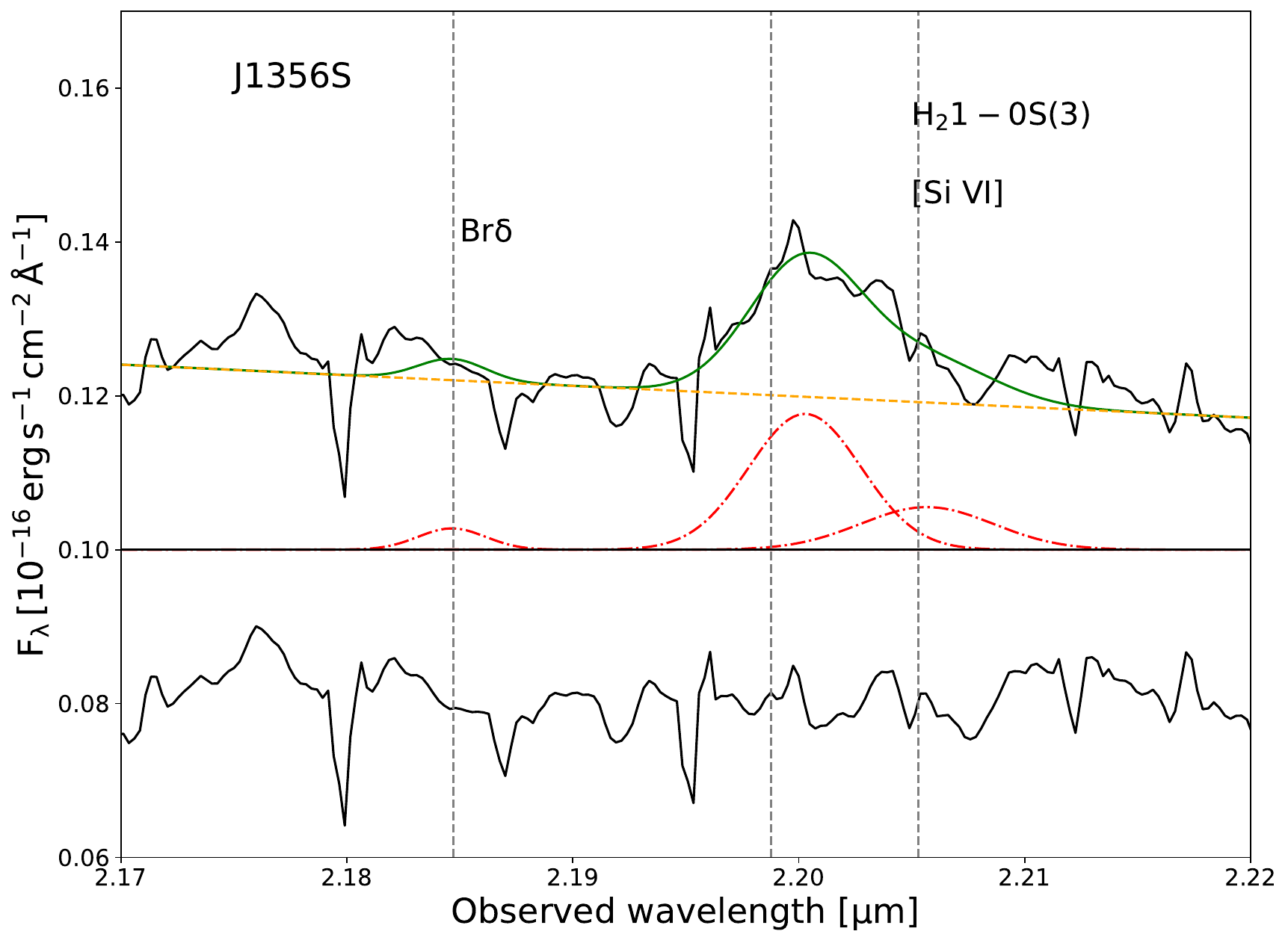}}
\resizebox{\hsize}{!}{
\includegraphics{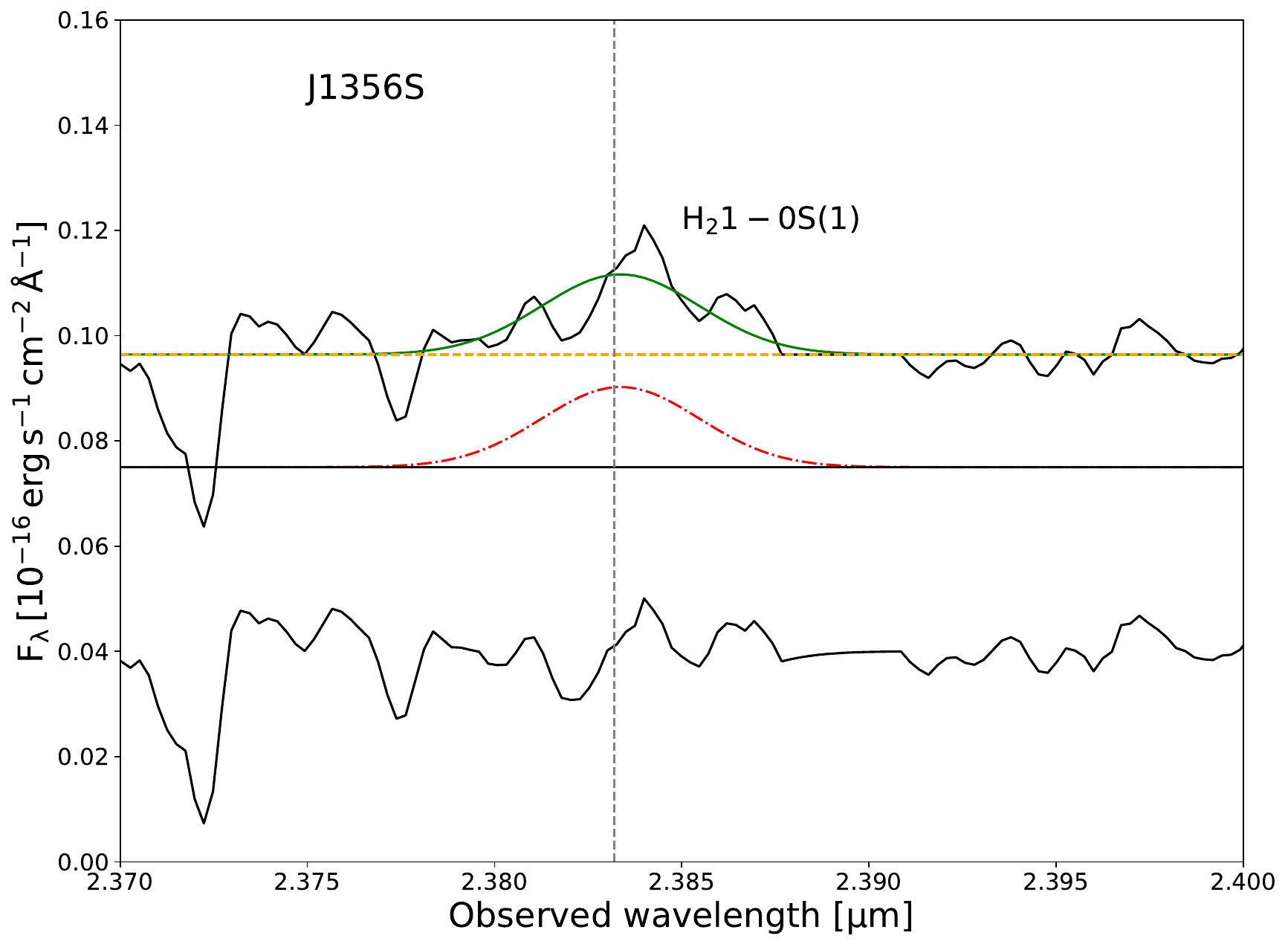}
\includegraphics{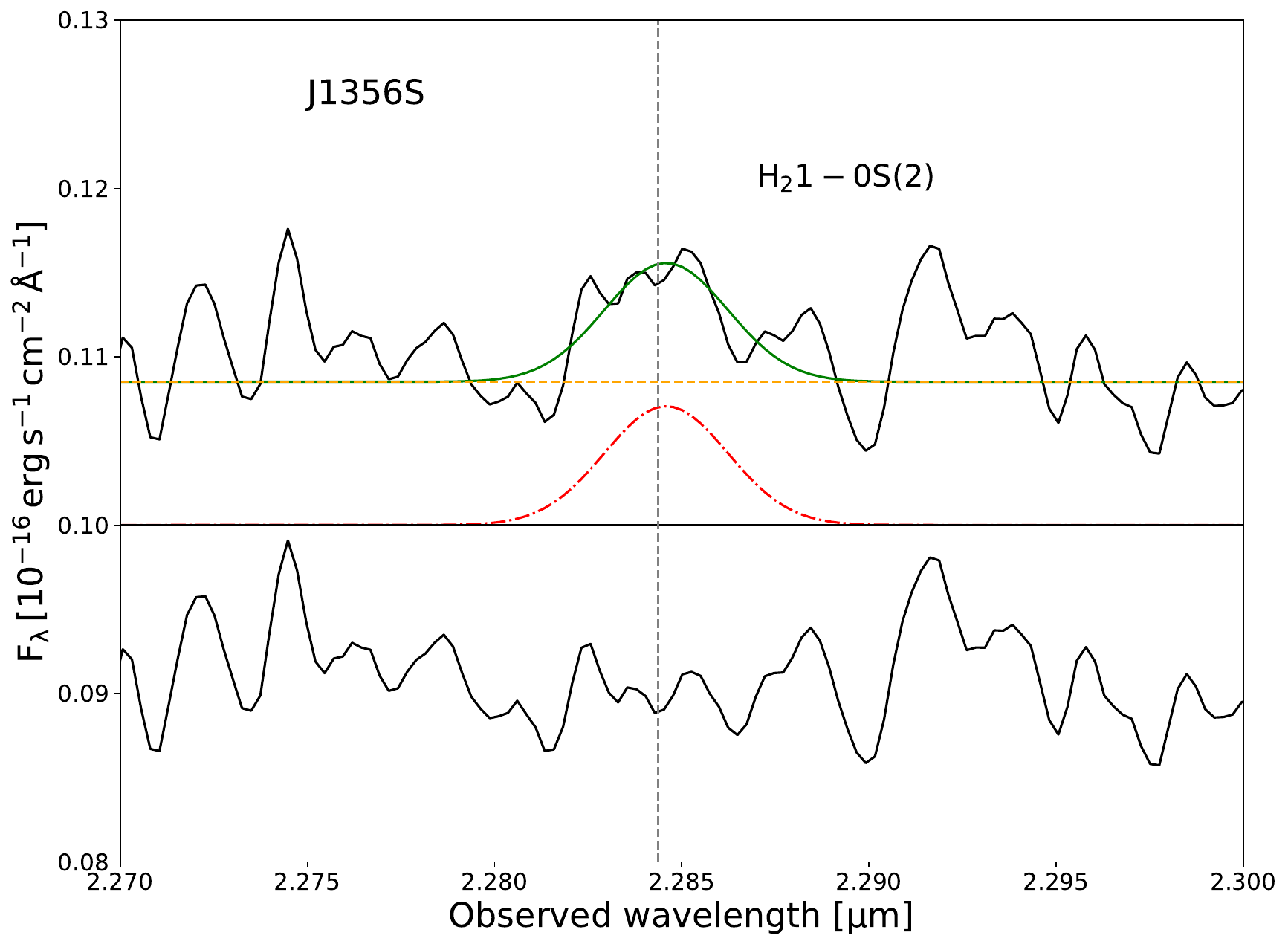}}
\caption{Same as Figure \ref{fig:nuclear_fits} and \ref{fig:H2_nuclear_fits} but for J1356S. 
%The \Brg~profile is not shown since this line remains undetected in J1356 south nucleus spectra.
}
\label{fig:nuclear-fit_app}
\end{figure*}

\section{\barolo~disc model}

In the following, we show the analog of the PVDs in Figure \ref{fig:PVD} but extracted from the stacked \hd cubes. The red contours refer to the \barolo~models described in Section \ref{sec:kiematics}.
Figures \ref{fig:J1430_chanmap} and \ref{fig:J1356_chanmap} show the channel maps produced by \barolo~by fitting the \hdone~data cubes of J1430 and J1356N. The color scale and the blue contours refer to the data, and the red contours correspond to \barolo~disc model. The cyan cross indicates the peak of the emission and was chosen as the center of the model. 
The velocity of the channel map is reported above in each panel. 
%J1356 shows complex gas kinematics, possibly due to the ongoing merger and/or to the interaction of the compact radio jet with the interstellar medium.
We note that J1430 shows clumps of emission with velocities of up to $\pm$480 \kms, and J1356 up to $\pm$370 \kms. 
%These high-velocity values agree with the velocity range chosen to integrate the flux of the outflowing gas; see Section \ref{sec:outflow}. 

\begin{figure*}
\resizebox{\hsize}{!}{
\includegraphics{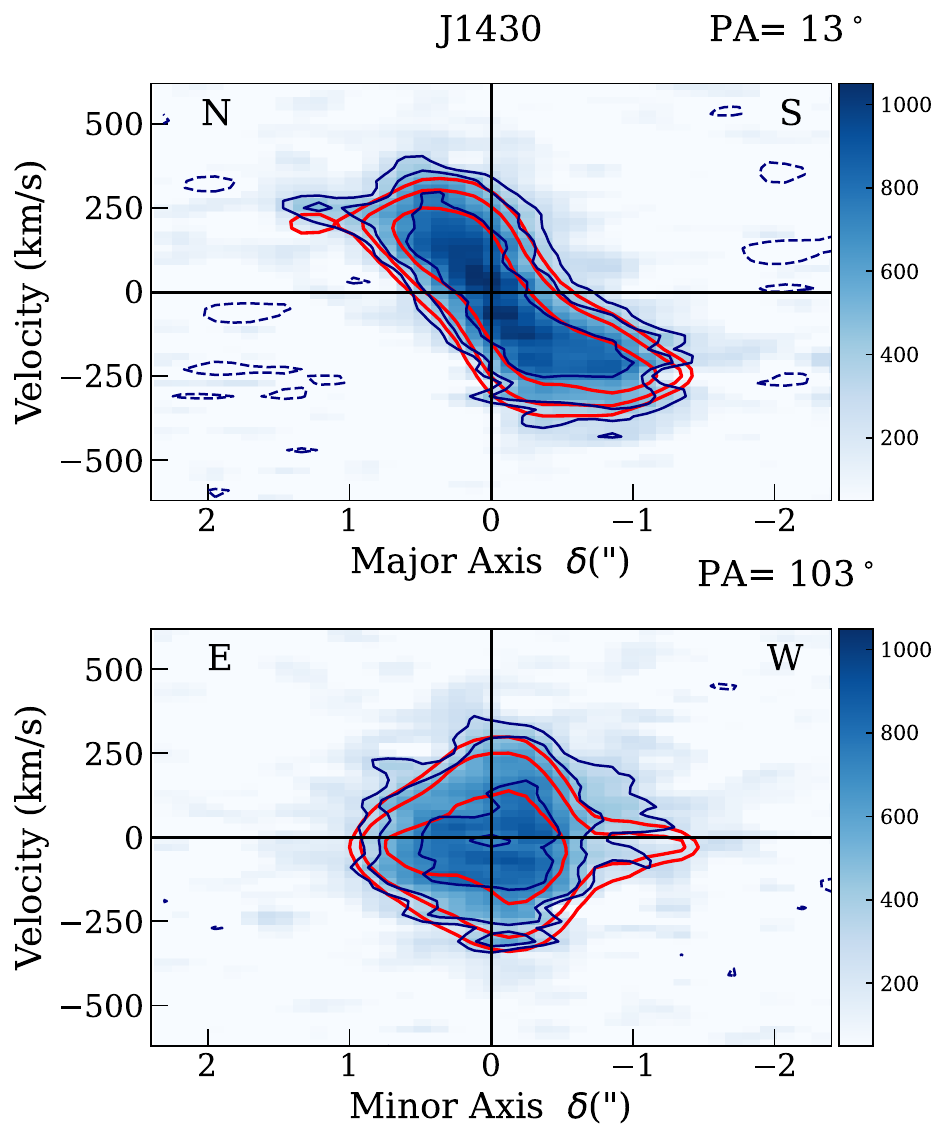}
\includegraphics{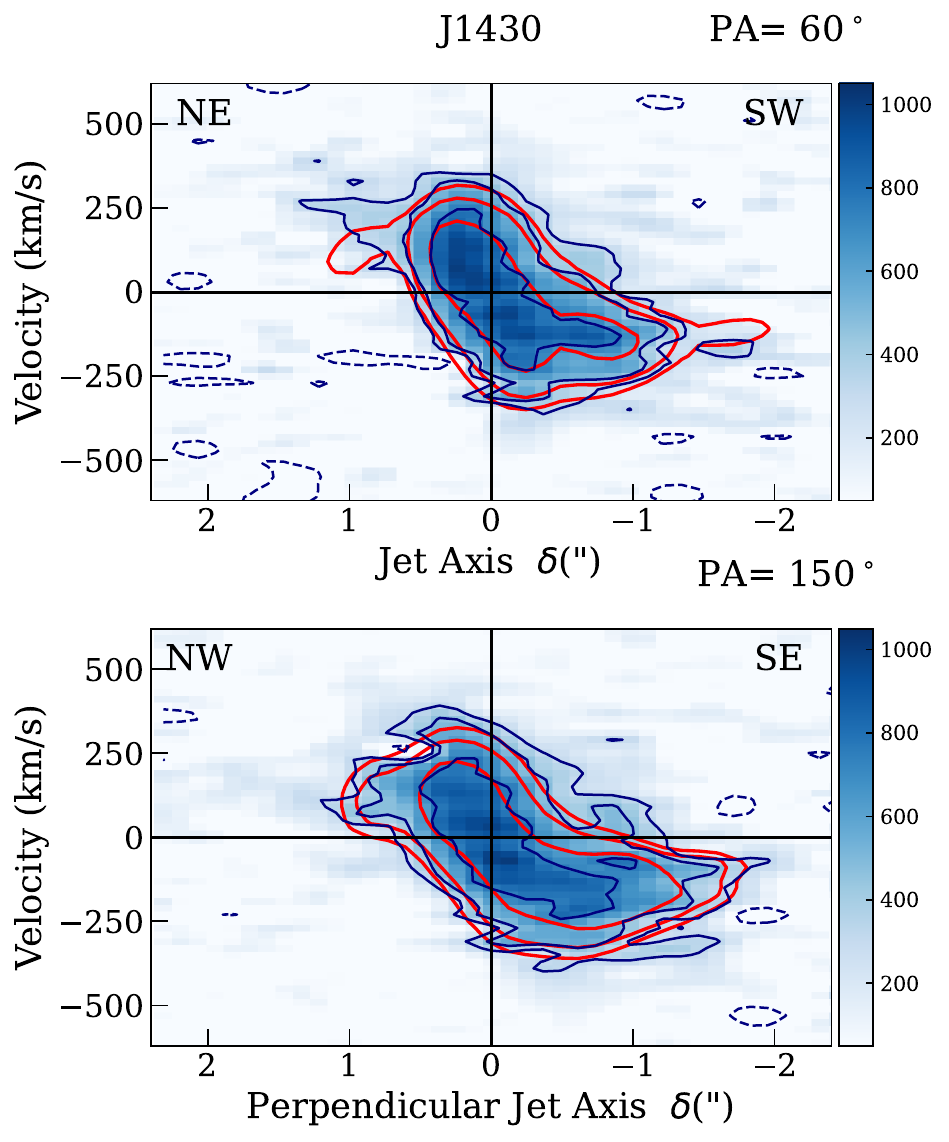}
\includegraphics{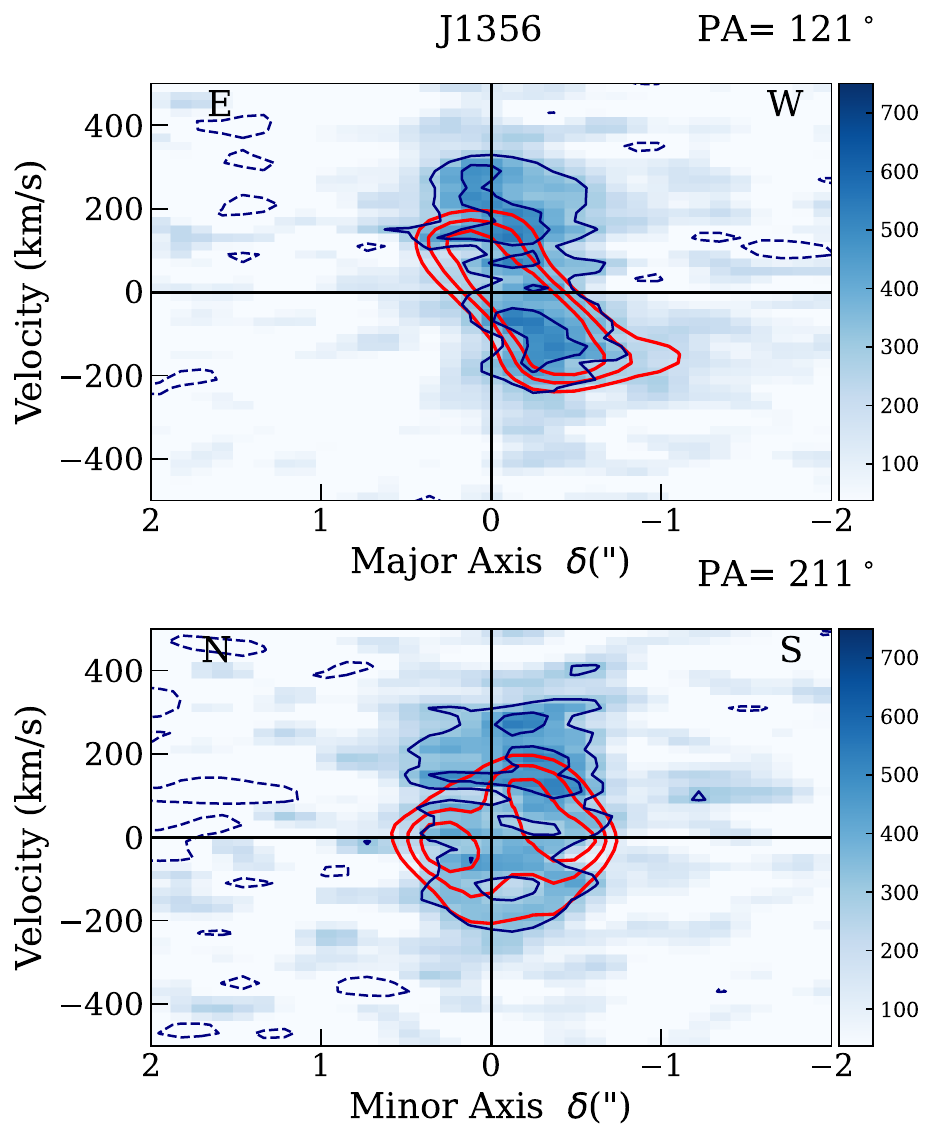}
}
\caption{Same as Figure \ref{fig:PVD} but for the \hd stacked data cubes. The red and blue contours are drawn at (2, 3, 5)$\sigma$ for J1430 and at (2, 3, 4)$\sigma$ for J1356, where $\sigma$ = 0.15 and $\sigma$ = 0.21 for J1430 and J1356, respectively.}
\label{fig:PVD_stacked}
\end{figure*}

\begin{figure*}
\resizebox{\hsize}{!}{
\includegraphics{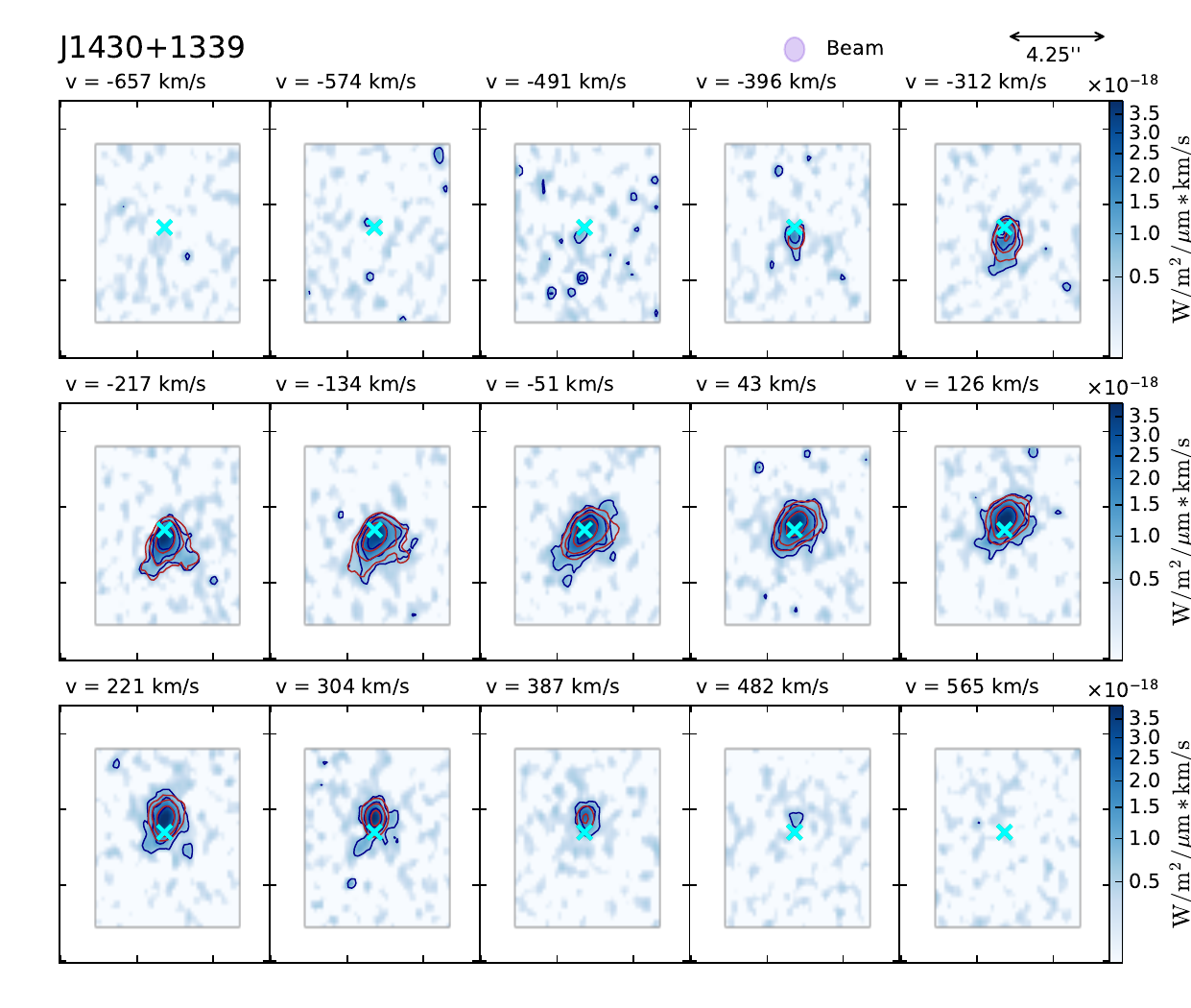}}
\caption{Channel maps of the \hdone~emission line of J1430. Color scale maps and blue contours trace the emission in SINFONI data cube while red contours trace the emission in the \barolo~model cube. Contours are drawn at (2, 4, 8, 16)$\sigma$.}
\label{fig:J1430_chanmap}
\end{figure*}

\begin{figure*}
\resizebox{\hsize}{!}{
\includegraphics{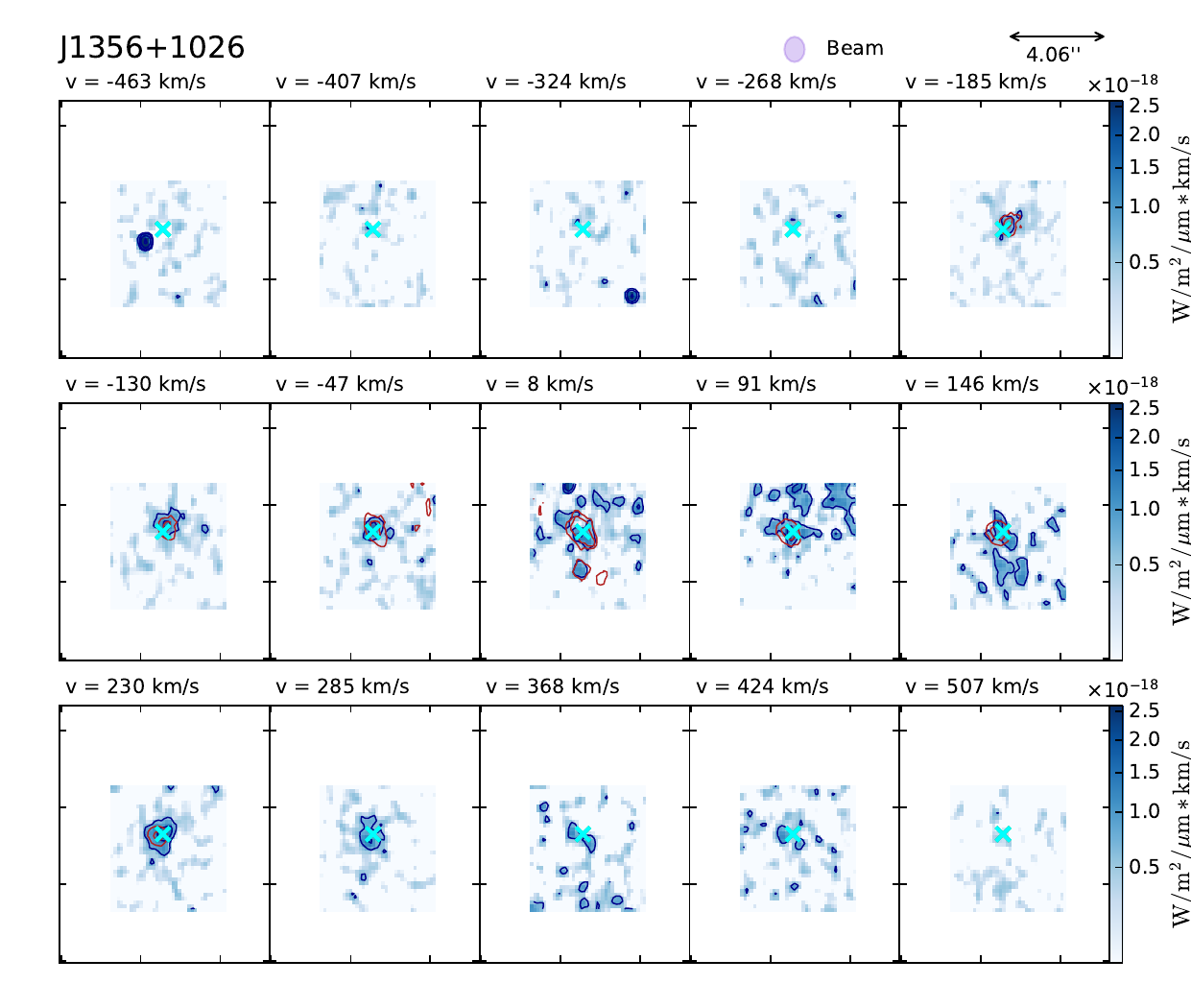}}
\caption{Same as Figure \ref{fig:J1430_chanmap} but for J1356. Contours are drawn at (2, 4, 8, 16)$\sigma$. }
\label{fig:J1356_chanmap}
\end{figure*}

\end{appendix}

\end{document}